\documentclass[twocolumn]{aastex631}

\usepackage{amsmath}
\usepackage{bm}

\graphicspath{{./}{Figures/}}
\begin{document}

\title{Characterising The Atmospheric Dynamics Of HD209458b-like Hot Jupiters Using AI Driven Image Recognition/Categorisation}

\author{F. Sainsbury-Martinez$^1$$^2$, P. Tremblin$^1$, M. Mancip$^1$, S. Donfack$^1$$^3$, E. Honore$^1$, M. Bourenane$^1$}

\affiliation{Université Paris-Saclay, UVSQ, CNRS, CEA, Maison de la Simulation, 91191, Gif-sur-Yvette, France, \\$^2$ School of Physics and Astronomy, University of Leeds, Leeds LS2 9JT, UK, \\$^3$ IDRIS, CNRS, 91403 Orsay, France}

\begin{abstract}
In-order to understand the results of recent observations of exoplanets, models have become increasingly complex. Unfortunately this increases both the computational cost and output size of said models. We intend to explore if AI-image-recognition can alleviate this burden. \\    
We used DYNAMICO to run a series of HD209458-like models with different orbital-radii. Training data for a number of features of interest was selected from the initial outputs of these models. This was used to { train a pair of multi-categorisation} convolutional-neural-networks {(CNN)}, which we applied to our outer-atmosphere-equilibrated models. \\ 
The features detected by our CNNs revealed that our models fall into two regimes: models with a shorter orbital-radii exhibit significant global mixing which shapes the entire atmospheres dynamics. { Whereas}, models with longer orbital-radii exhibit negligible mixing except at mid-pressures. Here, the initial non-detection of any trained features revealed a surprise: a night-side hot-spot. Analysis suggests that this occurs when rotational influence is sufficiently weak that divergent flows from the day-side to the night-side dominate over rotational-driven transport, such as the equatorial jet. \\
We suggest that image-classification may play an important role in future, computational, atmospheric studies. However special care must be paid to the data feed into the model, from the colourmap, to training the CNN on features with enough breadth and complexity that the CNN can learn to detect them. However, by using preliminary-studies and prior-models, this should be more than achievable for future exascale calculations, allowing for a significant reduction in future workloads and computational resources.
\end{abstract}


\section{Introduction} \label{sec:introduction}
Over the last decade models of Exoplanetary Atmospheres, particularly hot Jupiters, have become increasingly complex whilst also requiring increasingly large amounts of time to reach equilibrium (as deeper regions of the atmosphere are considered).\\
Briefly, this increase in model complexity (and hence computational resource requirements) has come about as a result of a desire to more accurately model atmospheres (specifically their chemical composition and radiative dynamics) in order to better match/recover/understand the high spectral resolution observations made possible by Hubble (e.g. \citealt{2000ApJ...529L..45C,2011ESS.....2.1206R}), JWST (e.g. \citealt{2022A&A...661A..80J,2022A&A...661A..81F,2022A&A...661A..82B,2022arXiv220713067P}), and other next-generation observing missions. 
As a result, whilst earlier (and longer time-scale) studies ignored chemistry and/or modelled radiative dynamics in an equilibrium sense (e.g. Newtonian Cooling - \citealt{Showman_2008,Rauscher_2010,2014GMD.....7.3059M}), { recently this has started to thanks to} next generation global circulation models (GCMs). These models now include explicit non-equilibrium chemistry, including cloud formation and condensation, and multi-banded, correlated-k, radiative transport schemes (see, e.g. \citealt{2016A&A...595A..36A,2021MNRAS.506.2695L,2022MNRAS.512.3759D}) which can simulate more accurate atmospheric feedback of evolving atmospheric dynamics. As a result, even when using high-performance, next generation, GCMS (such as LFRic - \citealt{2018arXiv180907267A,2020QJRMS.146.3917M}), both the time required to run the model as well as the total output data generated has increased exponentially. This increase in data requirements is further exasperated by recent studies (such as \citealt{2019A&A...632A.114S,2021A&A...656A.128S,10.1093/mnras/stad1905}) which have shown that accurately modelling both the outer atmosphere dynamics, as well as global observable features, also requires the model to include a (close to) equilibrium deep atmosphere (which is computationally expensive due to the long dynamical time-scales of deep atmospheric circulations).\\
Whilst some of this burden can be lifted via run-time data-pre-processing (e.g. the XIOS\footnote{http://forge.ipsl.jussieu.fr/ioserver} library, which can not only interpolate complex GCM grids onto a simple long-lat grid at run-time, but also perform temporal and spatial averages), this still leaves a lot of data that needs to be analysed. A time consuming endeavour which only becomes more so when paired with the need to monitor on-going models in order to either confirm their stability or determine when they have equilibrated. \\
Fortunately, recently developments in deep-learning (DL - a subset of general machine-learning methods, { which allows for input data and trained features to be encoded at various levels of abstraction}) driven image-classification provide a potential solution. Through the proper application of training data from GCM studies, it might be possible to train a DL-model to detect anticipated atmospheric features and dynamics, thus enabling a somewhat automated form of concurrent- or post-processing of simulation data. Note that we focus our investigation on image-classification models due to the maturity of algorithms in this area, driven by research into, for example, facial recognition, on-device image-recognition, and of course driverless cars.    \\

In this paper we will explore how DL driven image-classification can aid in our understanding of exoplanetary atmospheres. More specifically, we will explore image-classification for a series of HD209458b-like models based upon those first performed in \citet{2019A&A...632A.114S}, albeit with varying orbital radii (and hence surface irradiations and, synchronous, rotation rates) in order to expand upon the atmospheric features available for characterisation.  \\
The structure of this work is as follows: In \autoref{sec:Method} we introduce both the atmospheric models analysed here (which we calculate using the 3D GCM DYNAMICO - \citealt{gmd-8-3131-2015}) as well as the deep-learning model used for the post-processing image classification (which itself is based on a { multi-categorisation} convolutional neural-network implemented in TensorFlow - \citealt{tensorflow2015-whitepaper,tensorflow_developers_2023_8118033}). We follow this, in \autoref{sec:Results}, with our results: First we introduce some of the key features of exemplary HD209458b-like models, including the atmospheric features selected to train the DL-models. Note that that we have made the decision to use computer vision to analyse our results since a) this allows us to use the visualisations for both the models and our own analysis, and b) computer vision deep-learning algorithms are highly mature. 
Then, after applying the trained neural-networks to our atmospheric models, we explore their performance at tagging (i.e. identifying) atmospheric features of interest. This includes a discussion of how the failure of the neural-networks to detect any features at some pressures of our slowly-rotating model is in and of itself an interesting result which lead to the discovery of an uncommon/unusual atmosphere feature: a vertically coherent hot-spot on the night-side of our tidally-locked hot Jupiter. 
Finally we also discuss what features are easy/difficult for the network to detect, as well as how the format and breadth of the data being analysed impacts the final classifications - including factors ranging from the completeness of the training data-set, to the colourmaps used for each plot, which play an import role in transmitting information from the model to the CNN, with the choice of colourmaps playing a particularly outsized role in the detection of gradients in model data.
We finish, in \autoref{sec:Conclusions}, with concluding remarks, discussing the implications of our results and potential plans for future studies which will explore the atmospheric dynamics of variably rotating hot Jupiters in more detail.

\section{Methods} \label{sec:Method}
In order to investigate how image classification can aid in the concurrent- and post-processing of exoplanetary atmospheric models (particularly hot Jupiters), we must introduce both the atmospheric models we intend to analyse (\autoref{sec:HJ_models}) as well as the machine-learning model we will use to perform said analysis (\autoref{sec:AI}). 
\subsection{Hot Jupiter Models} \label{sec:HJ_models}
Following \citet{2019A&A...632A.114S,2021A&A...656A.128S} we use the GCM DYNAMICO to perform a series of 3D atmospheric models at different orbital radii (i.e. with different stellar-irradiation profiles and, tidally-locked, planetary rotation rates). Here we give a brief introduction to DYNAMICO, and its Newtonian Cooling approach to radiative transport (\autoref{sec:DYNAMICO}) before introducing the exact model setups considered here \autoref{tab:simulation_params}, including how the outer atmosphere, equilibrium, Newtonian Cooling profiles were calculated (\autoref{sec:ATMO}). 

\subsubsection{DYNAMICO} \label{sec:DYNAMICO}
DYNAMICO is a highly computationally efficient GCM that solves the primitive equations of meteorology ({see \citealt{Vallis17} for a review and \citealt{2014JAtS...71.4621D} for a more detailed discussion of the approach taken in DYNAMICO}) on a spherical, icosahedral, grid \citep{gmd-8-3131-2015}. It remains under-development as a next-generation dynamical core for Earth and planetary climate studies at the Laboratoire de Météorologie Dynamique and is publicly available\footnote{DYNAMICO is available at http://forge.ipsl.jussieu.fr/dynamico/wiki}.\\

In brief, DYNAMICO takes an energy-conserving Hamiltonian approach to solving the primitive equations of meteorology (see \citealt{2020SSRv..216..139S,2019ApJ...871...56M} for as discussion of the validity and limits to this approach). \\
Rather than the traditional latitude-longitude horizontal grid (which presents numerical issues near the poles due to singularities in the coordinate system - \citealt{WILLIAMSON2007}), DYNAMICO uses a staggered horizontal--icosahedral grid for which the total number of horizontal cells, $N,$ is defined by the number of subdivisions, $d,$ of each edge of the main spherical icosahedral\footnote{Specifically, to generate the grid we start with a sphere that consists of 20 spherical triangles (sharing 12 vertex, i.e. grid, points) and then we subdivide each side of each triangle $d$ times using the new points to generate a new grid of spherical triangles with $N$ total vertices. These vertices then form the icosahedral grid.}:
\begin{equation}
  N=10 d^2 + 2.
\end{equation}
In all the models considered here, we set the number of subdivisions to 30, which results in a total horizontal resolution of 9002 cells. This corresponds to an angular resolution of approximately $2.4^\circ$. Additionally, at run-time, the output data is passed to XIOS which converts the horizontal--icosahedral grid onto a regular lat-long grid, with a resolution of 90x180 (i.e $2^\circ$) in the lat/long directions respectively, in order to simplify analysis. \\
Vertically, DYNAMICO uses a pressure coordinate system whose levels can be defined by the user at runtime. In our models, this means 33 pressure levels that are linearly spaced in $\log\left(P\right)$ space  between $10^{-3}$ and $200$ bar. \\ 
Finally, the boundaries of our simulations are closed and stress-free with zero-energy transfer (i.e. the only means of energy injection and removal are the horizontal numerical dissipation, i.e. hyperviscosity, and the Newtonian Cooling thermal relaxation scheme - see \autoref{sec:ATMO}). \\

We introduce the aforementioned horizontal numerical dissipation in order to stabilise the system against the accumulation of grid-scale numerical noise. This takes the form of a horizontal hyperdiffusion filter with a fixed hyperviscosity and a dissipation timescale at the grid-scale, $\tau_{dissip}$, which acts to adjust the strength of the filtering (i.e. the longer the dissipation timescale, the weaker the dissipation strength). \\
For all the models presented here, we set a horizontal dissipation timescale of $\tau_{dissip}=2500$ following the arguments of \citet{2021A&A...656A.128S} - a series of test cases with both faster and slower numerical dissipation were considered, and other than in the most extreme rotation cases (where we found that rapid dissipation lead to model instabilities - hence our choice to set $\tau_{dissip}=2500$ rather than $\tau_{dissip}=1250$) the results were found to be essentially $\tau_{dissip}$ independent.  \\
Note that this hyperviscosity is not a direct equivalent of the physical viscosity of the planetary atmosphere, but rather can be viewed as a form of increased artificial dissipation that both enhances the stability of the model and somewhat accounts for sub-grid-scale dynamics. This approach, known as the large eddy approximation, has long been standard practice in both the stellar (e.g. \citealt{2005LRSP....2....1M}) and planetary (e.g \citealt{doi:10.1098/rsta.2008.0268}) atmospheric modelling communities. \\

Finally, since DYNAMICO (like many other GCMs) does not include a dynamic time-step, the time-step for each model had to be manually set. For the models considered here, we simply followed prior studies and set the time-step to 120 seconds. Note that test cases with shorter time-steps were explored, with little to no difference found in the observed dynamics.

\subsubsection{HD209458b-like Model Atmospheres} \label{sec:ATMO}
In our HD209458b-like atmospheric models, we do not directly model either the incident thermal radiation on the day-side or the thermal emission on the night-side of the exoplanet. This is due to the high computational cost of modelling radiative dynamics directly with current-generation GCMs and the preliminary-science status of next-generation GCMs (e.g. LFRic). Instead we use a simple thermal relaxation scheme to model these radiative effects, with a spatially varying equilibrium temperature profile, $T_{eq}$, and a radiative relaxation timescale, $\tau_{rad}$, that increases with pressure throughout the outer atmosphere. Specifically, we model the radiation by adding a source term to the temperature evolution equation which takes the form 
\begin{equation}
\frac{\partial T\left(P,\theta,\phi\right)}{\partial t} = - \frac{T\left(P,\theta,\phi\right)-T_{eq}\left(P,\theta,\phi\right)}{\tau_{rad}\left(P\right)} \,.
\end{equation}
This method is known as Newtonian cooling and has long been applied within the 3D GCM exoplanetary community (i.e. \citet{2002A&A...385..156G}, \citet{Showman_2008}, \citealt{Rauscher_2010}, \citet{2011ApJ...738...71S}, \citealt{2014GMD.....7.3059M}, \citealt{GUERLET2014110}, or \citealt{Mayne_2014}) when a more complete/complex treatment of the radiative dynamics is unfeasible. \\
However in order to use this approach we must first find/set $T_{eq}\left(P,\theta,\phi\right)$ for every orbital radii (i.e. every stellar insolation rate). At first glance, the simplest solution would be to use a 1D atmospheric model (like ATMO - \citealt{2015ApJ...804L..17T,2016A&A...594A..69D}) to calculate day-side and night-side profiles for each hot Jupiter model at each orbital radii. However, as discussed in \citet{2021A&A...656A.128S}, this technique has a number of downsides, resulting in overly strong outer atmosphere dynamics thanks to an exaggerated day/night temperature difference (which itself results from the 1D models lacking horizontal advection which cools the day-side and heats the nightside). Similarly, we cannot use the solution proposed in \citet{2021A&A...656A.128S} since these models do not represent real exoplanets and hence we lack observations of the advected/observed day/night temperature contrast. \\
\begin{table}[tbp]
\centering
\small
\begin{tabular}{c|c|ccc}
$a$ (AU) & $\Delta T_0$ (K) & \multicolumn{3}{|c}{$T_{eq}$ Profile Interp Points $\left(\frac{T_{eq}}{1 \textrm{K}},\frac{P}{1 \textrm{bar}}\right)$}                                                                \\ \hline
 0.021  & 800 & $(1300,10^{-6})$ & $(1800,10^{-2})$ & $(2600,10)$  \\
 0.192  & 260 & $(360,10^{-6})$ & $(500,10^{-2})$ & $(1400,10)$ \\
\end{tabular}
  \caption{Equilibrium temperature profiles for our `hot' and `cool' HD209458b-like atmospheric models.}
  \label{tab:equilbirum_temperature}
\end{table}
\begin{table}[tbp]
\centering
\small
\begin{tabular}{c|ccc}
$a$ (AU) & \multicolumn{3}{|c}{$\tau_{rad}$ Profile Interp Points $\left(\log\left(\frac{\tau}{1 \textrm{sec}}\right),\frac{P}{1 \textrm{bar}}\right)$}                                                                \\ \hline
 0.021  & $(1.0,10^{-6})$ & $(2.9,10^{-2})$ & $(7.5,10)$  \\
 0.192  & $(2.7,10^{-6})$ & $(4.2,0.1)$ & $(7.6,10)$ \\
\end{tabular}
  \caption{Newtonian cooling radiative timescale profiles for our `hot' and `cool' HD209458b-like atmospheric models.}
  \label{tab:radiative_timescale}
\end{table}

\begin{figure*}
\gridline{\fig{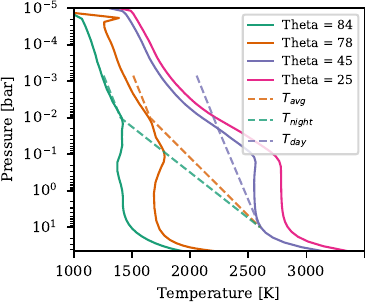}{0.45\textwidth}{a) `Hot': 1D and equilibrium T-P profiles  \label{fig:T_P_0021}}
          \fig{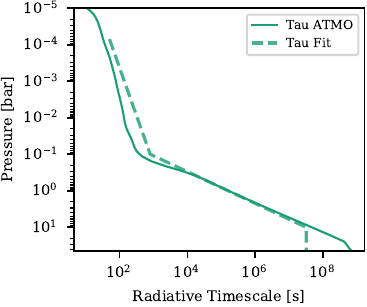}{0.45\textwidth}{b) `Hot': radiative timescales   \label{fig:Radiative_0021}}}
\gridline{\fig{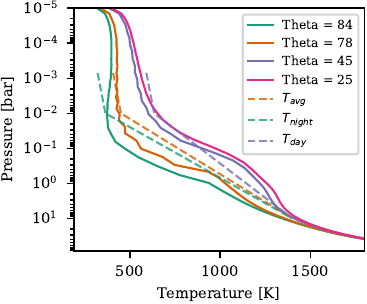}{0.45\textwidth}{c) `Cool': 1D and equilibrium T-P profiles  \label{fig:T_P_0192}}
          \fig{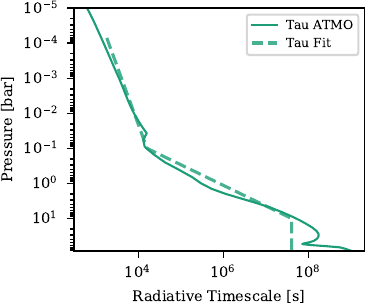}{0.45\textwidth}{d) `Cool': radiative timescales   \label{fig:Radiative_0192}}}
\caption{Equilibrium temperature-pressure (left) and radiative timescale (right) profiles for the Newtonian Cooling schemes used in our  exemplary `hot' 
(top) and `cool' (bottom)  HD209458b-like atmospheric models. Note that in the above plots, the background solid lines correspond to the 1D ATMO models upon which the equilibrium profiles (dashed lines) are based/extrapolated.  \label{fig:1D_equilibirum_profiles}}
\end{figure*}

As such we must find an alternate method to define $T_{eq}\left(P,\theta,\phi\right)$. Fortunately, since we are interested in the general dynamics of the atmosphere at different orbital radii rather than specific features in comparison to observations, we can take a slightly { parametrised} approach to deriving { an approximate} $T_{eq}\left(P,\theta,\phi\right)$ profile from 1D atmospheric models. This approach can be understood as follows: Using ATMO we run a series of 1D models of HD209458b at 12 different latitudes between the sub-stellar point (which is the point of the tidally-locked planets atmosphere which is closest to the host star) and the anti-stellar point (i.e. the point on the cold night-side furthest from the host star) - these models used HD209458-like stellar irradiation profiles based upon the work of \citet{2003IAUS..210P.A20C}. { We then try to match which 1D models best match the known HD209458b temperature-pressure profile at different points in the atmosphere. Specifically, and in order to recreate the $T_{eq}\left(P,\theta,\phi\right)$ profile from \citet{2019A&A...632A.114S}, we try to match the day-side, equilibrium, and night-side temperature in the outer atmosphere (i.e. reproducing the observed day/night temperature contrast at $10^{-2}$ bar) and the radiative-advective boundary temperature in the deep atmosphere (i.e. the temperature at 10 bar, where the day-night temperature difference is assumed to vanish.) The result is that the temperature at 10 bar is best fit with a 1D profile with a irradiation angle of 45 degrees (i.e. halfway between the substellar point and the terminator) and the outer atmosphere day/equilibrium/night-side temperatures (at $10^{-2}$ bar) are best fit by models with irradiation angles of 25/78/84 degrees respectively. As a consequence of this, and in order to estimate/calculate $T_{eq}\left(P,\theta,\phi\right)$ for our HD209458b-like models at different orbital radii, specifically radii of $0.021$AU and $0.192$AU, we next run the aforementioned 1D models for each of the orbital radii of interest. With this, assuming that the same points in the 3D and 1D models continue to coincide, we can now extract the estimated day/equilibrium/night-side and convergence temperatures, and use them to create parametrised $T_{eq}\left(P,\theta,\phi\right)$ profiles, as shown in \autoref{fig:1D_equilibirum_profiles} and \autoref{tab:equilbirum_temperature}. At the same time, we can also calculate the radiative timescale by perturbing  one of the temperature-pressure profiles (specifically the profile with a 45 degree irradiation angle) at every pressure level and measuring the time taken for the profile to settle (see \citealt{Showman_2008} for details about this approach). We plot the profile calculated by ATMO, as well as its linear in $\log(P)$ parametrisation, in \autoref{fig:Radiative_0021}/d, with the values at the interpolation points given in \autoref{tab:radiative_timescale}.\\}
Note that in the aforementioned fits, $T_{eq}\left(P,\theta,\phi\right)$, has been split into two pressure-dependent components. { This was done for compatibility reasons with DYANMICOs implication of Newtonian cooling.} The first is simply the 1D equilibrium profile, $T_{eq-1D}\left(P\right)$, which we define using a series of linear in $\log(P)$ space interpolations (with interpolation points given in \autoref{tab:equilbirum_temperature}). The second is the day/night temperature difference $\Delta{T}(P)$ which takes the form:
\begin{align}
\Delta T (P) &=\left\{ \begin{array}{ll}
  \Delta T_0 & \textrm{if } P<10^{-2} \\
 \Delta T_0 \log (P/10^{-2}) & \textrm{if } 10^{-2}  < P < 10 \\
 0 & \textrm{if } P > 10.  \end{array}
\right. \, \label{eq:deltaT}
\end{align}
Taken together, $T_{eq}\left(P,\theta,\phi\right)$, is then given by a position dependent combination of the two profiles:
\begin{align}
T_{eq}\left(P,\theta,\phi\right) &= T_{eq-1D}\left(P\right) - \frac{\Delta{T}(P)}{2} \notag \\
&+ \Delta{T}(P) \cos\left(\theta\right)
\max \left[ 0, \cos (\phi - \pi) \right] \, .
 \label{eq:Teq}
\end{align} 
\\
\begin{table}[tb!]
  \centering
  \footnotesize
  \def\arraystretch{1.0}
  \begin{tabular}{c|c|cc}
    Quantity (units) & Description & `Hot' HJ & `Cool' HJ \\
    \hline \hline
    dt ($s$) & Time-step & \multicolumn{2}{c}{120} \\
    $N_z$ & No. of Pressure Levels & \multicolumn{2}{c}{33} \\
    $d$ & No. of Sub-divisions & \multicolumn{2}{c}{30} \\
    $N^\circ$ & Angular Resolution & \multicolumn{2}{c}{$~2.4^\circ$}\\
    $P_{top}$ (bar) & Pressure at Top  & \multicolumn{2}{c}{$7 \times 10^{-3}$}  \\
    $P_{bottom}$ (bar) & Pressure at Bottom & \multicolumn{2}{c}{200}  \\
    $g$ ($ms^{-1}$) & Gravity & \multicolumn{2}{c}{8.0} \\
    $R_{HJ}$ (m) & Planetary Radius  & \multicolumn{2}{c}{$10^8$} \\
    $a$ (au) & Orbital Radius & $0.021$ & 0.192 \\
    $\Omega$ ($s^{-1}$) & Angular Rotation Rate & $7 \times 10^{-5}$ & $2.54 \times 10^{-6}$ \\
    $c_p$ ($jkg^{-1}K^{-1}$) & Specific Heat & \multicolumn{2}{c}{13226.5} \\
    $\mathcal{R}$ ($jkg^{-1}K^{-1}$) & Ideal Gas Constant & \multicolumn{2}{c}{3779.0} \\
    $T_{init}$  (K)  & Init Adiabatic T @ 10b & 1400 & 2600 \\
  \end{tabular}
  \caption{Parameters for HD209458b-like simulations}
  \label{tab:simulation_params}
\end{table}

\subsection{Deep-Learning Setup} \label{sec:AI}
\begin{figure*}[tbp] %
\begin{centering}
\includegraphics[width=0.99\textwidth]{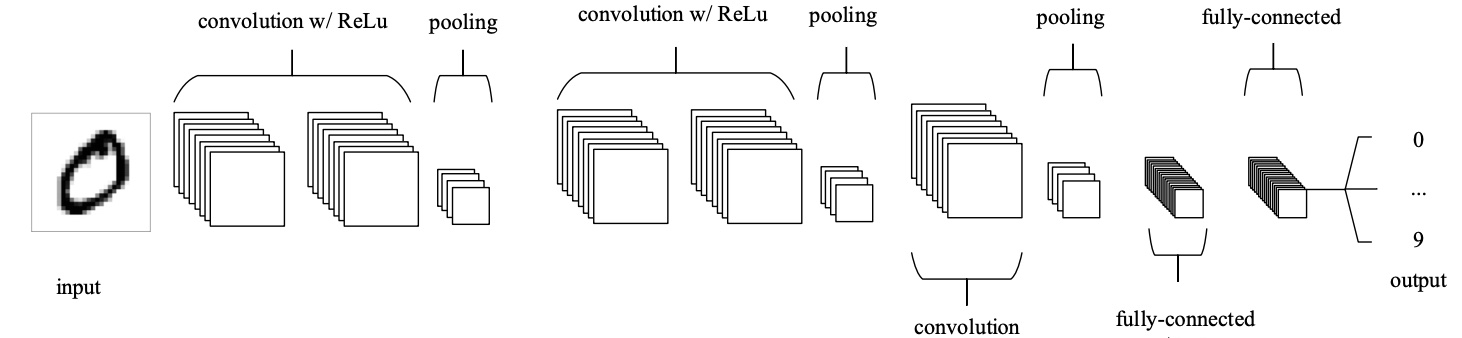}
\caption{ An example CNN layer structure used for modern optical character recognition. Note the use of stacked, and hence non-linear, convolution layers which are separated by dimensionality-reducing pooling layers, and which eventually feed into fully connected neural-network layers which perform the final image classification.
Reproduced from \citet{2015arXiv151108458O}.  \label{fig:CNN} }
\end{centering}
\end{figure*}
\begin{figure}[tbp] %
\begin{centering}
\includegraphics[width=0.90\columnwidth]{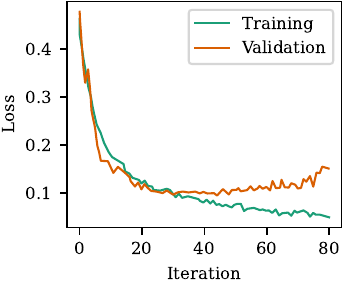}
\caption{Convergence tests for our multi-categorisation CNN, i.e. thermal features CNN, showing the convergence of the loss function during both training (green) and validation (orange). The training is considered complete/optimised after 30 iterations since this is the point at which overfitting starts to occur, as represented by a regression in validation accuracy (i.e. increase in loss).   \label{fig:model_comparison} }
\end{centering}
\end{figure}
In order to autonomously analyse the outputs of our HD209458b-like atmospheric models, and hence enable the run-time detection of key/interesting atmospheric features, we decided to make use of a convolutional neural network, i.e. a CNN, (specifically the Keras CNN which is implemented as part of TensorFlow). The approach we take here is based upon the work of \citet{DeepLearningforSpatiallyExplicitPredictionofSynopticScaleFronts}, who used a deep-learning model to identify large-scale (i.e. synoptic-scale) fronts in meteorological data, modify so as to be able to detect multiple different events (tags) from each image. { A more indepth discussion of the structure of the lightweight CNN considered here is given in the Appendix, here we give a broader overview.}\\
Briefly a convolutional neural network is a type of DL algorithm (which themselves are a subset of machine-learning data analysis tools) that is particularly well suited to image recognition, and whose design was inspired by how neurons in the visual cortex behave \citep{726791}. The key feature which separates a CNN from other image classification algorithms is, as the name would suggest, the inclusion of convolution layers. These are neural-network layers whose primary purpose is the extraction and isolation of both low-level and high-level features from input data, resulting in a reduction in the size of the data set and thus allowing for the final, reduced, image classification to be performed by more conventional, that is to say fully-connected, neural-network layers. 
This reduction in the size of the data-set to be worked on is crucial because, for the type of images we are interested in classifying, the resolution means that each node (which are the equivalent of neurons) in a fully-connected neural-network layer would contain far too many weights (i.e. learnable parameters) to fit in even the largest GPUs memory. Consider, for example, a fully connected neural-network layer analysing a 128x128 colour image, to maintain complete connection, each {\it individual} node would have to contain 49152 weights, ballooning the memory and computational footprint of the network. \\
Briefly, this decrease in data dimensionality takes place via alternating convolution layers (which do the actual feature analysis and detection) with pooling layers in between (which reduce the dimensionality of the data). Note that, as the dimensionality of the data is reduced, the complexity of the upcoming convolutional layers is typically increased. See \autoref{fig:CNN} for an example of how the layers in a typical OCR CNN are arranged, { \autoref{fig:Ai_flowchart} for a flowchart showing how the layers in our CNN are arranged}, and \citet{fukushima1982neocognitron,lecun1989backpropagation,yann1998,NIPS2012_c399862d,2015arXiv151108458O,8308186} for a more in-depth discussion of both the origins of CNNs, as well their specialised neural-network layers. \\
But how does the CNN model know what features to extract? Via training. Training, specifically supervised training, is the process by which a `blank' neural network is provided a data-set in which the expected tags (i.e. classifications) for each image are known - in our case the total training data set consisted of { at least} 50 images for each tag, although in some cases those images were artificially generated, via an over-sampling approach, in-order to expand the available data set (see \autoref{sec:AI_categories_new} for details of how and why this was done). { Note that, due to the configuration of our models, the bottom 10 (P$>\sim25$ bar) and top 10 (P$<\sim10^{-2}$ bar) layers where not considered when creating the training set, the former because the deep atmosphere had not equilibrated at the time, and the latter because we already had a wealth of features representing the very outer atmospheres dynamics (an unadvected day-side hotspot which we refer to as locked).} The network is then evolved, via a sequence
of forward and backward data propagation that uses a gradient descent algorithm to optimize a stochastic differentiable manifold (i.e. a surface of differential equations) to solve a very non-linear problem, until it is able to reproduce the set of known tags with its own set of assigned tags. For all the models discussed here this training was complete after approximately 30 epochs (i.e. iterations) with either no improvement or a regression in model accuracy (i.e. the ability of the model to reproduce the training data identified tags without being guided) if the model was evolved further - a phenomenon known as over-fitting. { For the CNN(s) discussed here our initial set of hand selected/identified data was split into separate training and validation pools, with 80\% of the tagged images being used to train the CNN and 20\% of images used for validation. The results of this validation procedure can be seen in \autoref{fig:model_comparison}, which revels that the trained CNN(s) are able to recover approximately 90\% of expected features in the validation data set after 30 epochs of evolution, after which accuracy regressed.    } \\
After this rather computationally expensive training process, { we were left with a pair of multi-classification CNNs, one for thermal features and one for wind features,} that could rapidly analyse the results of new/evolved simulations. We discuss this analysis, as well as potential limitations/pitfalls (and how to overcome them) below.

\section{Results} \label{sec:Results}
In order to explore the the use of machine learning and image classification for the autonomous detection and characterisation of hot Jupiter atmospheres, we ran a large series of synchronously rotating, HD209458b-like, hot Jupiter atmospheric models with orbital radii between $0.012\textrm{au}$ and $0.334\textrm{au}$. { This orbital radii range was selected to span from near the inner edge of the distribution of known exoplanets (where irradiation is strong and synchronous rotation is rapid) out towards the upper limit of the highly irradiated and synchronously rotating regime (which occurs when tidal effects are weak enough that the synchronisation time approaches the system age). Within this range, orbital radii were selected to be a whole fraction/multiple of the orbital radius of HD209458b. } \\
Early outputs of these { variable orbital-radii} models were then manually tagged with atmospheric features of interest (see \autoref{sec:AI_categories_init}/\autoref{sec:AI_categories_new}), after which this set of tagged data was used to train the CNNs. Finally, after the simulations had been run further and the outer atmospheres ($P<1$bar) had equilibrated, the full time-series of temporally averaged ( in order to remove small-scale fluctuations) horizontal wind and temperature maps was feed into the CNNs and a series of { multi-}categorisation maps was generated - here we focus on this process, and its results, for two HD209458b-like models in different rotation/insolation regimes, one `hot', with a short orbital radius and hence strong surface irradiation, and one `cool', with a longer orbital radius and hence weak surface irradiation. We also briefly explore the ability of our trained CNN to detect atmospheric features for the adiabatically initialised model of \citet{2019A&A...632A.114S} - a model which is expected to contain examples of all the atmospheric features of interest. Finally, we also discuss the importance of selecting proper training/{ validation} data for a CNN, including a discussion of how the initial categorisation of our hot Jupiter atmospheres lead to the discovery of an uncommon/unusual, but robust, atmospheric feature on the night-sides of our `cool' regime HD209458b-like atmospheric models. 
\subsection{HD209458b-like Atmospheric Models} \label{sec:hd209-like-models}

\begin{figure*}
\gridline{\fig{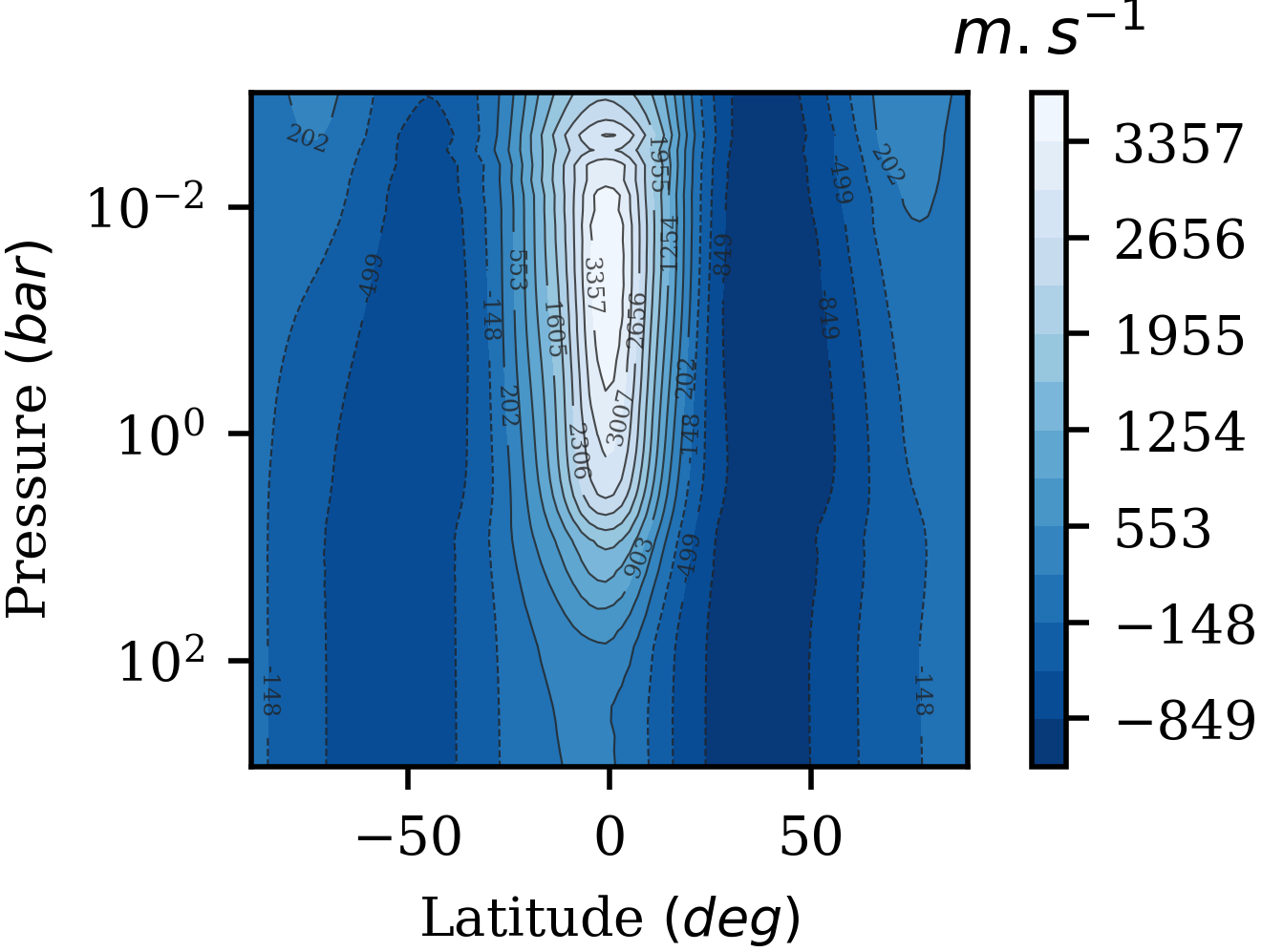}{0.45\textwidth}{a) `Hot': zonal wind}
          \fig{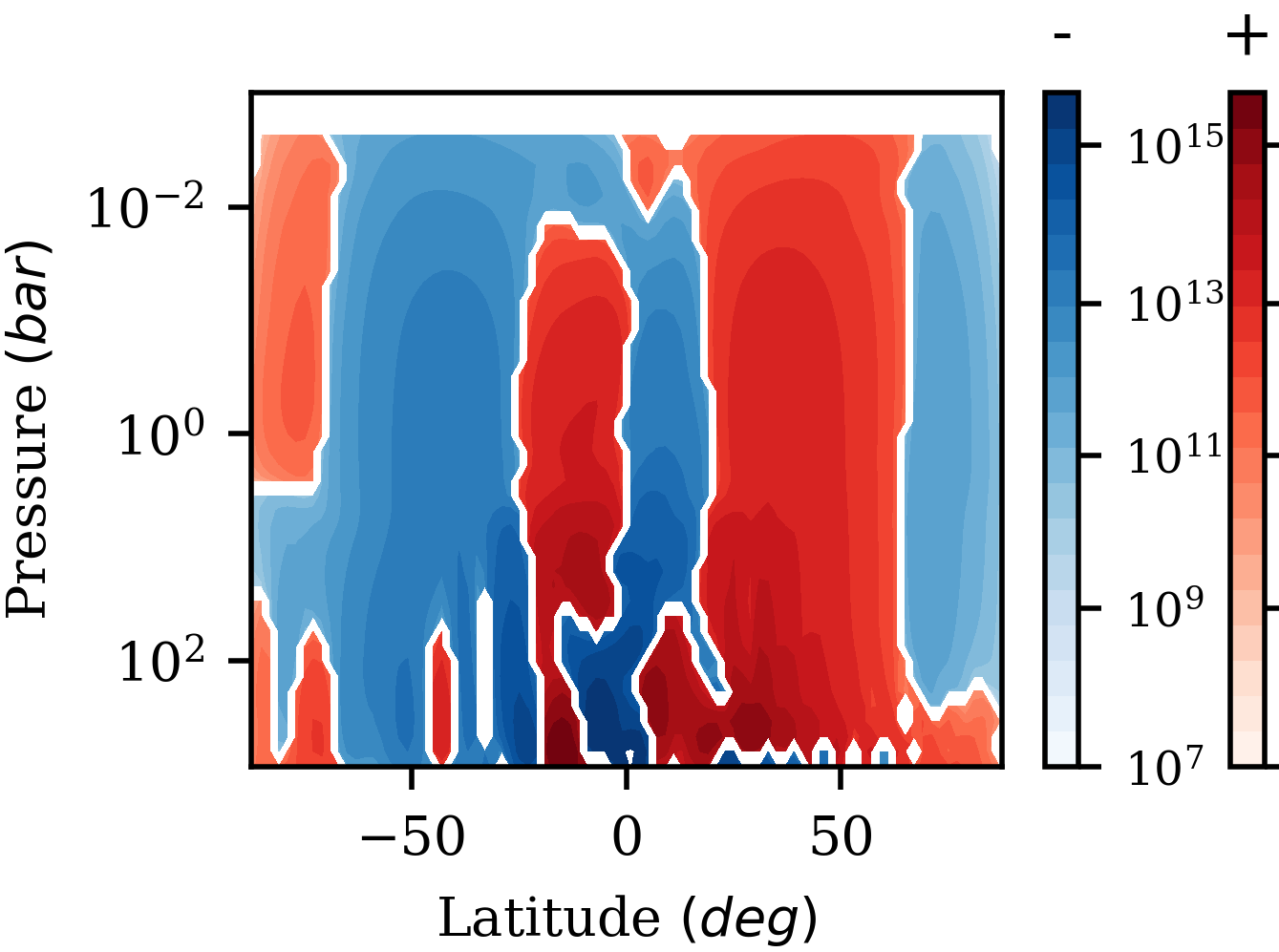}{0.45\textwidth}{b) `Hot': meridional circulation streamfunction}}
\gridline{\fig{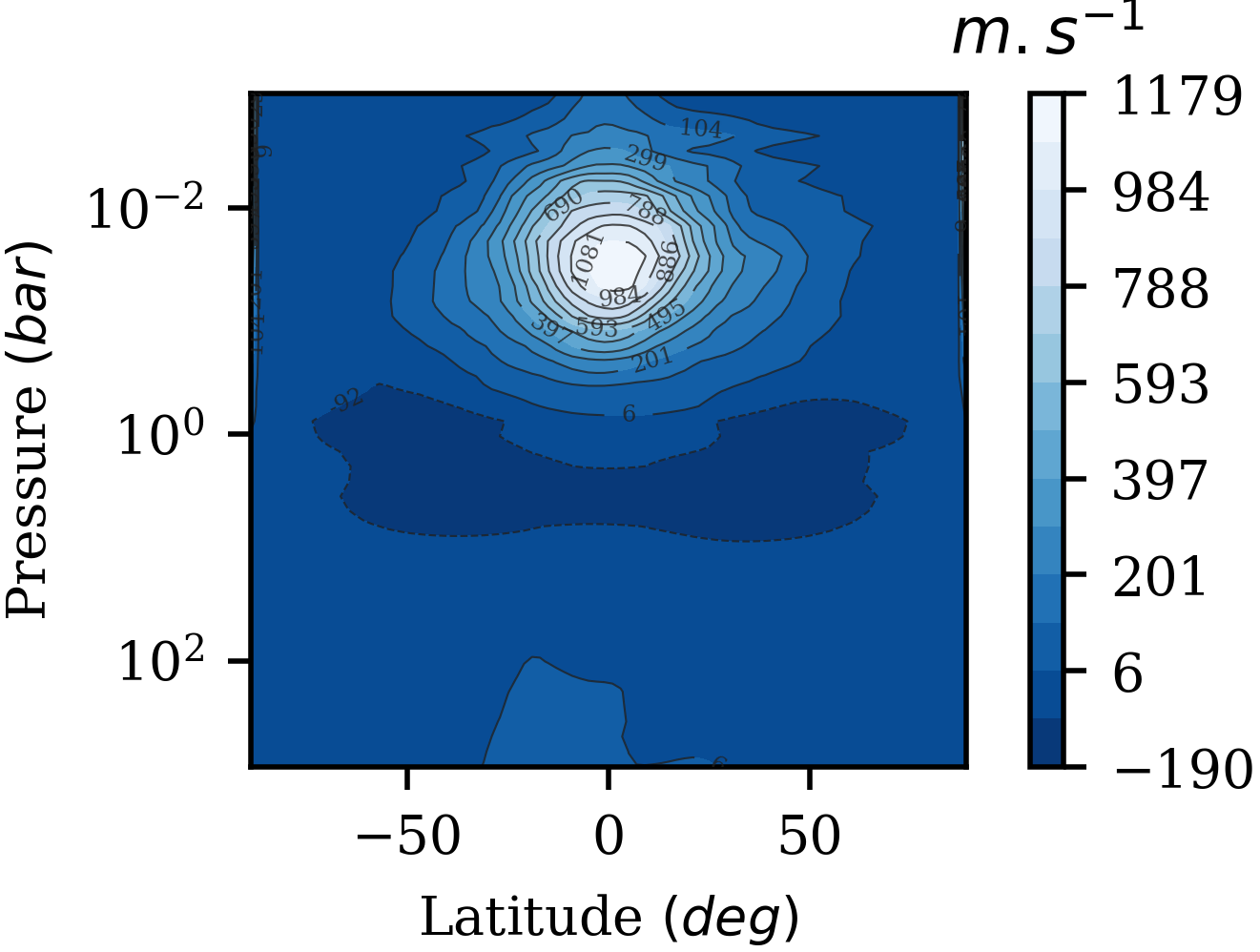}{0.45\textwidth}{c) `Cool': zonal wind}
          \fig{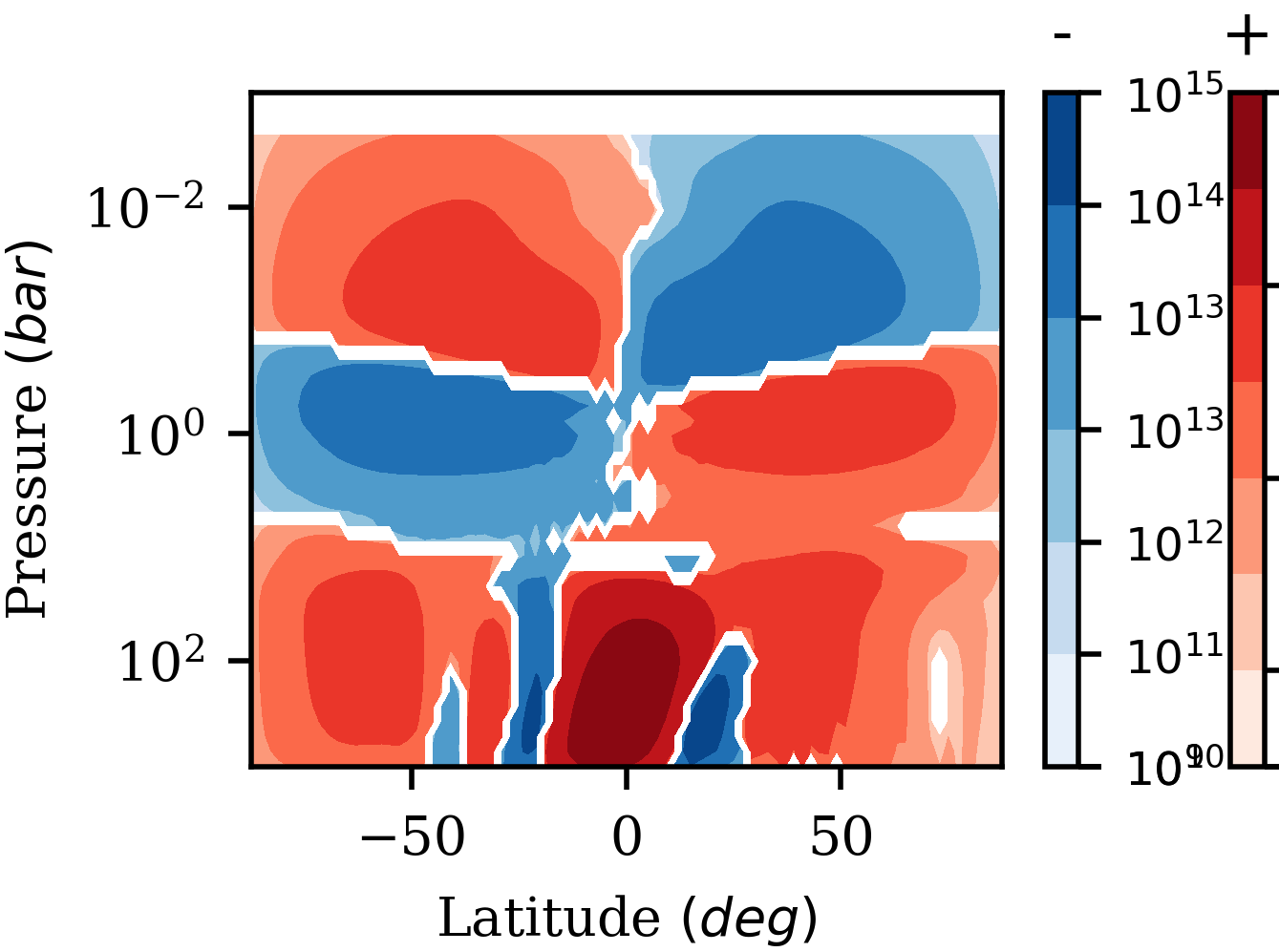}{0.45\textwidth}{d) `Cool': meridional circulation streamfunction}}
\caption{Longitudinally and temporally averaged zonal wind (left) and meridional circulation streamfunction (right) for both our `hot' 
(top) and `cool' (bottom)  HD209458b-like atmospheric models. In the zonal wind maps, positive quantities correspond to eastward flows, whilst in the meridional circulation profiles clockwise circulations are shown in red and anti-clockwise in blue. Additionally the meridional circulation streamfunction is plotted on a log scale. \label{fig:Zonal_wind_streamfunction}}
\end{figure*}

For the sake of brevity we have chosen to focus our analysis and discussion on two HD209458b-like models which, together, exhibit all the key features found in the broader set of models explored. Specifically we have chosen one model in each of the primary dynamical regimes observed: a short-orbital radius, $a=0.021\textrm{au}$, model (labelled `hot') in which the dynamics are strongly influenced by both the strong stellar irradiation and significant rotational effects (i.e. strong Coriolis forcing), and a long-orbital radius, $a=0.192\textrm{au}$, model (labelled 'cool') in which the stellar insolation is significantly weakened, and the dynamics are much less rotationally influenced, leading to a weaker equatorial jet and a more divergence driven day/night wind.  \\

The difference in dynamics between these two primary regimes of interest can easily be identified when exploring the zonal-mean (i.e. longitudinal-mean) zonal wind and meridional (i.e. vertical and latitudinal) streamfunction (where the meridional streamfunction is a measure of mass circulations on the meridional/latitudinal plane). We plot both of these quantities in \autoref{fig:Zonal_wind_streamfunction} for our two exemplary HD209458b-like atmospheric models.\\

Starting with the zonal wind (left) we find that, at steady-state, our `hot' model (top) maintains a strong, deep, super-rotating (i.e. easterly) equatorial jet which is braced by significantly weaker, mid-latitude, westerly counterflows. This is reminiscent of a mix of the wind structure found when modelling HD209458b and SDSS1411b with DYNAMICO \citep{2019A&A...632A.114S,2021A&A...656A.128S} and matches with the leading theory for the driving mechanism of super-rotating jets in hot Jupiters: the pumping of easterly angular momentum from mid-latitudes to the equator by standing Rossby and Kelvin waves, resulting in easterly acceleration at the equator and westerly acceleration at mid-latitudes \citep{2011ApJ...738...71S,2014ApJ...793..141T}.  \\
Moving onto the slower rotating, and more weakly radiatively driven, `cool' model (bottom), the zonal-mean zonal wind profile reveals that whilst a super-rotating `jet' appears to have once again formed, it is both significantly more vertically constrained and slower than the jet found in either the `hot' model or prior HD209458b models. We will explore why these differences occur in more detail below (\autoref{fig:Zonal_wind_streamfunction}), but briefly it can be linked to the relative strength of divergence driven (i.e. day-night) in comparison to rotationally driven (i.e. Rossby and Kevin wave driven) flows - as divergent/rotational flows become stronger/weaker, respectively, the zonal-mean zonal jet weakens since divergent flows have little to no east/west preference, leading to them broadly cancelling out in the zonal-mean, and rotational winds are key to driving a super-rotating jet due to their role in angular momentum transport. \\

In turn these very different zonal winds drive very different meridional circulations. In our `hot' model, the meridional circulation streamfunction reveals a pair of narrow clockwise (northern hemisphere - negative latitudes) and anti-clockwise (southern hemisphere) circulation cells which drive a downflow at the equator and, in conjunction with weaker high-latitude circulation cells, upflows at low to mid latitudes - specifically those latitudes at which the zonal wind transitions from being primarily easterly to 
 westerly. In turn, as discussed in \citet{2019A&A...632A.114S}, this circulation drives a significant vertical heat flux leading to the heating of the deep, advective, adiabat, and hence radius inflation (see \autoref{fig:Longitudinal_T}a) thanks to the increased internal entropy \citep{2017ApJ...841...30T,2019A&A...632A.114S}. \\
On the other hand, in our `cool' model, the circulation is rather different. Not only are the circulations cells much wider latitudinally, with each cell occupying an entire hemisphere, but the circulation direction of the cells changes with pressure. In the outer atmosphere we have a clockwise/anti-clockwise circulation cell in the northern/southern hemisphere, respectively, which drives a downflow at the equator (with balancing upflows at the poles). Whereas at higher pressures (down to 10 bar), we find that the sense of the circulations has changed, resulting in a slight upflow at the equator. The pressure at which this change in circulation regime occurs matches the pressure at which the zonal-mean zonal jet vanishes, which reinforces the conclusion of \citet{2017ApJ...841...30T,2019A&A...632A.114S} that flows associated with the equatorial jet are responsible for driving vertical flows that advect mass/temperature/enthalpy. This also explains why we see little to no heating of the deep, advective, adiabat (\autoref{fig:Longitudinal_T}b), and hence would expect to observe little to no radius inflation for a planet at a similar orbital radius and stellar insolation.\\

{ \subsubsection{A Note About Deep Atmosphere Equilibrium}}

{ It is important to note that,} unlike \citet{2019A&A...632A.114S} and \citet{2021A&A...656A.128S}, the deep atmosphere circulation profiles considered here reveal significant time-variability. This is simply a result of the deep atmosphere only being initialised close to its adiabatic steady-state and the high computational cost of running the models long enough for the dynamically slow deep atmospheres to completely thermally equilibrate - although this should not affect our results here as our focus in on the use of AI for analysis, and not the steady state dynamics in the deep atmosphere, the detailed analysis of which we leave to future studies. \\ 
{ However, whilst} attempts have been made to solve the problem of the high computational-cost of resolving the steady-state deep atmospheres, these solutions come with their own problems. For example, \citet{2022A&A...666L..11S}, attempted to find the steady-state deep atmosphere of the ultra-hot-Jupiter WASP-76b by calculating the observed cooling at 650-bar in their model, including its rate of change, and then using this to extrapolate towards the steady-state T-P profile everywhere in the atmosphere (that is to say at pressures lower than 650 bar), which they find to be 'cold', implying that potential temperature advection is not driving any deep heating. However, this approach has a number of caveats which mean that extrapolation is generally a bad idea: To start, the pressure dependence of radiative heating, and radiative dynamics more generally, means that the 650 bar cooling rate is unlikely to be representative of the cooling rate at lower pressures. Specifically, at lower pressures, one needs to assess whether the atmosphere has reached equilibrium with the radiative dynamics or with the advection of potential temperature. An assessment which cannot be made via by the dynamics at 650 bars, especially since even this region has not yet reached equilibrium. 
Furthermore, the model was initialised significantly hotter than the expected, inflated, steady state, resulting in an enhanced deep cooling rate driven by the deep atmospheres need to expel excess energy. In turn, this deep cooling will drive very different deep circulations (as seen by \citet{2019A&A...632A.114S} when they modified the equilibrium temperature of the outer atmosphere of a previously equilibrated model), which take time to evolve/settle back to deep heating once the atmosphere has cooled - in fact, evidence of this shift to deep heating can be seen in Figure 2 of \citet{2022A&A...666L..11S}: the extrapolated model shows signs of heating that is slowly pushing deeper with time, driven by potential temperature advection, and which will likely lead to a steady state somewhere between the two models at equilibrium.  As such, we advise against estimating the deep atmospheres temperature-pressure profile via extrapolation, and instead suggest that future studies focus upon next-generation GCMs which will be efficient enough to model the equilibrium state of the deep atmosphere within a reasonable, computational, timescale. And which will benefit greatly from pairing with a concurrent, AI-analysis, model. 

\begin{figure*}
\gridline{\fig{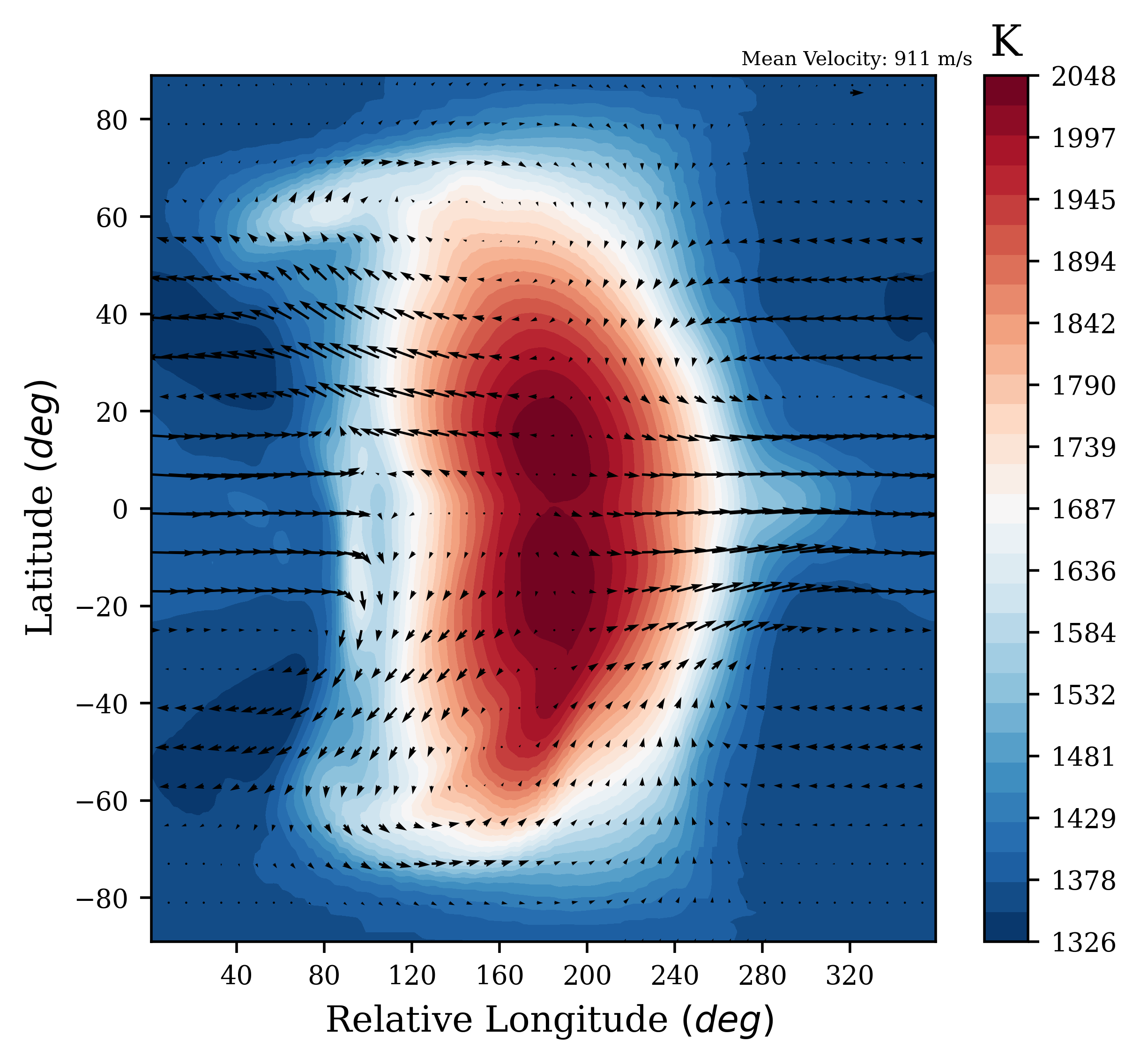}{0.25\textwidth}{a) `Hot': 0.0026 bar - `Locked'}
          \fig{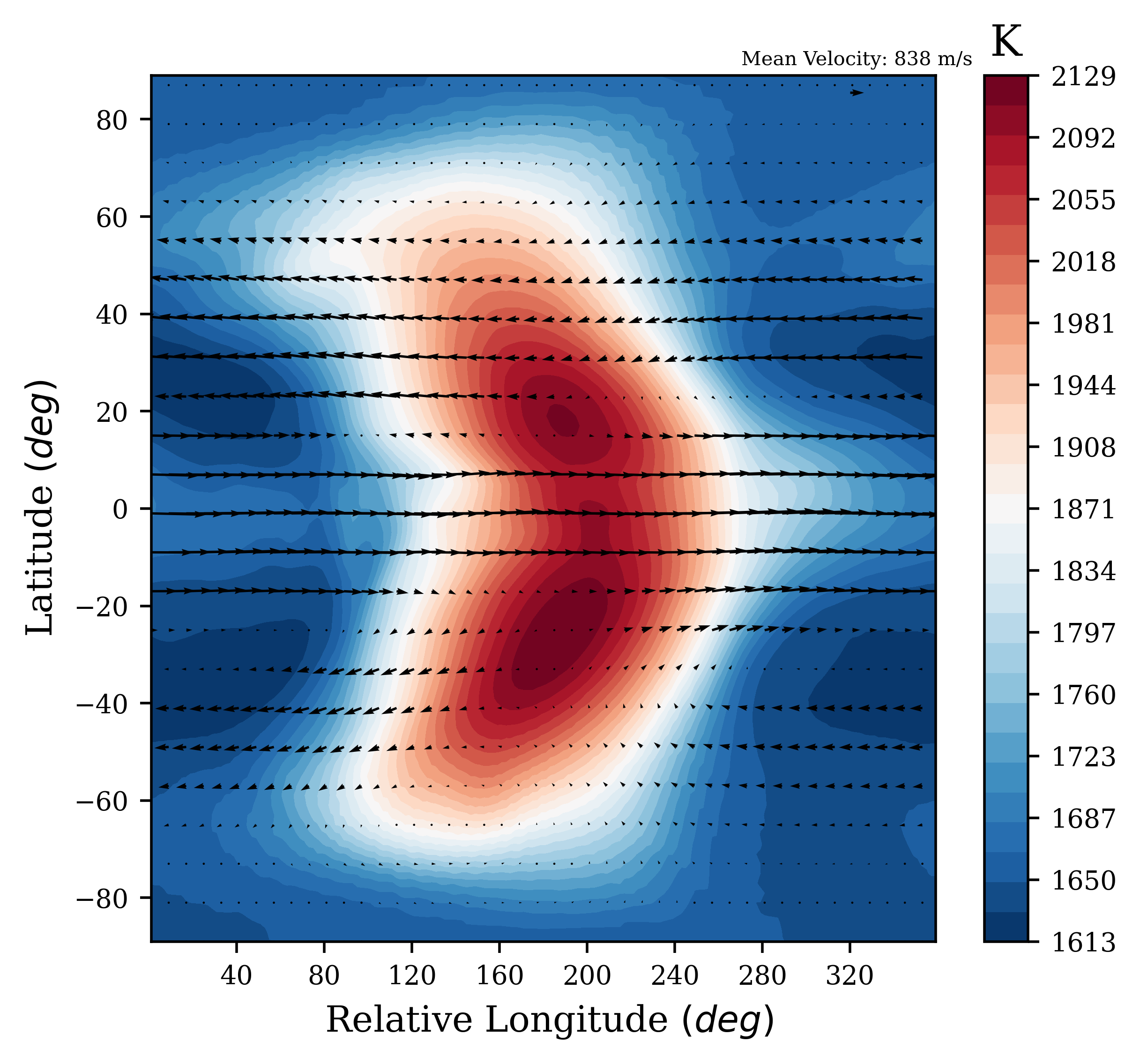}{0.25\textwidth}{b) `Hot': 0.016 bar - 'Butterfly'}
          \fig{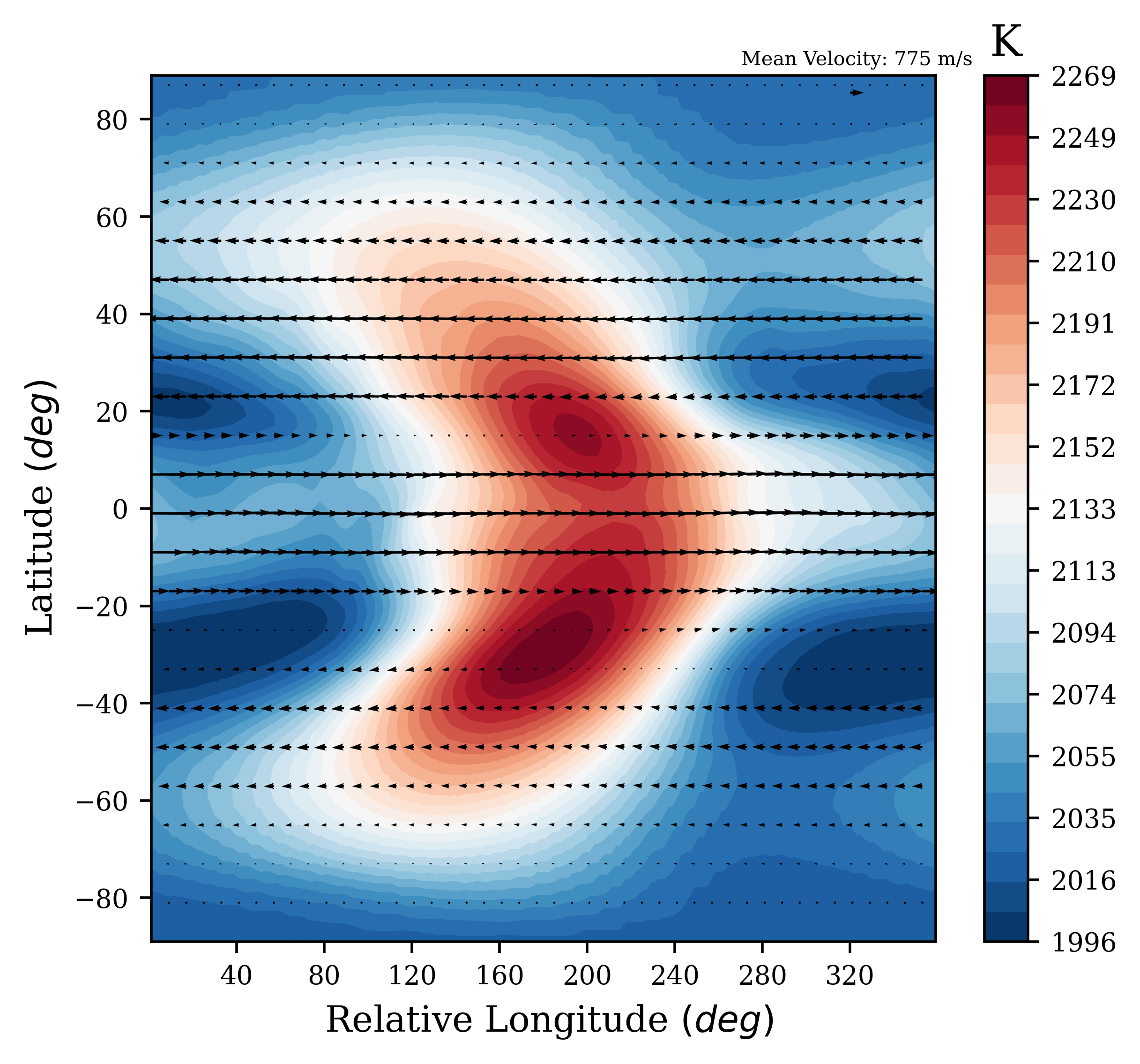}{0.25\textwidth}{c) `Hot': 0.2 bar - 'Butterfly'}
          \fig{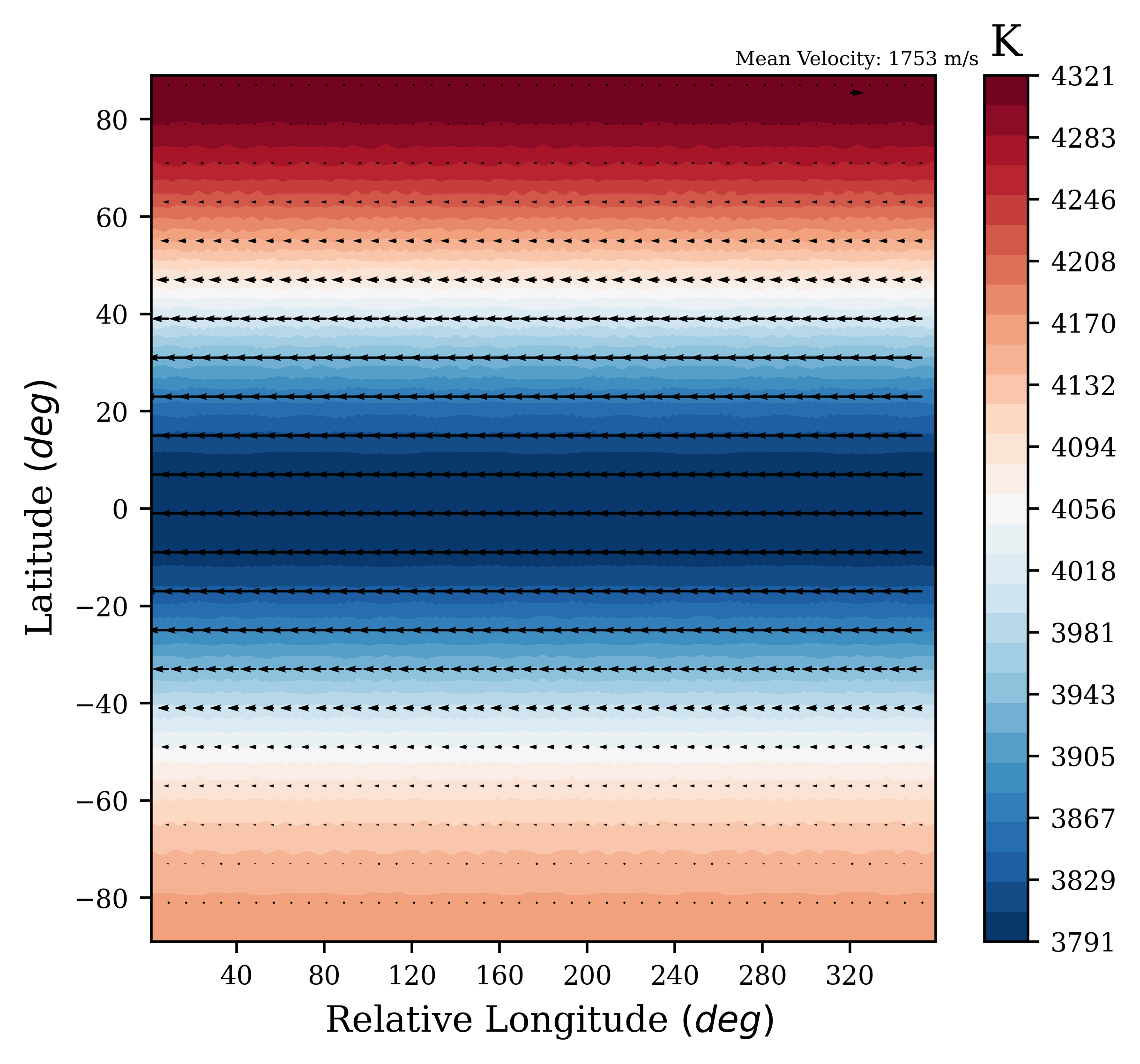}{0.25\textwidth}{d) `Hot': 40 bar - 'Banded'}}
\gridline{\fig{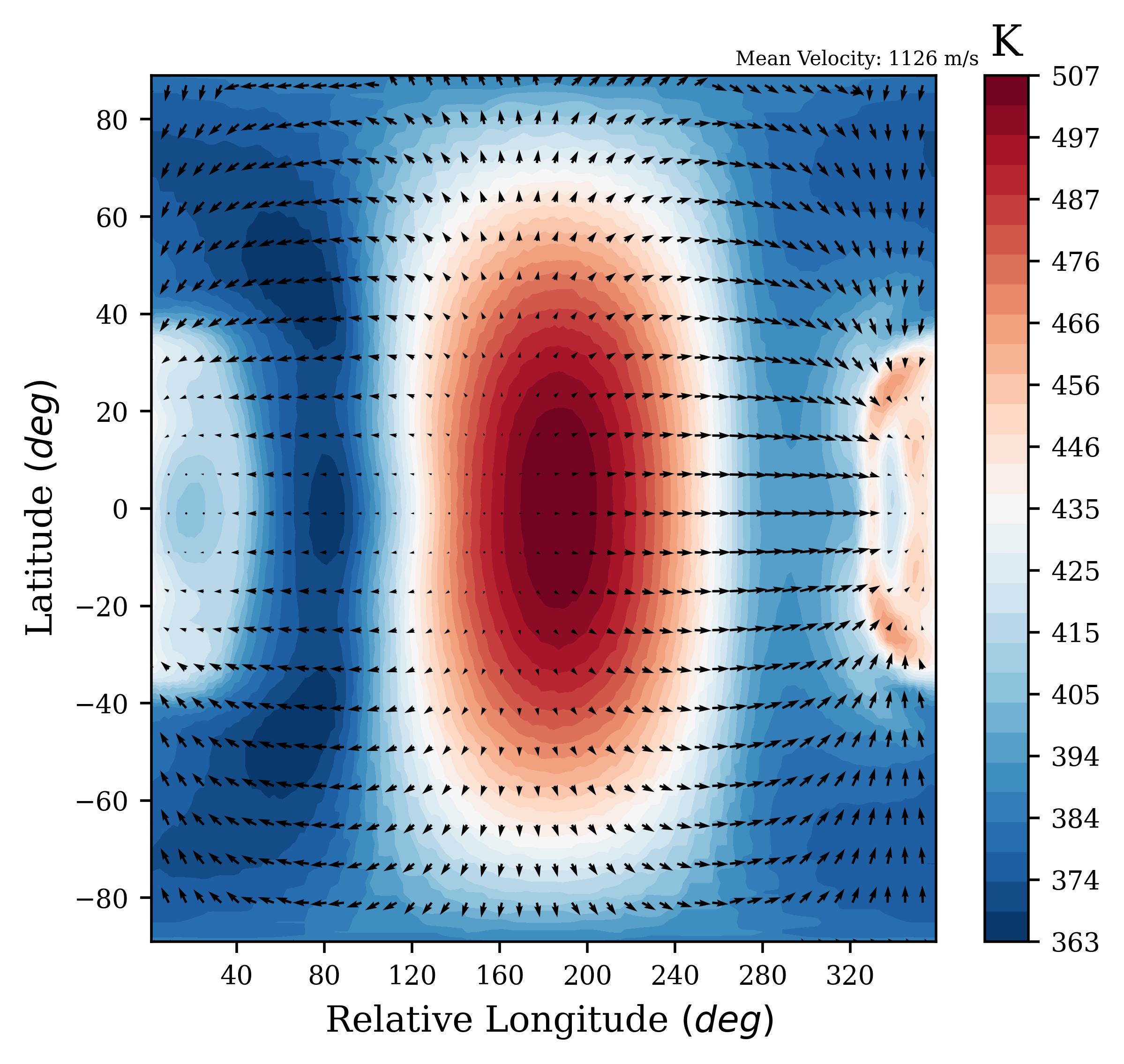}{0.25\textwidth}{e) `Cool': 0.0026 bar - 'Locked'}
          \fig{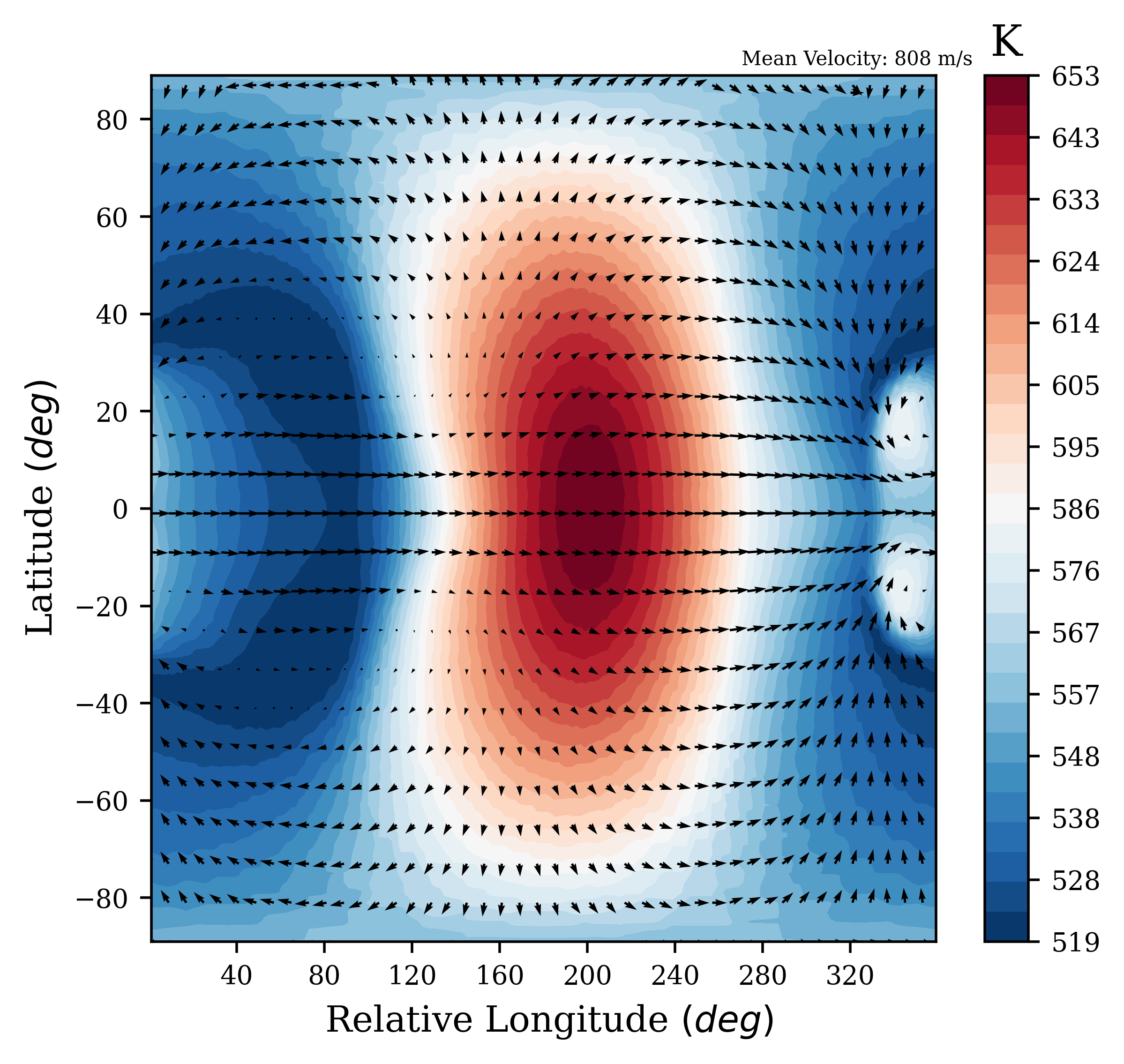}{0.25\textwidth}{f) `Cool': 0.016 bar - 'Locked'}
          \fig{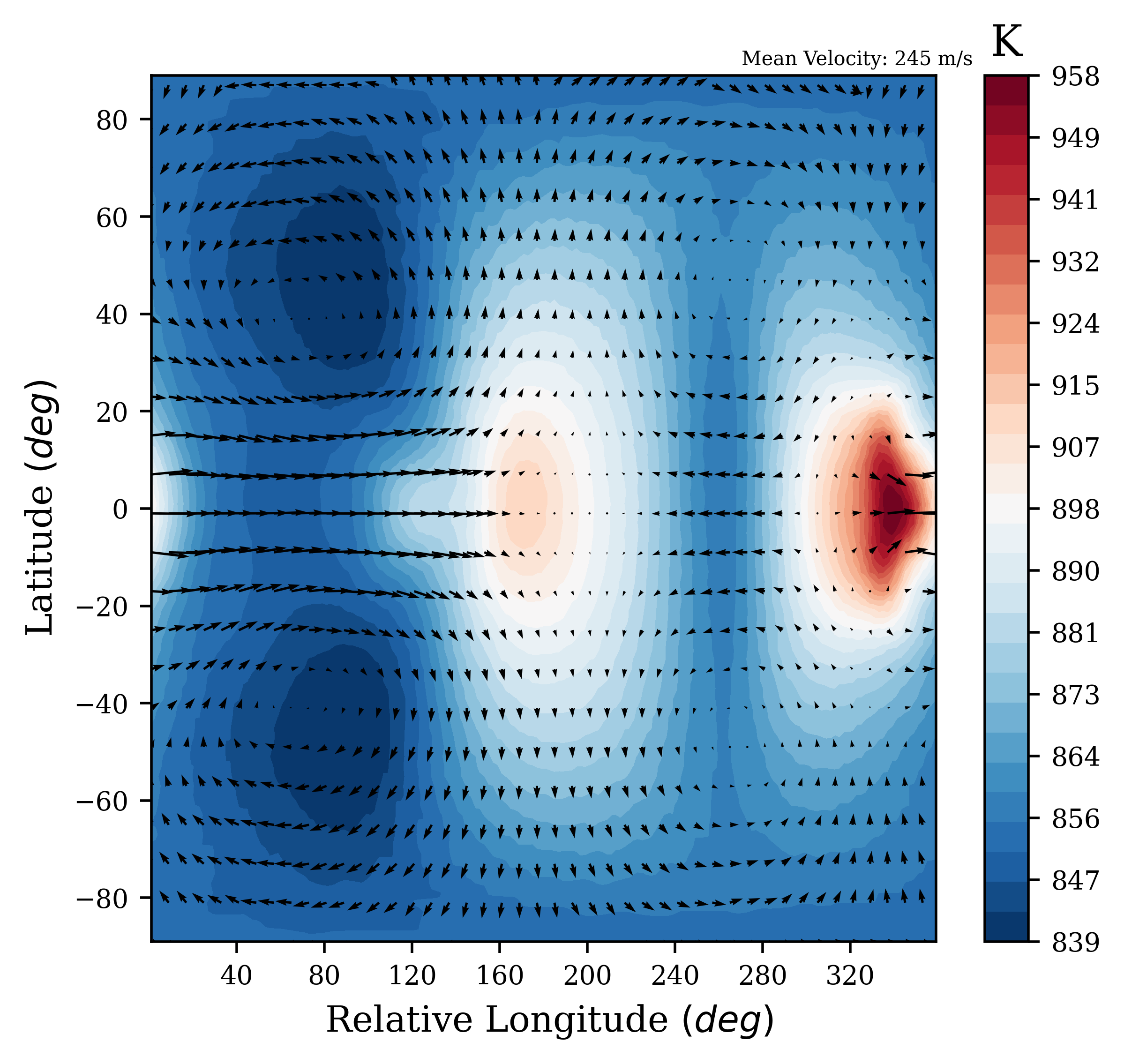}{0.25\textwidth}{g) `Cool': 0.2 bar - 'Inversion'}
          \fig{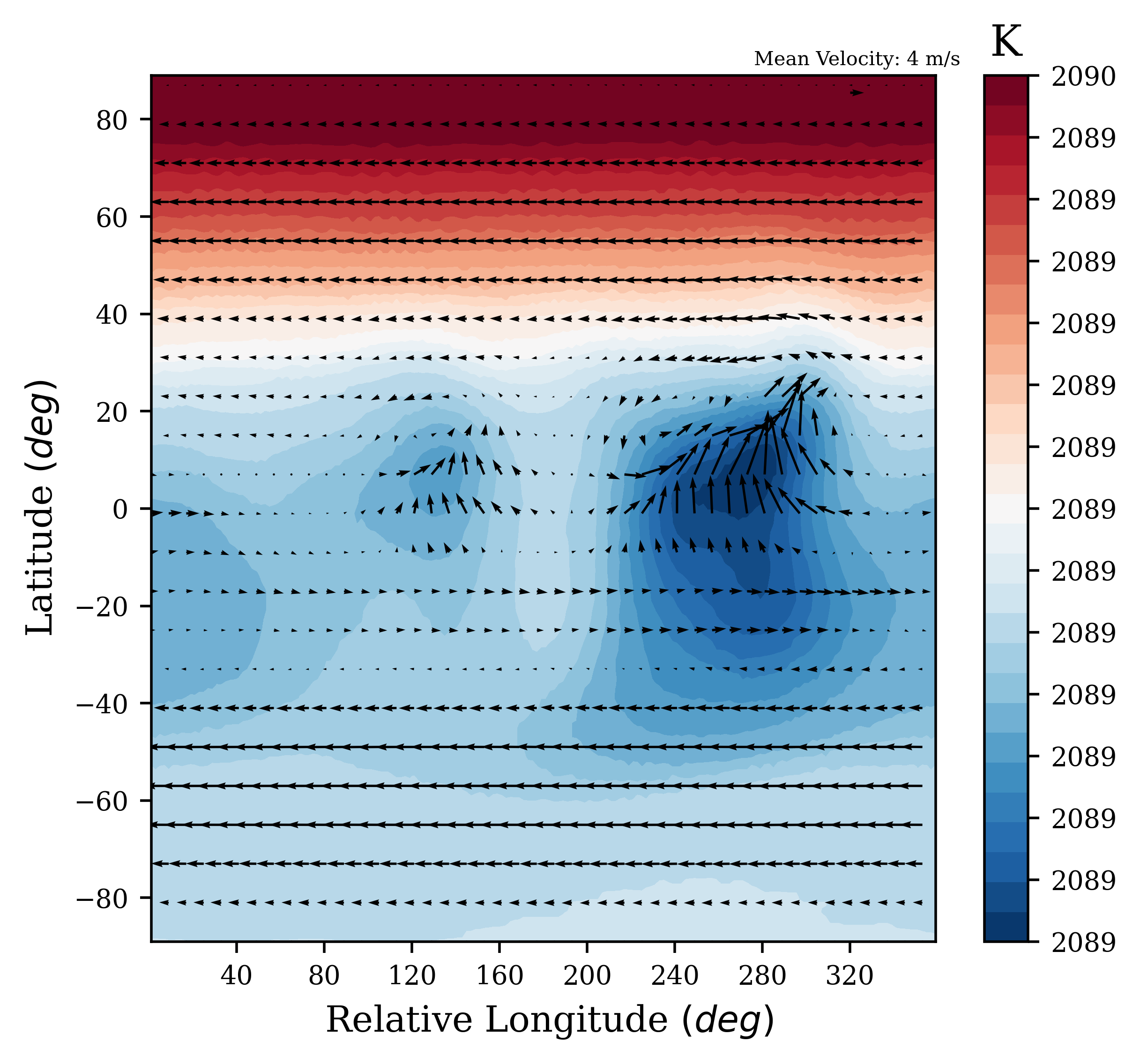}{0.25\textwidth}{h) `Cool': 40 bar - 'Asymmetric'}}
\caption{Temporally averaged zonal wind (arrows) and temperature profile (map) at four different pressures levels (0.0026 bar - left, 0.016 bar - middle left, 0.2 bar - middle right, and 10 bar - right) for both our `hot' (top) and `cool' (bottom) HD209458b-like atmospheric models. Each profile has been labelled with the (class of) tag assigned to it by the CNN. \label{fig:Wind_Temp}}
\end{figure*}

\begin{figure}[tbp] %
\begin{centering}
\includegraphics[width=0.95\columnwidth]{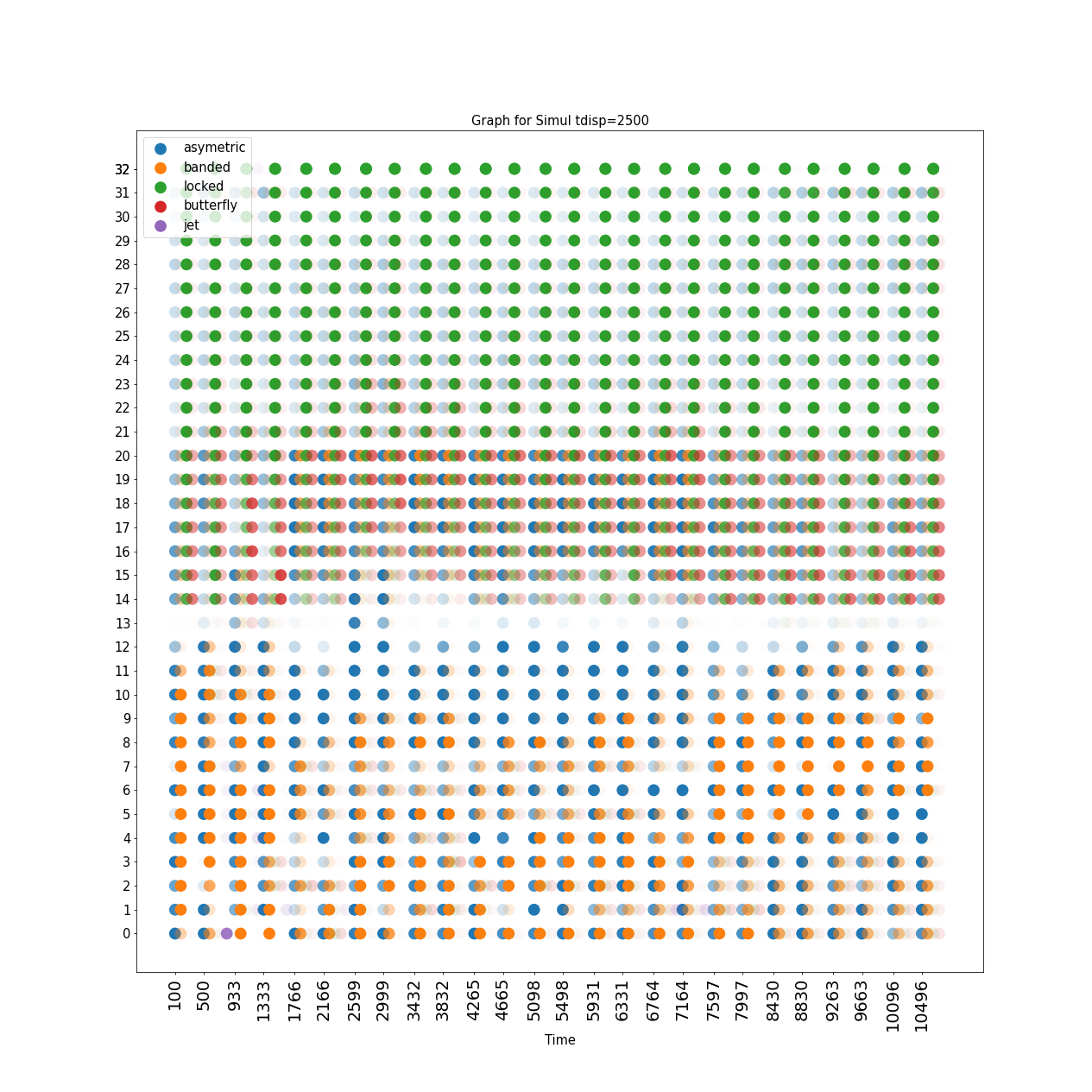}
\caption{ AI { multi-}categorisation map for the HD209458b atmospheric model of \citet{2019A&A...632A.114S}. Here we show the categories detected (with the opacity of each detection point representing the strength of the detection) at each pressure level (where an increase in pressure level corresponds to a decrease in actual pressure as we move to higher altitudes) against time (where t=0 corresponds to solid-body, adiabatic, initialisation, and each latter point corresponds to the centre of the temporal mean). The categories in question correspond to the detection of a north-south asymmetric thermal structure in blue, a banded (i.e. horizontally homogenised) thermal structure in orange, a radiative-dominated thermal structure (driven by tidal-locking) in green, a butterfly thermal structure (i.e. eastward equatorial advection flanked by slight, off-equator, westward advection) in red, and an equatorial (wind) jet in purple. Note that this model was run at lower resolution than the other models considered here, which has slightly impacted the ability of the CNN to discriminate between different atmospheric features. \label{fig:Categorisation_0048} }
\end{centering}
\end{figure}
\begin{figure}[tbp] %
\begin{centering}
\includegraphics[width=0.95\columnwidth]{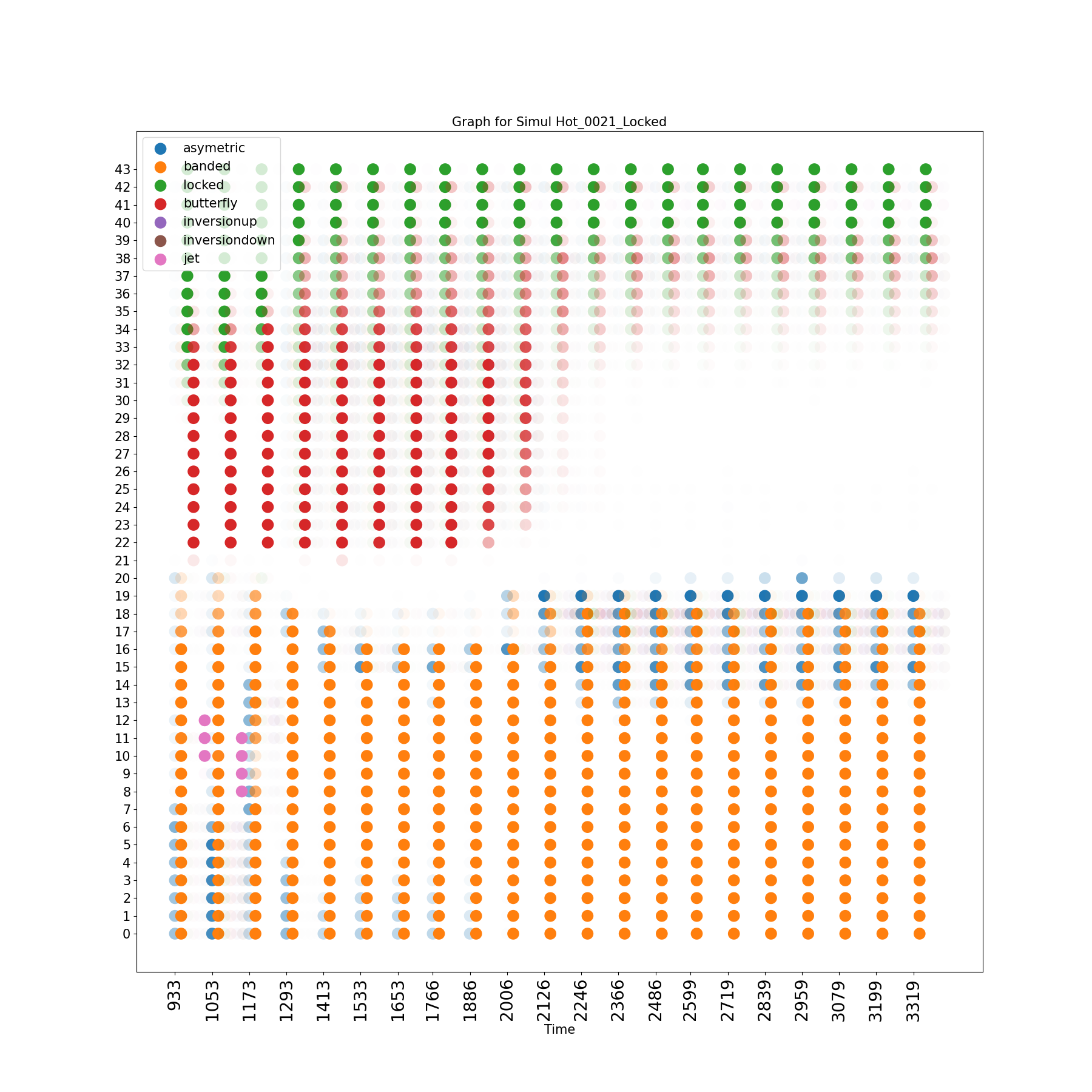}
\caption{ AI { multi-}categorisation map for our `hot' HD209458b-like atmospheric model. Here we show the categories detected (with the opacity of each detection point representing the strength of the detection) at each pressure level (where an increase in pressure level corresponds to a decrease in actual pressure as we move to higher altitudes) against time (where t=0 corresponds to solid-body, adiabatic, initialisation, and each latter point corresponds to the centre of the temporal mean). The categories in question correspond to the detection of a north-south asymmetric thermal structure in blue, a banded (i.e. horizontally homogenised) thermal structure in orange, a radiative-dominated thermal structure (driven by tidal-locking) in green, a butterfly thermal structure (i.e. eastward equatorial advection flanked by slight, off-equator, westward advection) in red, a vertical thermal inversion in purple (increasing T with P) and brown (decreasing T with P), and an equatorial (wind) jet in pink. \label{fig:Categorisation_0021} }
\end{centering}
\end{figure}

\begin{figure*}
\gridline{\fig{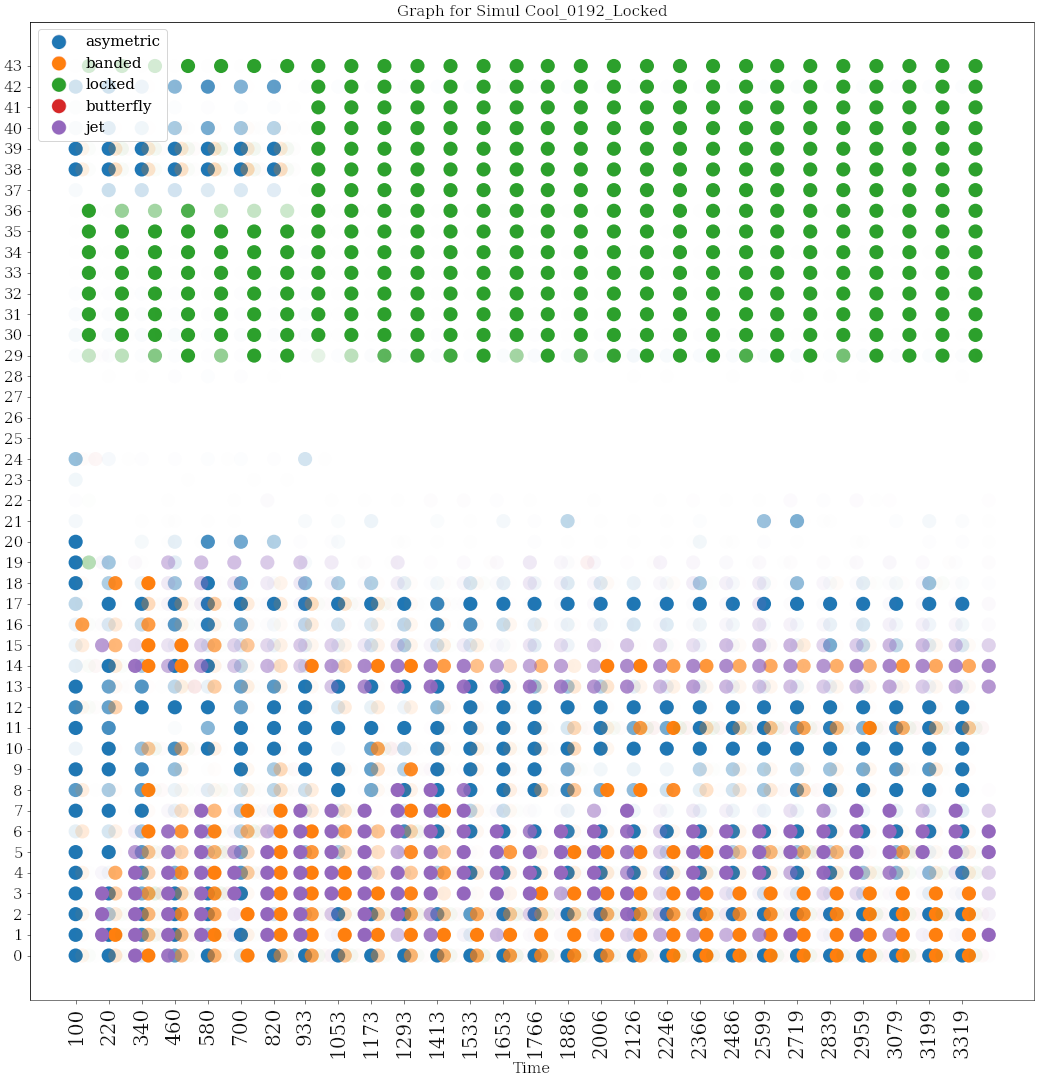}{0.49\textwidth}{a) Without Inversion Detection}
          \fig{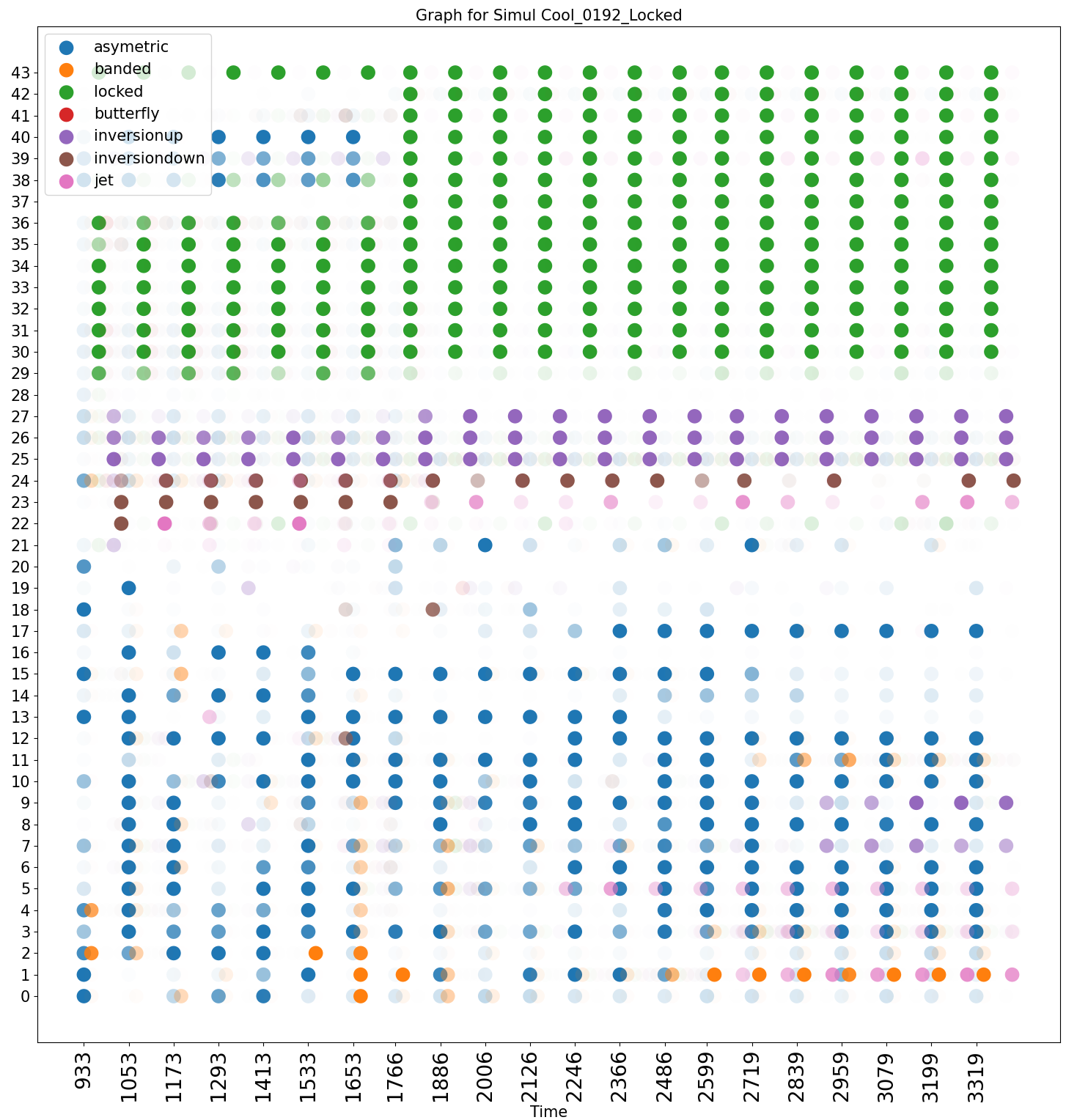}{0.484\textwidth}{b) With Inversion Detection}}
\caption{AI { multi-}categorisation map(s) for our `cool' HD209458b-like atmospheric model. Here we show the categories detected (with the opacity of each detection point representing the strength of the detection) at each pressure level (where an increase in pressure level corresponds to a decrease in actual pressure as we move to higher altitudes) against time (where t=0 corresponds to solid-body, adiabatic, initialisation, and each latter point corresponds to the centre of the temporal mean) for two different { multi-categorisation} CNNs, one without thermal inversion detection (left) and one with (right). { Note that the CNN that lacked thermal inversion detection also differs from the CNN used throughout the rest of this work in a number of other regards being an earlier iteration of the final model. This includes changes to the colour map, resolution, and averaging period of the input data}.  The categories in question correspond to the detection of a north-south asymmetric thermal structure in blue, a banded (i.e. horizontally homogenised) thermal structure in orange, a radiative-dominated thermal structure (driven by tidal-locking) in green, a butterfly thermal structure (i.e. eastward equatorial advection flanked by slight, off-equator, westward advection) in red, a vertical thermal inversion { (in the RHS figure)} in purple (increasing T with P) and brown (decreasing T with P), and an equatorial (wind) jet in { purple (LHS) or pink (RHS)}. \label{fig:Categorisation_0192}}
\end{figure*}
\subsection{Initial Data Tags, Training, and Results} \label{sec:AI_categories_init}
Whilst interesting, the aforementioned zonal-mean dynamics are not what we intend to explore with our { neural-networks}. Instead, we plan to look at the pressure dependent horizontal wind and temperature profiles that combine to give these zonal-mean flows, and which contain many more interesting, and unique, features for the CNN(s) to detect. \\
Initially, we decided to train our CNNs to detect the presence of: { day-side hot-spots in which the zonal winds have caused significant horizontal thermal advection (and whose shape is typically referred to to as a butterfly in the hot Jupiter community - e.g. \autoref{fig:Wind_Temp}b/c), longitudinally homogenised and latitudinally symmetric thermal bands (e.g. \autoref{fig:Wind_Temp}d), day-side hot-spots in which radiative affects dominate over advective dynamics (i.e. an irradiative hot-spot which has not been significantly advected by horizontal winds - e.g. \autoref{fig:Wind_Temp}e and \autoref{fig:Wind_Temp}a to a lesser extent), latitudinally asymmetric thermal structures (see, e.g. \autoref{fig:Wind_Temp}h), and, finally, }the presence of a equatorial zonal jet (see the arrows which trace the horizontal wind in, for example, \autoref{fig:Wind_Temp}b/c), although, as discussed in \autoref{sec:AI_Characterisation_2}, this final categorisation{ , and hence the wind-based CNN more generally,} did not pan out for a number of reasons.\\
This training was performed using early outputs from our complete simulation data-set, with orbital radii of between $0.012\textrm{au}$ and $0.334\textrm{au}$, which was hand labelled/tagged such as to produce at least 50 examples of each feature of interest.  \\

Once the training was complete, but before we explored how rotation impacts the detected atmospheric features, we first investigated how our trained neural-networks would behave when applied to a model which has manually been confirmed to contain examples of all of the current atmospheric features of interest, and from which no training { or validation} data was extracted. \\
Specifically, we consider the adiabatically initialised HD209458b model of \citet{2019A&A...632A.114S}. As shown in \autoref{fig:Categorisation_0048}, { and as anticipated, the thermal CNN successfully identifies all of the expected atmospheric features, whilst on the other hand the zonal jet, and the zonal wind more generally, proved to be highly intractable} (\autoref{sec:AI_Characterisation_2}). { The identified features include:} a outer atmosphere which is dominated by the radiatively driven (tidally-locked) day-side hot-spot, a mid-atmosphere which shows signs of an advected hot-spot (i.e. a thermal butterfly), and a deep atmosphere which shows a mix of asymmetric and banded thermal structures, likely as a result of the ongoing deep heating in the model as it warms from its slightly cooler than steady-state adiabatic start.\\
{ Note that, by testing against a model from which no training or validation data was extracted, we are able to test the portability/generality of our trained CNN(s). However, it is important to note that the model we consider was also calculated using DYANMICO, using a very similar setup for HD209458b. For future studies, we suggest that a more diverse range of models should be considered when testing generalisability. This could include testing the CNN(s) on the outputs of models run with different GCMs or on the outputs of models calculated using the same GCM, but for rather different objects, such as the brown Dwarf models of \citet{2021A&A...656A.128S}.} \\

Having confirmed that our CNN(s) can recover all the thermal atmospheric features for which they { were} trained, { but when applied to a model from which no training data was taken,} we next move on to exploring how rotation influences the detected dynamics. To do this, we consider our two exemplary models, which we ran until their outer atmospheres had reached equilibrium: i.e. for $\sim40$ Earth-years of simulation time. We then applied the trained CNNs to the full data-set, using temporal averaging window of between 100-150 outputs in order to reduce the effects of small-scale oscillations on the final characterisation. Note that we used broader averaging windows for the HD209458b-like model shown above (\autoref{fig:Categorisation_0048}) due to the longer time-series available for that model. \\

As was the case for the zonal-mean dynamics, we find that the horizontal dynamics, and hence the detected atmospheric features, changes significantly with rotation rate. Specifically, we find little crossover in the features detected in our exemplary `hot' and `cool' models: \\
Starting with the `hot' regime, whose detected characteristics are shown in \autoref{fig:Categorisation_0021}, we find that the dynamics are dominated by three features: at very low pressures (which are slightly higher near initialisation, before advection kicks in) the thermal structure is dominated by a (tidally) locked day-side hot-spot. However this changes as we move towards higher pressures, with the advective time-scale becoming comparable to (and eventually faster than) the radiative time-scale, leading to significant horizontal advection and hence a transition to a thermal butterfly. Note that this detection of a butterfly tag does eventually vanish at later times, but as we discuss in \autoref{sec:AI_Characterisation_2}, this is a problem with our initial training data set not properly accounting for the influence of very rapid rotation on horizontal thermal advection. 
Finally, the deep atmosphere is generally dominated by latitudinally symmetric, and longitudinally homogenous, thermal bands, which indicate that it is generally well homogenised longitudinally, likely thanks to the zonal jet. Note however, that we do occasionally find that the CNN assigns the asymmetric tag to the outer deep atmosphere - this is due to ongoing heating of the lowest pressure regions of the deep atmosphere.   \\
On the other hand, analysis of our `cool' regime model, as shown in \autoref{fig:Categorisation_0192}a/b, reveals a rather different set of identified dynamics. Here we again find that the outer atmosphere is dominated by a tidally-locked, day-side, hot-spot, but unlike in our `hot' model, this does not transition into a thermal butterfly with increasing pressure. Instead, we find that a) the locked-profile extends significantly deeper than in the `hot' model, except near initialisation where the radiative-forcing time-scale dominates the dynamics, and b) this detection transitions into a region of non-detection within the mid-atmosphere (i.e. between pressure levels 20/21 ($\sim 1\textrm{bar}$) and 29 ($\sim 0.05\textrm{bar}$)). The former effect can likely be explained by the weaker role that zonal winds play in the atmospheric dynamics of `cool' regime models (\autoref{fig:Zonal_wind_streamfunction}c and \autoref{sec:analysis_1}), whilst we explore the later lack of detected feature in more detail in \autoref{sec:AI_categories_new}. Finally, in the deep atmosphere, we find that detected dynamics are generally dominated by weakly asymmetric thermal bands. Note that, unlike in the `hot' regime, these asymmetric structures are not being driven by deep heating, instead analysis (\autoref{sec:analysis_1}) suggests that it occurs because the deep atmosphere is rather quiescent (\autoref{fig:Zonal_wind_streamfunction}c), with very weak vertical heat transport, leading to persistent, but weak, horizontal temperature gradients in the deep atmosphere. \\
{ Note that the differences between the detected features in \autoref{fig:Categorisation_0192}a and b are due to the preliminary nature of the models that lacked thermal inversion detection (see below - \autoref{sec:AI_categories_new}), with the preliminary models using a different colormap and averaging scheme to the final CNN models presented in \autoref{sec:AI}.} \\

We explore why these differences in detected atmospheric features occurs in \autoref{sec:analysis_1}.


\begin{figure*}
\gridline{\fig{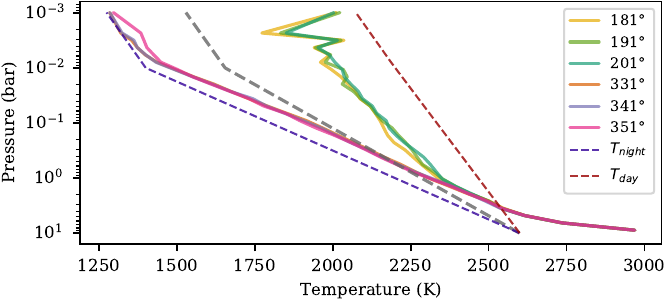}{0.7\textwidth}{a) `Hot' atmospheric model  \label{fig:Longitudinal_T_0021}}}
\gridline{\fig{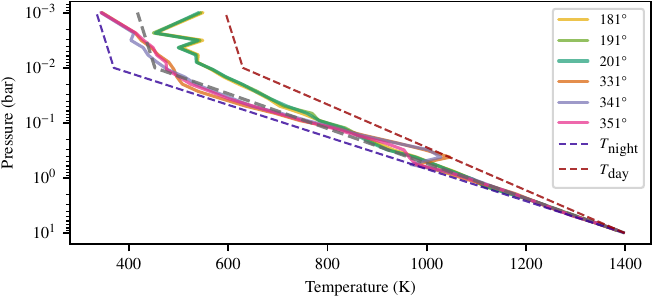}{0.7\textwidth}{b) `Cool' atmospheric model \label{fig:Longitudinal_T_0192}}}
\caption{Latitudinally and temporally averaged (at the equator) T-P profiles for our `hot' (top) and `cool' (bottom)  HD209458b-like atmospheric models. Each plot includes profiles from 6 different longitudes, ranging from the sub-stellar point (whose equilibrium profile is shown in red) eastwards to the anti-stellar point (whose equilibrium profile is shown in blue) on the night-side. Note that we have excluded the deep atmospheres ($P>10\textrm{bar}$) from these plots since it has not fully equilibrated.  \label{fig:Longitudinal_T}}
\end{figure*}

\begin{figure*}
\gridline{
\fig{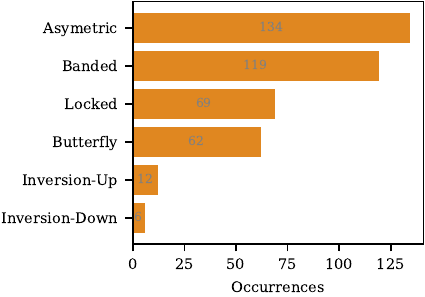}{0.45\textwidth}{a) Distribution of Training Data Before Oversampling}
\fig{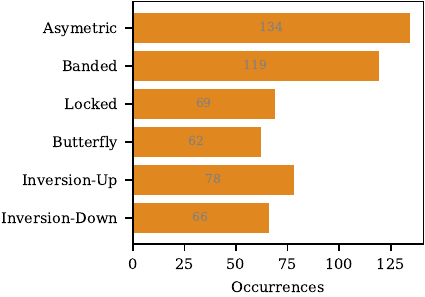}{0.45\textwidth}{b) Distribution of Training Data After Oversampling}}
\caption{Distribution of (thermal) training/validation data before (top) and after (bottom) applying an oversampling technique to the inversion data-set so as to generate artificial training data. The total count of items in each training data set is shown in grey. Note that we do not include the amount of training data used for the jet detection for two reasons: a) unlike all the other features that where trained on the thermal structure, the jet tag was trained in the horizontal wind, and b) the jet detection model did not make it into our final analysis due to problems with detecting highly symmetric structures (\autoref{sec:AI_Characterisation_2}). \label{fig:Oversampling}}
\end{figure*}

\subsection{Updated Data Tags: A Night-Side Thermal Inversion} \label{sec:AI_categories_new}
In an effort to understand the region of non-detection identified for our `cool' atmospheric models (\autoref{fig:Categorisation_0192}a), we elected to explore this region of the atmosphere in more detail, and if appropriate, update our { thermal CNN} with what we find. As shown in \autoref{fig:Wind_Temp}f/g, our analysis reveals that, once we are deep enough for advective transport to start to dominate over radiative forcing (via Newtonian Cooling), rather than a butterfly-like thermal structure on the day-side, as found in the classical hot Jupiter regime (e.g. \autoref{fig:Wind_Temp}c) we instead find that a hotspot has formed on the cold-night side, slightly to the west of the anti-stellar point (see \autoref{fig:Wind_Temp}f). 
This was of particular interest since it presence may have significant implications for both the atmospheric dynamics and observable features of more weakly irradiated Jupiters (so called `warm' Jupiters). Furthermore, it came as somewhat of a surprise since prior studies (e.g. \citealt{2019A&A...632A.114S}), other HD209458b-like models (e.g. \autoref{fig:Categorisation_0021}), as well as our initial training tags suggested that weak (due to the slower rotation rate and weaker surface irradiation at higher orbital radii) butterfly-like features should have been detected at these pressure levels. One of the main potential impacts of this night-side hot-spot is its associated thermal inversion, which can be clearly seen in longitudinally sliced temperature-pressure profiles of the `cool' model \autoref{fig:Longitudinal_T}b. Briefly, as shown here, a thermal inversion occurs when the temperature profile switches from cooling as the pressure decreases to warming with decreasing pressure - this is similar to the stratosphere on Earth and can have significant affects on observed atmospheric dynamics and chemistry (see, for example, \citealt{Hubeny_2003,2008ApJ...678.1419F,Spiegel_2009,2009ApJ...701L..20Z,Madhusudhan_2010,2015ApJ...813...47M,2017AJ....154..158B,2018ApJ...866...27L,2019MNRAS.485.5817G} for a discussion of thermal inversions in highly-irradiated hot Jupiters). \\
Of course, since this feature was not anticipated, our initial training data did not include it, hence the blank regions on the { multi-}categorisation maps. As such, and in order to better explore how wide-spread this feature is, we updated our training data-set { (and the underlying thermal CNN)} to include a pair of additional tags designed to cross-correlate temperature structures on the night-side and hence detect hot-spots: one indicating a night-side hot-spot which cools as the pressure decreases (`inversion-down') and one indicating the opposite (`inversion-up'). \\
However adding these new tags to the CNNs was not a simple matter since we had but a few examples of this phenomenon to use as training data (\autoref{fig:Oversampling}a). To resolve this, and thus properly train our CNNs to detect night-side hot-spots/thermal inversions, we turned to interpolative oversampling: that is to say we used interpolation to created a series of artificial tagged images from our limited sample of hand-tagged examples. For the `inversion-up' tag, we generate three artificial tagged images per input image, whilst for the `inversion-down' tag, which was significantly more numerous in our hand-labelled data than the `inversion-up' tag, we merely generated a single artificial tagged image per input image. Consequently, we now had at least 50 examples of each new feature (\autoref{fig:Oversampling}b) which could be used to update our { multi-categorisation thermal CNN} to detect the new feature of interest. \\

The results this updated analysis can be seen, for our `cool' regime model, in \autoref{fig:Categorisation_0192}b, which reveals that the mid-atmosphere region of non-detection has been replaced with both `inversion-up' and `inversion-down' tags, which when taken together indicate the peak of the night-side hot-spot (and also the pressure level at which the hottest point shifts from the day-side to the night-side). Further analysis of our models reveals that this feature is both robustly, but uniquely, present in the  `cool' regime. For example, our 'hot' regime model reveals no detected inversion tags, as expected from our visual analysis. But why does this feature only occur in our `cool' models? As we discuss below, it appears to be linked to the relative influence of rotation on the atmospheric dynamics and thermal energy transport. 

\begin{figure*}
\gridline{\fig{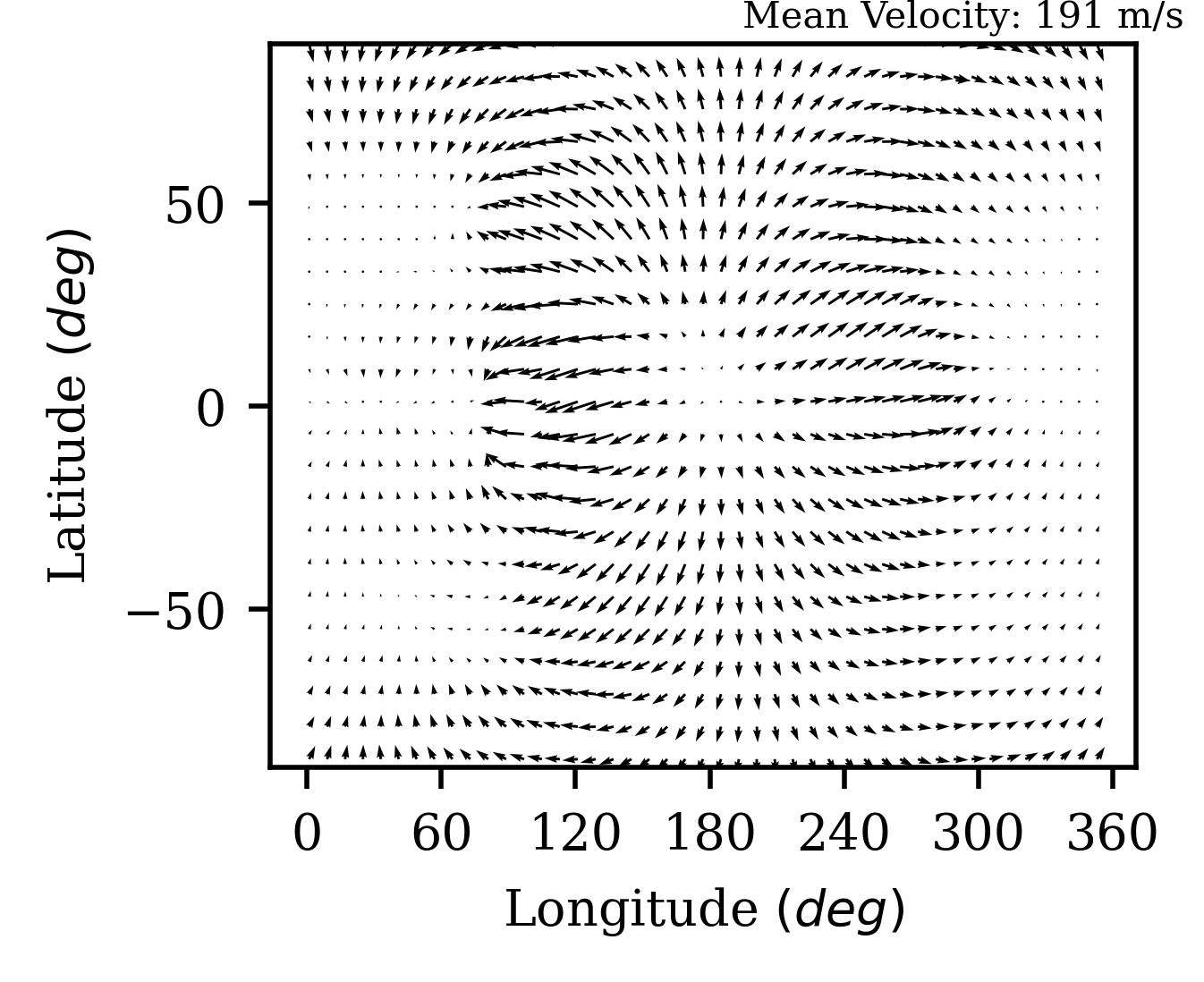}{0.3\textwidth}{a) `Hot': Divergent}
          \fig{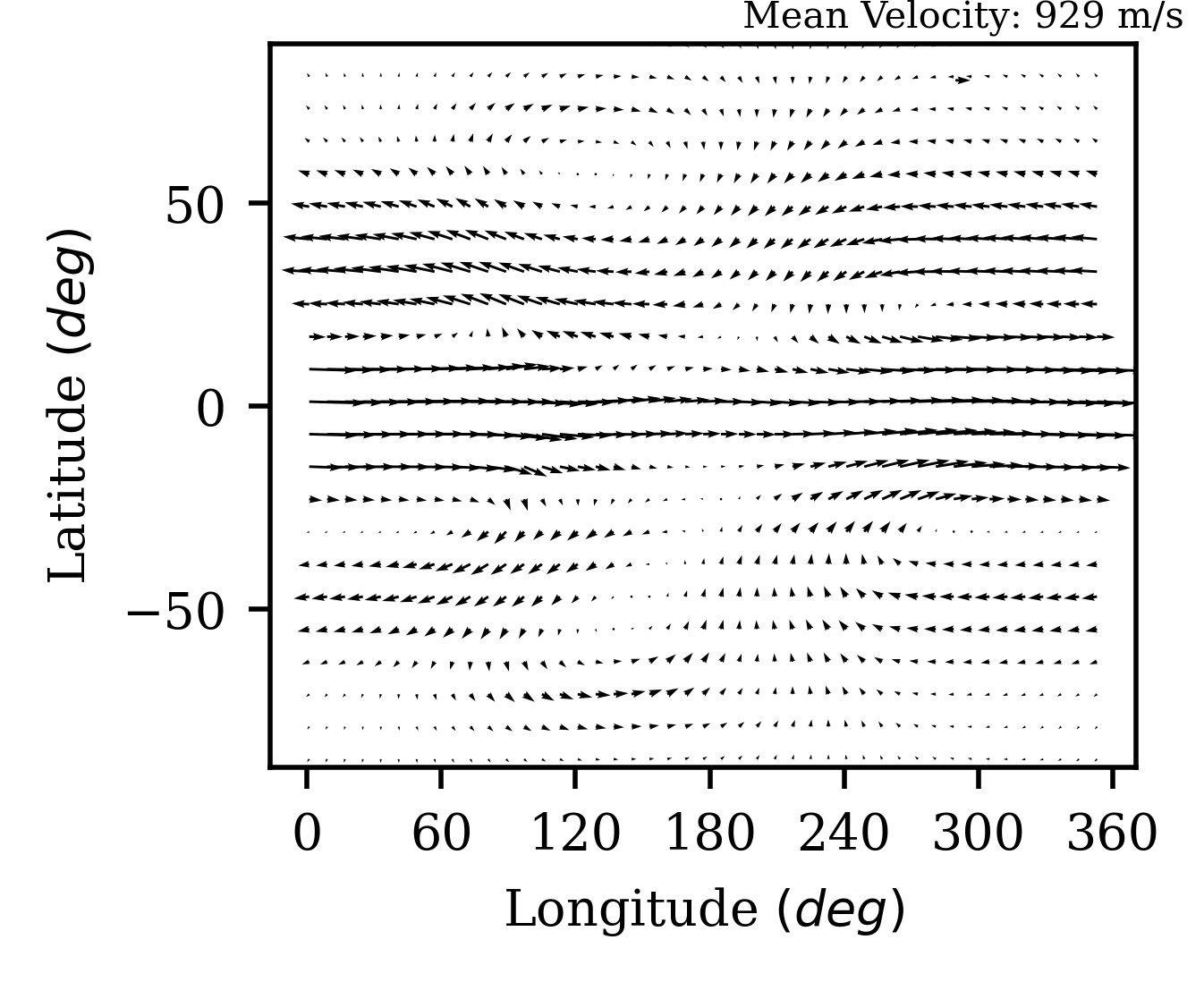}{0.3\textwidth}{b) `Hot': Rotational}
          \fig{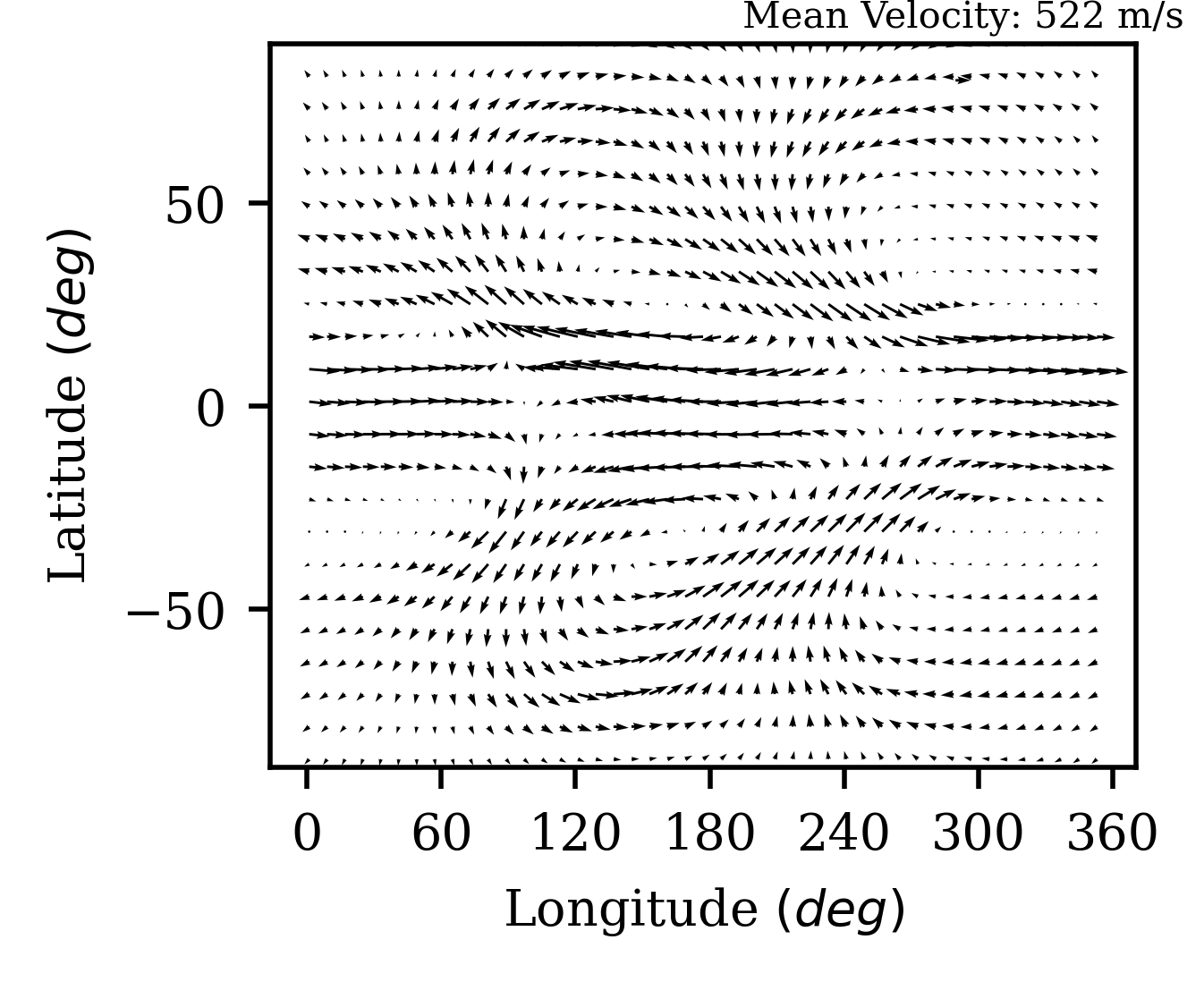}{0.3\textwidth}{c) `Hot': Eddy}}
\gridline{\fig{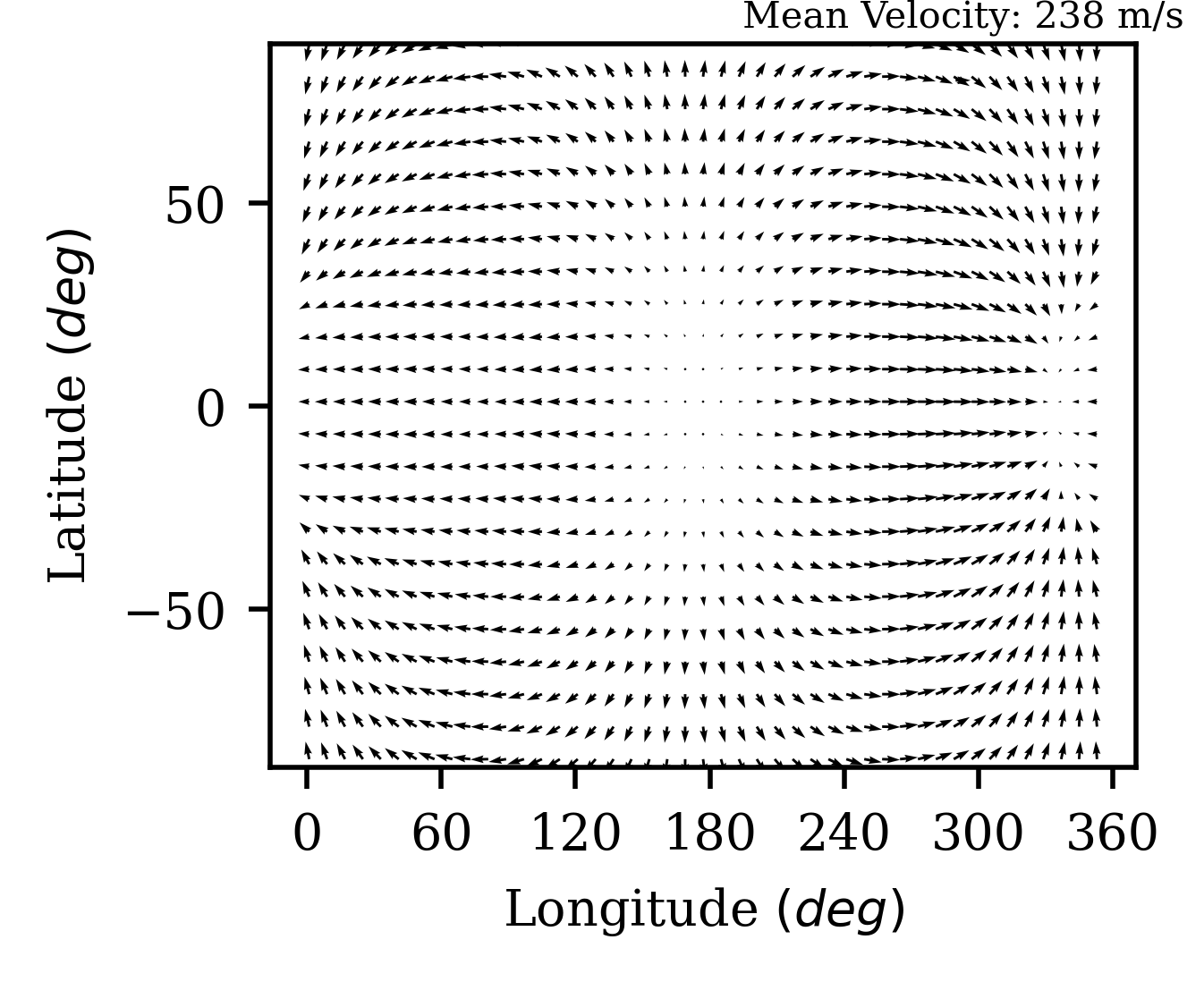}{0.3\textwidth}{d) `Cool': Divergent}
          \fig{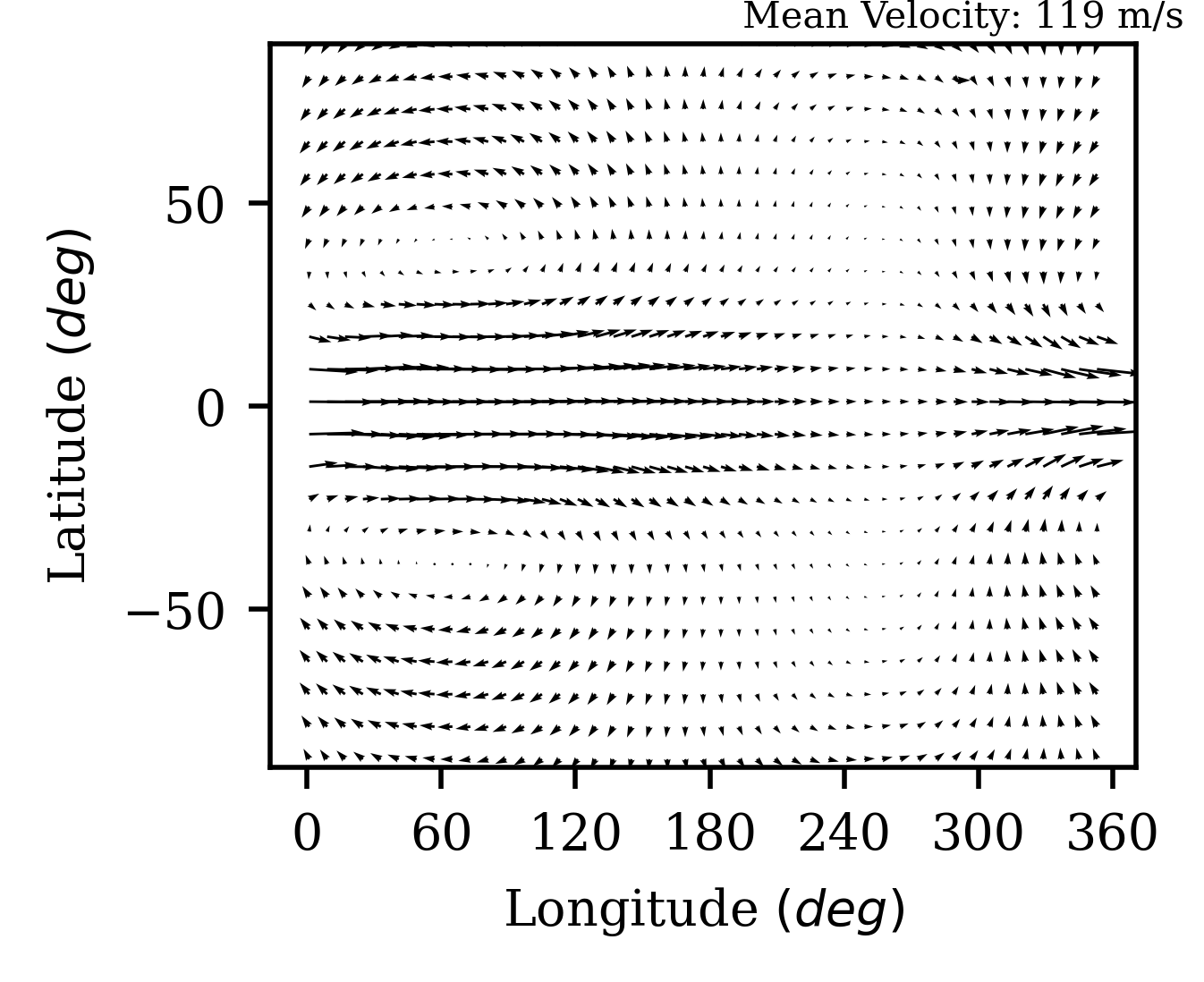}{0.3\textwidth}{e) `Cool': Rotational}
          \fig{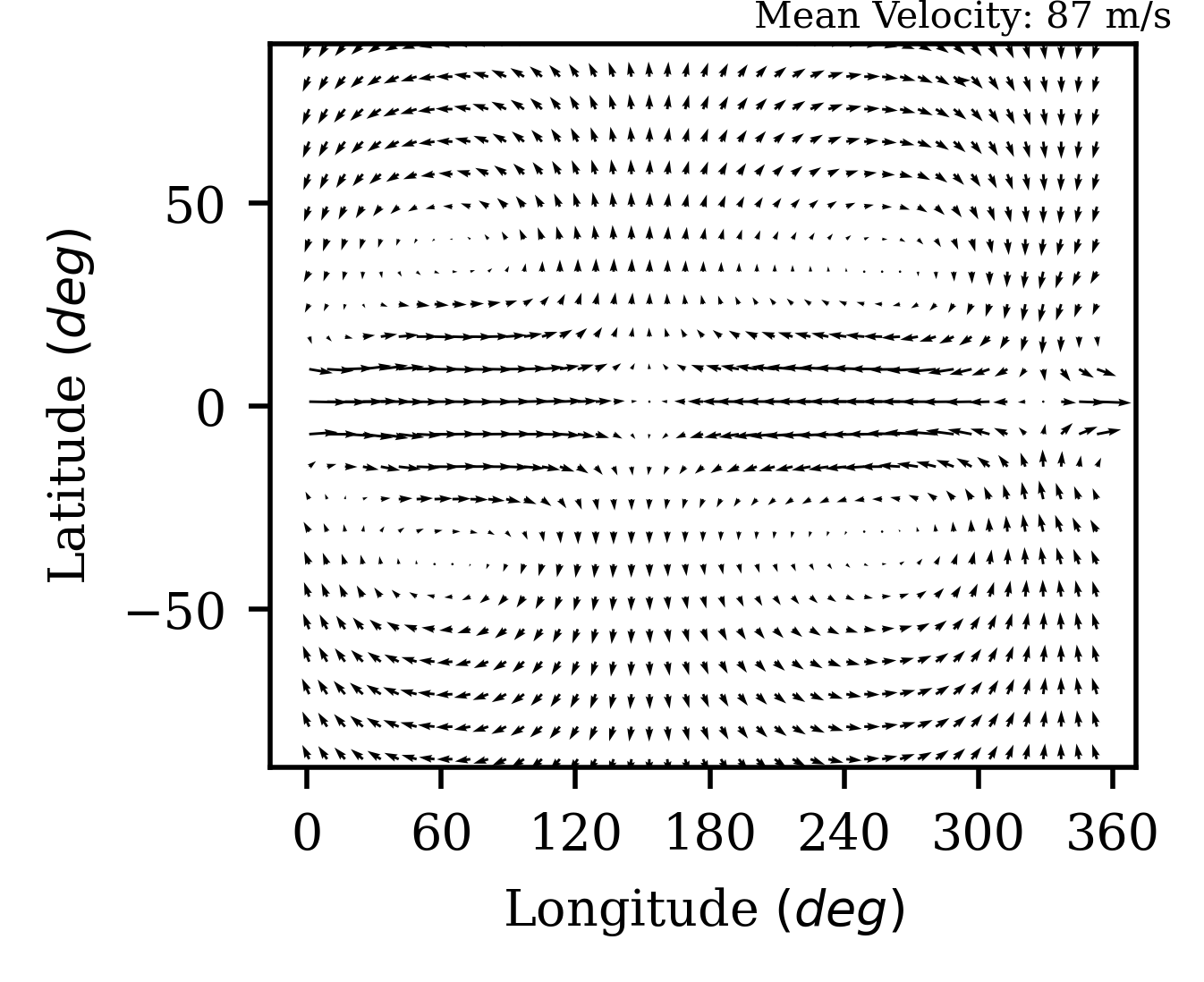}{0.3\textwidth}{f) `Cool': Eddy}}
\caption{Helmholtz decomposition of the radially and temporally averaged horizontal wind for both our `hot' (top) and `cool' (bottom) HD209458b-like atmospheric models. To the left we plot the divergent component ($\bm{u}_{div}$) of the Helmholtz decomposition, in the middle, the rotational component ($\bm{u}_{rot}$), and on the right the eddy component ($\bm{u}_{eddy} = \bm{u}_{rot} - \left<\bm{u}_{rot}\right>$) of the rotational component of the wind. A video version of this plot is available online: \url{https://www.youtube.com/watch?v=s-7AXs5_owg} and \url{https://www.youtube.com/watch?v=toDpbqer2e4} \label{fig:Helmholtz}}
\end{figure*}

\subsection{How Atmospheric Dynamics Impacts the Observed Features} \label{sec:analysis_1}

As previously alluded to, the HD209458b-like atmospheric models we consider here fall into two distinct regimes depending upon their, tidally-locked, orbital radii (i.e. surface irradiation and rotation rate). At short orbital radii (i.e. the `hot' regime), the zonal-mean atmospheric dynamics are dominated by a strong, super-rotating, equatorial jet which extends deep into the atmosphere (\autoref{fig:Zonal_wind_streamfunction}a), driving significant downflows (\autoref{fig:Zonal_wind_streamfunction}b) that result in strong vertical mixing, and hence deep heating (and radius inflation - \citealt{2017ApJ...841...30T,2019A&A...632A.114S}). On the other hand, at longer orbital radii (i.e. the `cool' regime), this is no longer the case and instead the zonal-mean zonal jet is significantly weaker and shallower (\autoref{fig:Zonal_wind_streamfunction}c), and thus associated with significantly reduced vertical mixing (\autoref{fig:Zonal_wind_streamfunction}d) which in turn drives little to no deep heating. \\

However, in order to fully understand the dynamics identified by our CNNs, we must look at more than just the zonal-mean dynamics. Specifically we are interested in the differences in horizontal wind between the `hot' and `cool' regimes, and how these differences affect horizontal and vertical energy (enthalpy) transport, leading to the various atmospheric features identified by our CNNs, particularly the night-side hot-spot found in the `cool' regime. \\
To that end, we next explore the Helmholtz decomposition of the zonal wind, a decomposition which has previously been used to study both the atmosphere of Earth \citep{dutton1986ceaseless} and hot Jupiters \citep{2021PNAS..11822705H}. Briefly, a Helmholtz decomposition can be used to split the horizontal wind at each pressure level, $\bm{u} = (u,v)$, into divergent (i.e. `vorticity free') and rotational (i.e. `divergence free') components \citep{dutton1986ceaseless}: 
\begin{align}
  \bm{u} &= \bm{u}_{div} + \bm{u}_{rot} \\
  &= -\bm{\nabla}\chi + \bm{k}\times\bm{\nabla}\psi,
\end{align}
where $\chi$ is the velocity potential function, $\psi$ is the velocity streamfunction, and both can be linked to the divergence $\delta$ / vorticity $w$ directly:
\begin{align}
  \nabla^{2}\chi &= \delta\\
  \nabla^{2}\psi &= w.
\end{align}
Additionally, in order to further isolate the equatorial zonal jet from other wind dynamics, we further split the rotational component, $\bm{u}_{rot}$ into a zonal-mean component $\bm{u}_{zonal}$ and an eddy component $\bm{u}_{eddy}$:
\begin{align}
  \bm{u}_{zonal} &= \left<\bm{u}_{rot}\right>\\
  \bm{u}_{eddy} &= \bm{u}_{rot} - \bm{u}_{zonal},
\end{align}
where $\left<\right>$ indicates the zonal-mean.\\
As for what these components represent: $\bm{u}_{div}$ represents flows that diverge 
from the hot-spot on the day-side and converge on the cold night-side, forming a closed cycle when combined with the upwelling below the day-side hot-spot and the downwelling on the nightside; $\bm{u}_{rot}$ represents dynamics 
driven by angular momentum transport via stationary Rossby and Kelvin waves - in typical hot Jupiters these standing waves transport angular momentum from mid-latitudes to the equator, resulting in slight westward flows at mid-latitudes and a super-rotating jet at the equator \citep{2011ApJ...738...71S}; and finally, as mentioned above, $\bm{u}_{eddy}$ and $\bm{u}_{zonal}$ are used to split $\bm{u}_{rot}$, allowing for us to explore the transport by standing waves in cases where the presence of a super-rotating equatorial jet would completely dominate the dynamics. \\
In \autoref{fig:Helmholtz}, we plot $\bm{u}_{div}$ (left), $\bm{u}_{rot}$ (centre), and $\bm{u}_{eddy}$ (right), radially averaged over the outer atmosphere, for both our `hot' (top) and `cool' (bottom) atmospheric models. We also, online, give a 3D view of each component of the horizontal wind for our `hot' atmospheric model$^b$.  \\
Starting with said `hot' atmospheric model, we can clearly see that by magnitude the rotational component ($\left|\bm{u}_{rot}\right| = 929 ms^{-1}$) significantly dominates over the divergent component ($\left|\bm{u}_{div}\right| = 191 ms^{-1}$). When combined with the difference between the rotational and eddy/zonal components of the wind, this suggests that the main driving force behind the horizontal dynamics seen here is, as expected, the presence of a strong equatorial jet. Further, as suggested by \citet{2011ApJ...738...71S}, and confirmed by both $\bm{u}_{rot}$ and $\bm{u}_{eddy}$, the super-rotating equatorial jet appears to be driven by standing Rossby and Kelvin waves. More specifically, we find a m=1 standing wave pattern which has become significantly tilted from west to east by a combination of both a strong Coriolis effect at high latitudes (thanks to rapid rotation), and equatorial, eastwards, angular momentum transport. Finally the divergent component of the wind plays a much more minor role, only transporting energy from the day-side hot-spot towards the terminators and poles, with a slight, rotationally influenced, preference for easterly flows, and little to no transport on the cold night-side. This wind balance is typical of hot Jupiters, and leads to the primarily equatorial heat transport that we discuss below, { heat transport} which is responsible for many of the the atmospheric features detected by our { thermal feature CNN, in particular the advected butterfly}. \\
On the other hand, the wind dynamics in the `cool' regime are rather different. Not only does the divergent component of the wind ($\left|\bm{u}_{div}\right| = 238 ms^{-1}$) dominate over the rotational component ($\left|\bm{u}_{rot}\right| = 119 ms^{-1}$), the difference between the rotational component and the eddy wind is small ($\left|\bm{u}_{eddy}\right| = 87 ms^{-1}$). Taken together, this suggests that, unlike in the `hot', or even classical hot Jupiter, regimes, the primary driver of this regimes horizontal dynamics is a divergent flow of material from the day-side hot-spot towards the colder night-side which dominates over a much weaker equatorial jet. However, the mechanism by which this very weak equatorial jet forms remains the same: $\bm{u}_{rot}$ and $\bm{u}_{eddy}$ both reveal a m=1 standing wave pattern which is slightly shifted west of the sub-stellar point, and which, thanks to the relatively weak influence that rotation has on the dynamics, is essentially untilted. Finally, is also interesting to note that the divergent component of the wind converges on the night-side just west of the anti-stellar point, nearly exactly where the hotspot can be found in \autoref{fig:Wind_Temp}e/f/g - as we discuss below, this is not a coincidence.  \\ 

\begin{figure*}
\gridline{\fig{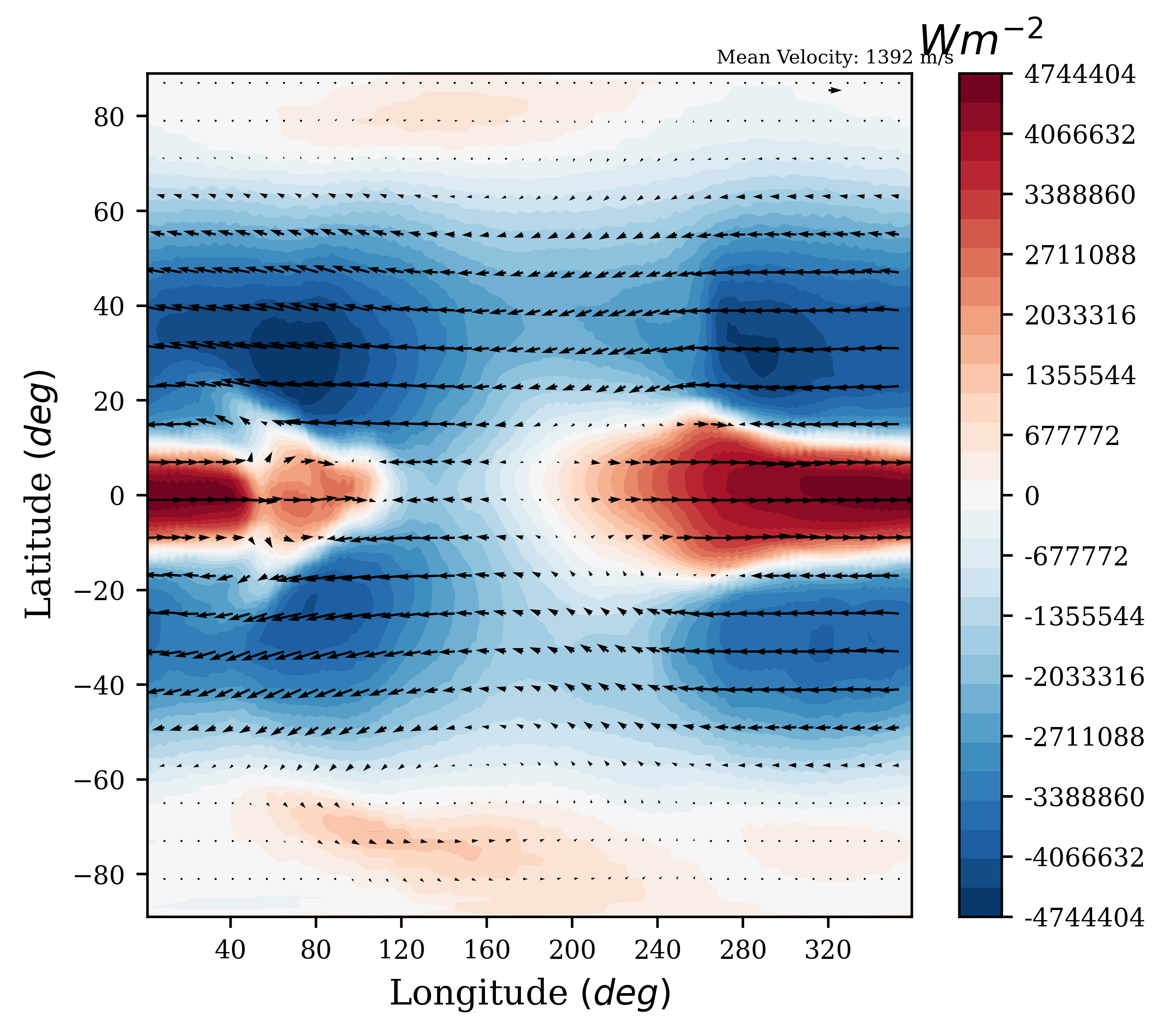}{0.24\textwidth}{a) `hot': Zonal - 0.0026}
          \fig{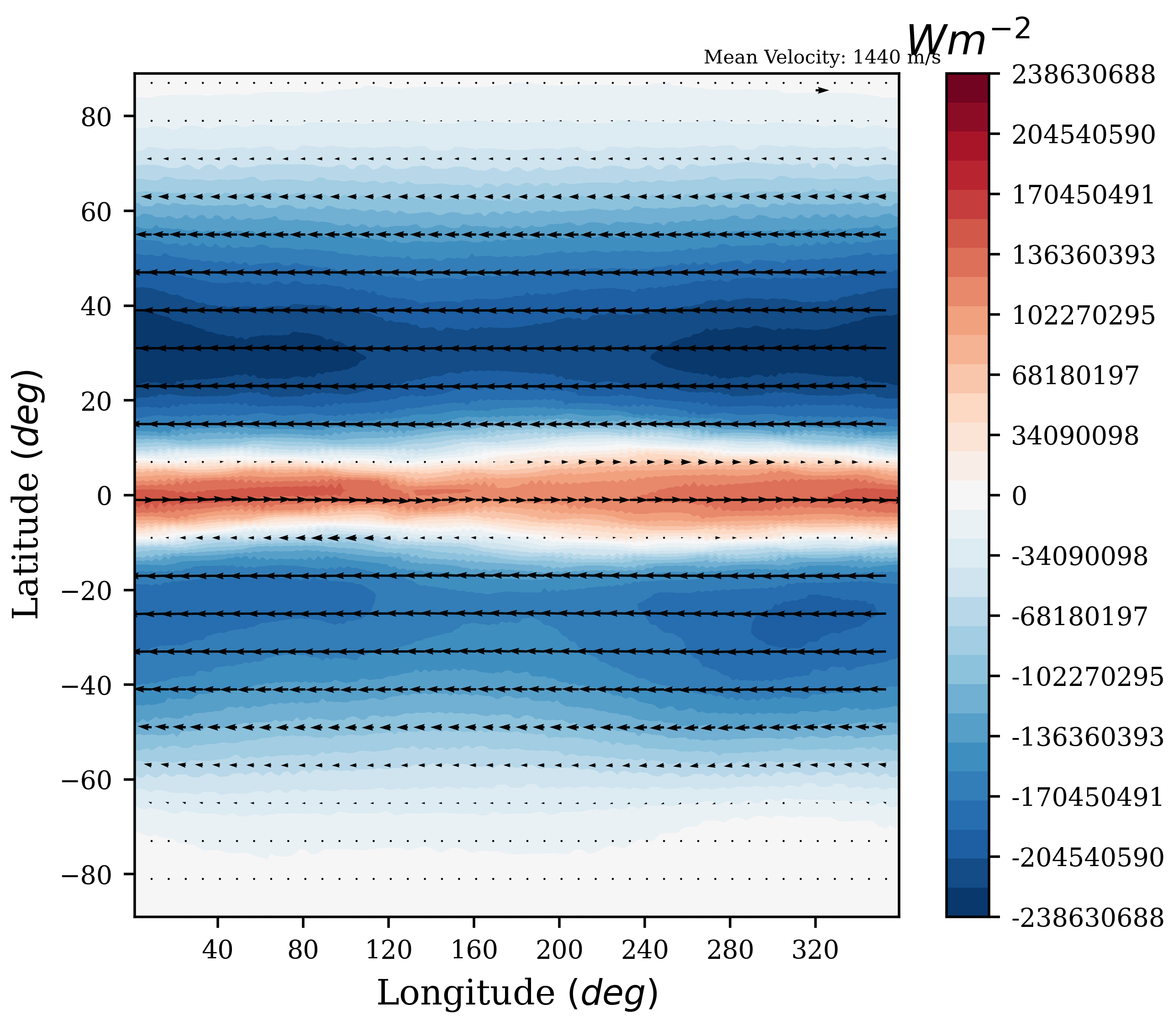}{0.24\textwidth}{b) `hot': Zonal - 0.2}
          \fig{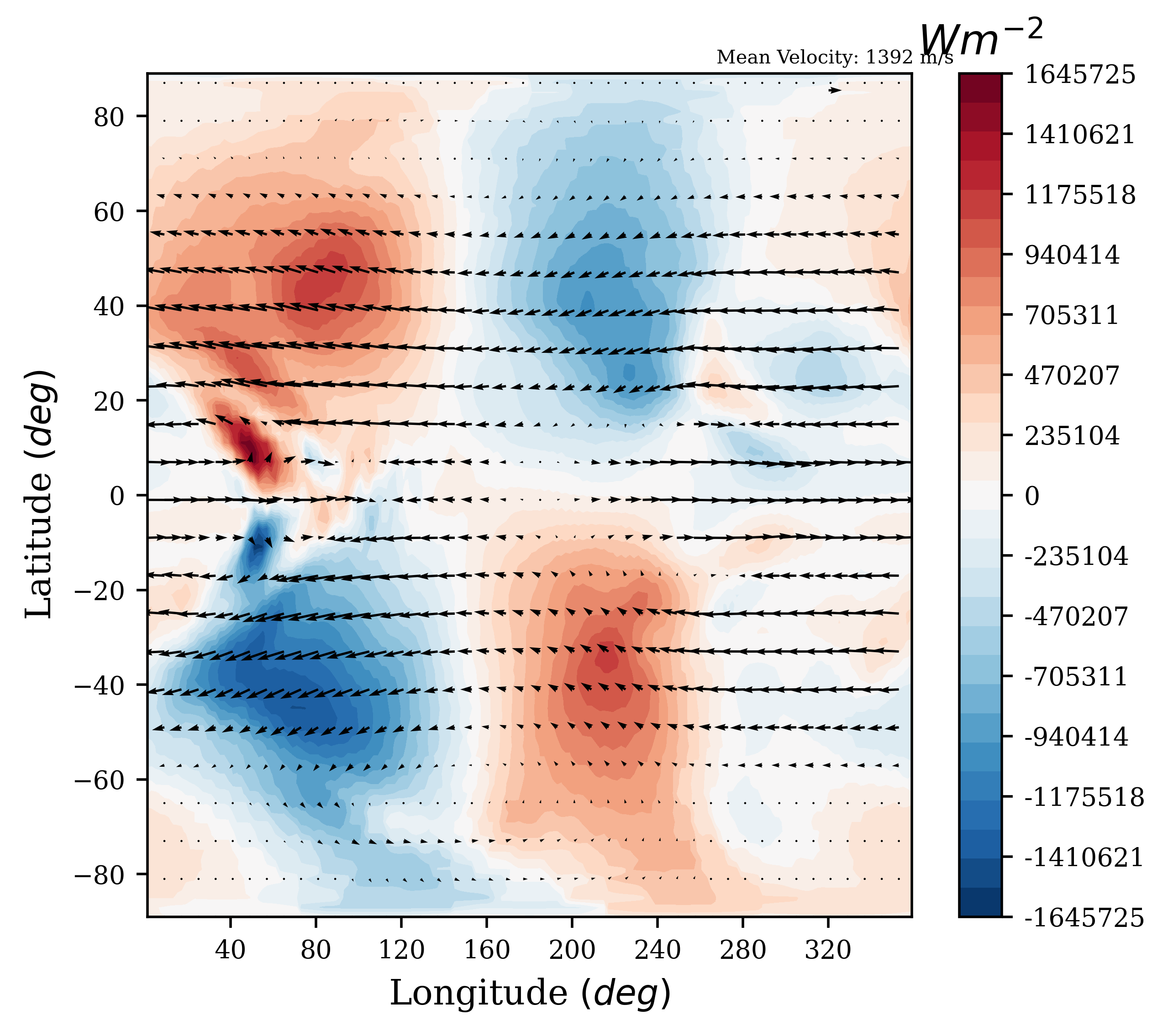}{0.24\textwidth}{c) `hot': Latitudinal - 0.0026}
          \fig{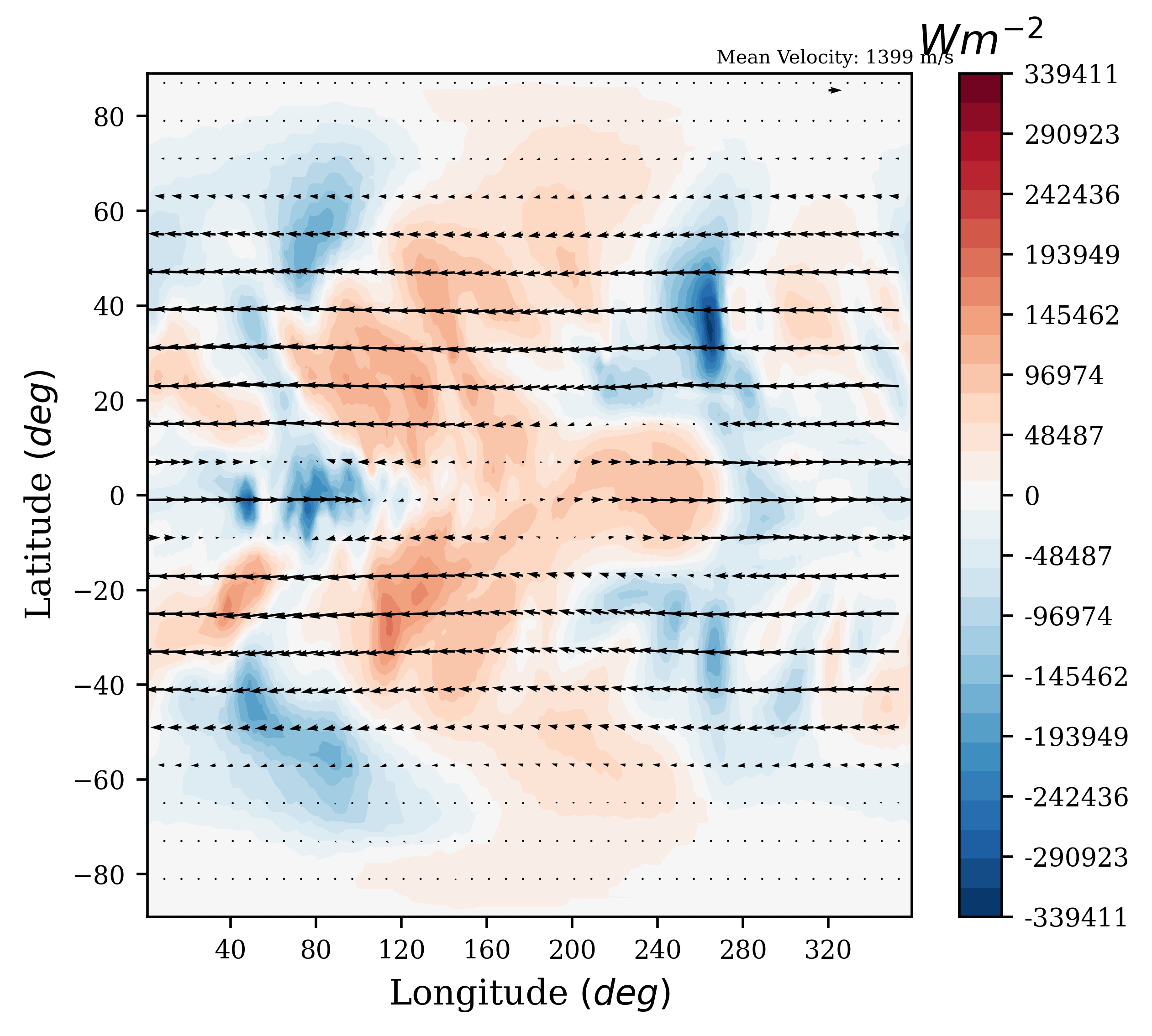}{0.24\textwidth}{d) `hot': Vertical - 0.016}}
\gridline{\fig{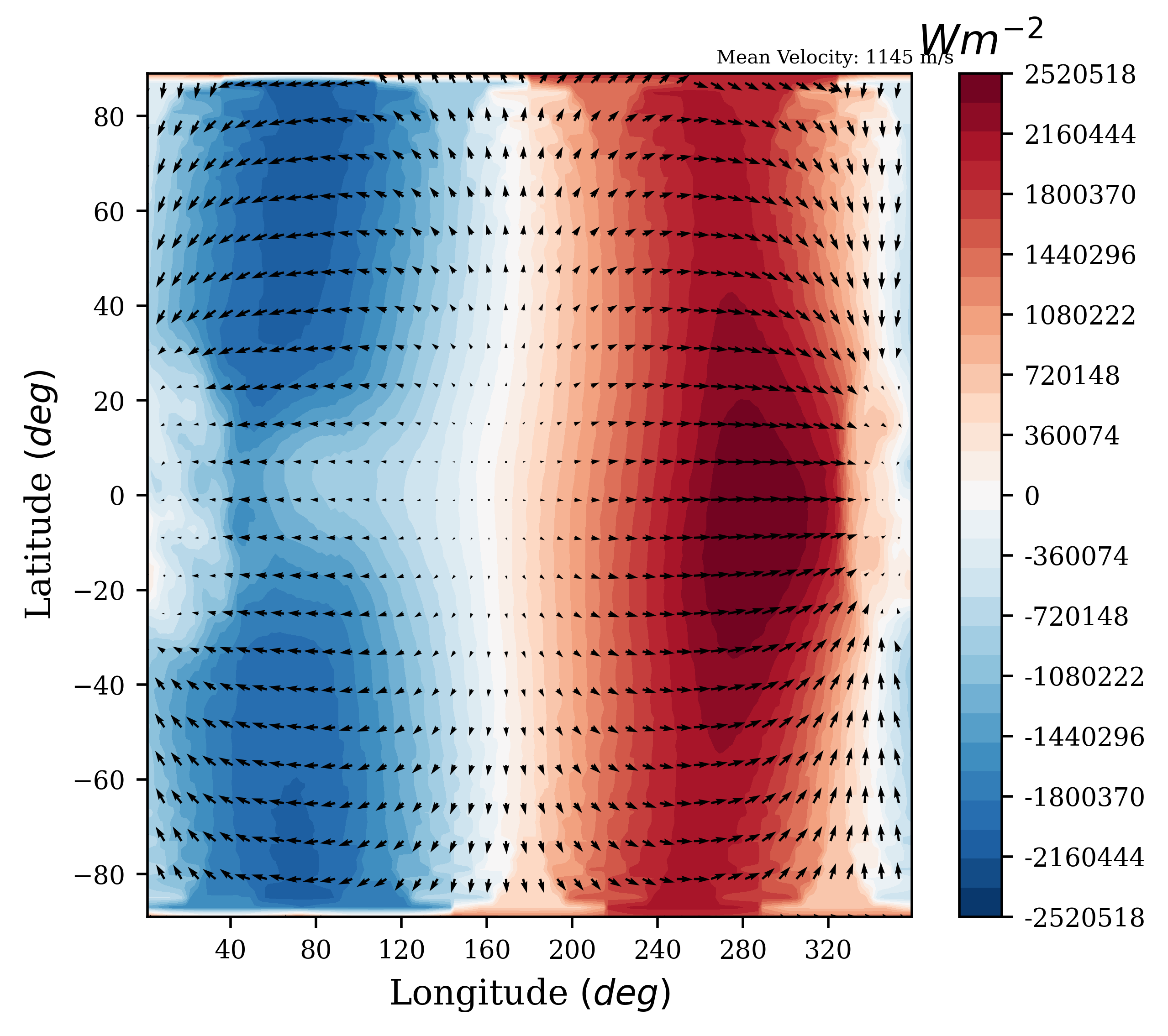}{0.24\textwidth}{e) `cool': Zonal - 0.0026}
          \fig{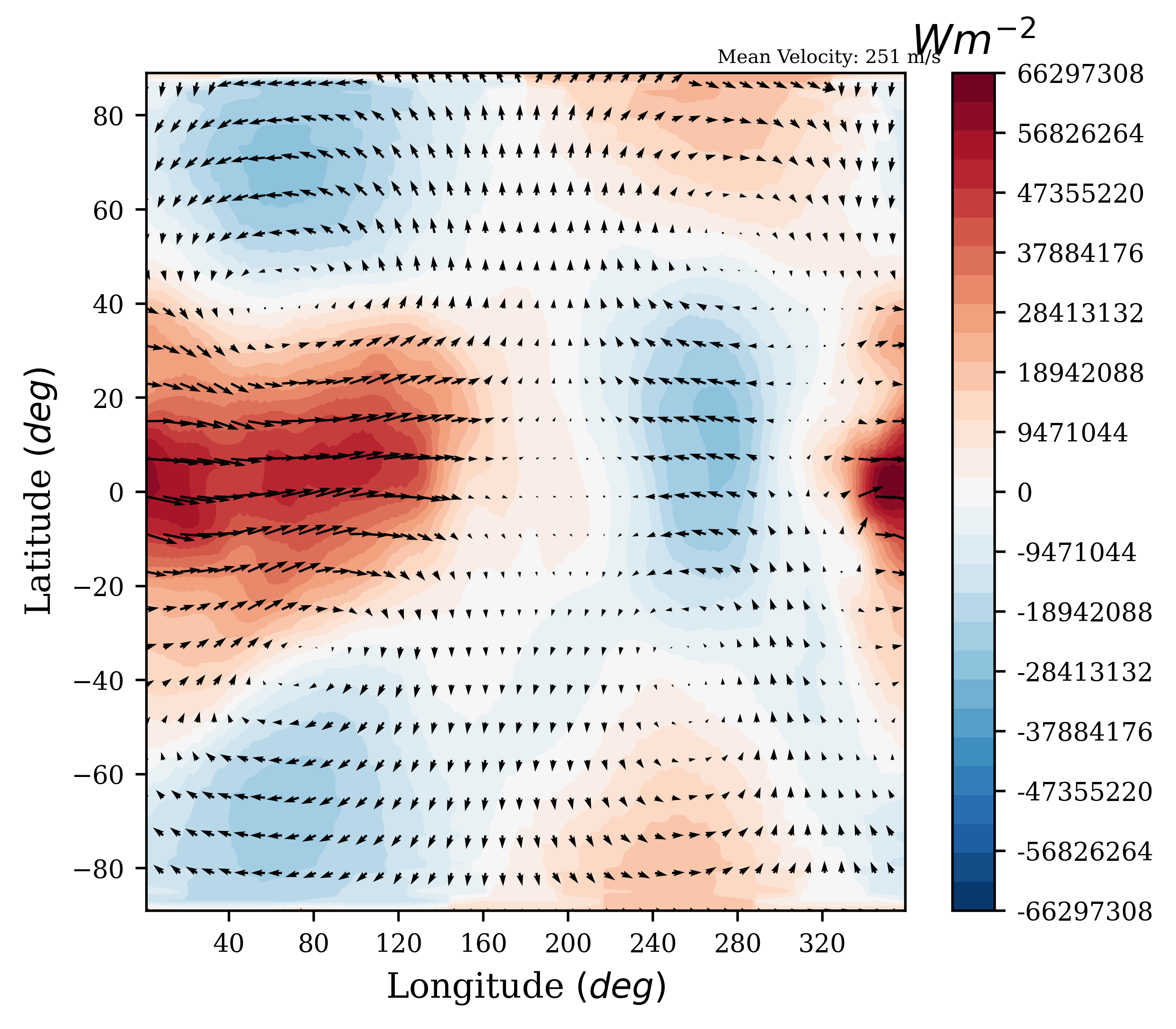}{0.24\textwidth}{f) `cool': Zonal - 0.2}
          \fig{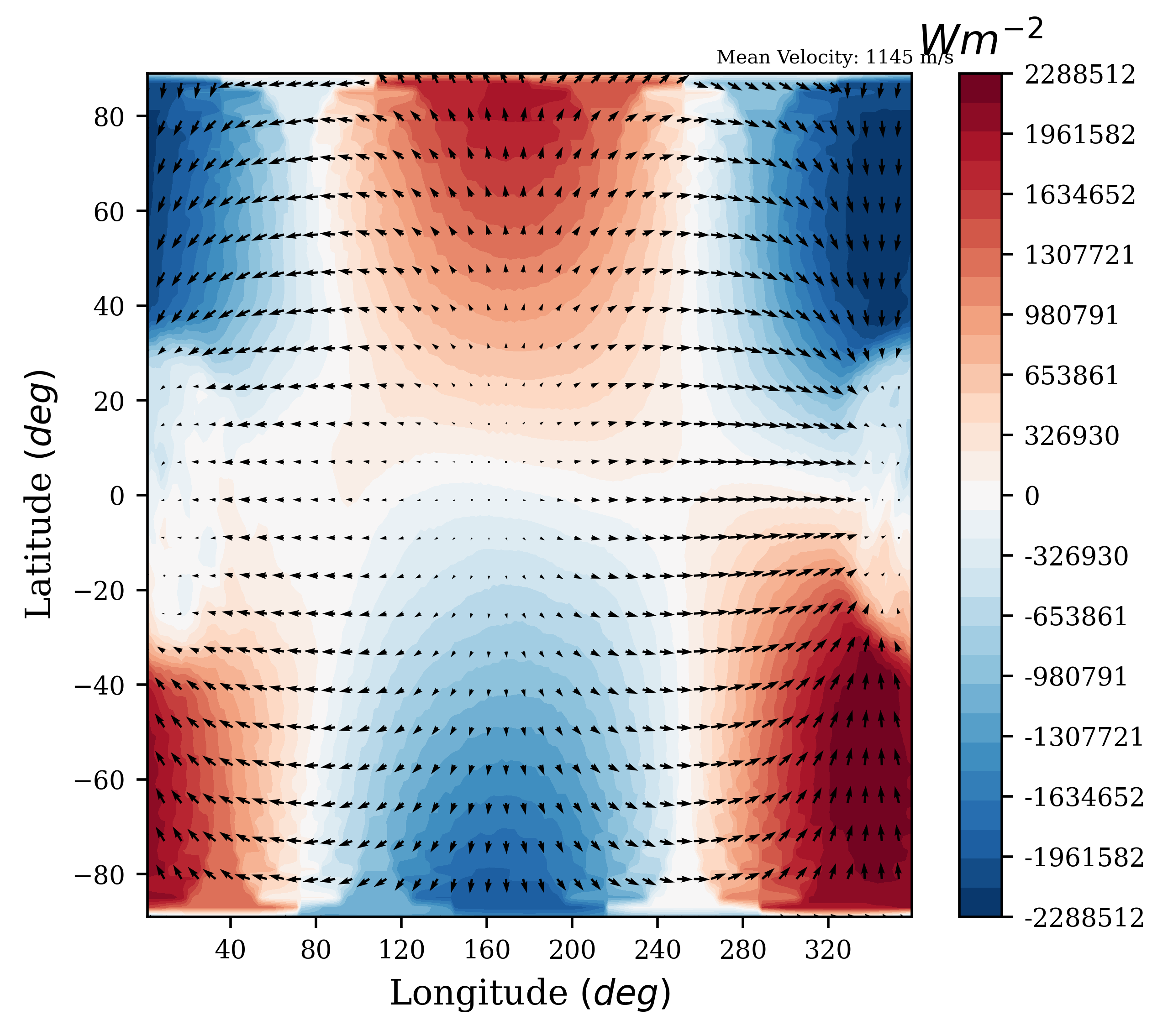}{0.24\textwidth}{g) `cool': Latitudinal - 0.0026}
          \fig{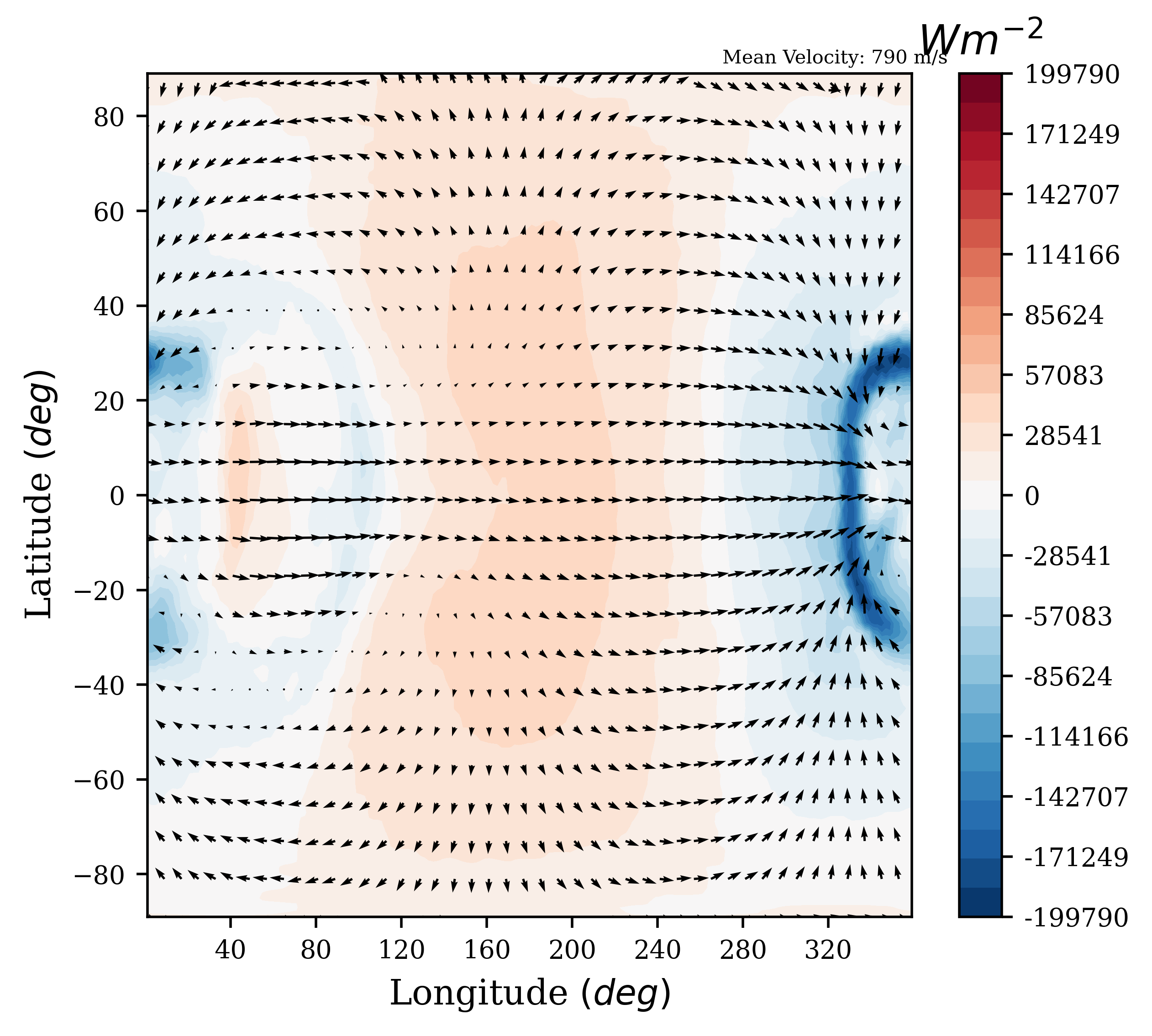}{0.24\textwidth}{d) `cool': Vertical - 0.016}}

\caption{ A selection of maps showing the temporally-averaged Zonal, Meridional, or Vertical Enthalpy transport at selected pressure levels for our exemplary `hot' (top) and `cool' (bottom)  HD209458b-like atmospheric models. Here, positive (red) fluxes represent eastward/polar/outwards flows for the the zonal/meridional/vertical enthalpy flux maps respectively. Note: We include two maps for the zonal enthalpy transport, at different pressure levels, in order to emphasise how the zonal advection changes with both height and orbital radius.\label{fig:Enthalpy_Maps} }
\end{figure*}

Given that our two models are in very different dynamical regimes, with very different horizontal wind structures, we next explore how these differences are reflected in the horizontal and vertical energy transport.  More specifically, in \autoref{fig:Enthalpy_Maps}, we explore the zonal, latitudinal, and vertical Enthalpy flux transport $E(u,v,w)=\rho*cp*T*\bm{u}(u,v,w)$ (where $\rho$ is the density, $T$ is the temperature, and $cp$ is the specific heat) at select pressures which were chosen in order to emphasise the differences in outer-atmosphere energetics/dynamics. \\

Starting in the `hot' regime, we find that, at all but the lowest of pressures where the radiative time-scale is very short (and hence advection is suppressed), the zonal enthalpy transport (e.g. \autoref{fig:Enthalpy_Maps}a/b) is dominated by strong eastward advection at the equator and significant westward advection off-equator. This advection can explain the thermal structure identified in the outer atmosphere by the CNNs: zonal advection leads to a shift in the day-side hotspot towards the east at the equator, and towards the west at mid-latitudes, leading to the well known butterfly-like structure (\autoref{fig:Wind_Temp}b/c) identified by the CNNs (\autoref{fig:Categorisation_0021}). \\
Moving onto the latitudinal enthalpy advection (\autoref{fig:Enthalpy_Maps}c), we find that it is strongly correlated with the eddy component of the Helmholtz wind decomposition. For example, by comparing this transport with \autoref{fig:Helmholtz}c, we see that the poleward and equatorward transport aligns well with the tilted standing wave pattern. This includes a peak in latitudinal enthalpy transport that occurs near the equator and west of the sub-stellar point, and which corresponds to a similar convergence/divergence point found in the eddy wind component. This correlation is reinforced by the relative magnitude of the latitudinal advection in comparison to the zonal transport - much like the eddy wind is much slower on average than the rotational wind (which includes the zonal jet), the latitudinal heat transport is much weaker than the zonal advection. \\
Interestingly, this difference in energy transport remains as we move into the deep atmosphere. Here, whilst we do see some slight signs of an asymmetric thermal structure around 10 bar, deeper than this, the strong vertical heat transport and significant longitudinal mixing has resulted in a deep atmosphere in which {\it zonal} temperature differences have almost completely vanished (\autoref{fig:Wind_Temp}d), leaving bands of temperature that vary latitudinally, with an enhancement in temperature near the poles thanks to the off-equator downflows (\autoref{fig:Enthalpy_Maps}d). Note: It is possible, and likely \citep{2019A&A...632A.114S}, that with enough time, the deep atmosphere will eventually mix further, reducing the latitudinal temperature differences and resulting in a deep atmosphere that is mostly horizontally homogenised, and hence may be identified by the CNNs as asymmetric if any small scale, residual, temperature variations are present.\\
An example of this vertical enthalpy transport, in the equilibrated outer atmosphere, can be seen in \autoref{fig:Enthalpy_Maps}d, where red/blue flows indicated upward/downward transport respectively. Here we see a slight upwelling on the day-side which can be linked to the hot-spot, surrounded by downwelling near the terminators, on the night-side, and at higher latitudes. This transport extends deep into the atmosphere, growing stronger as the pressure (and hence also density on the material being transported) increases, explaining the observed deep heating (\autoref{fig:Longitudinal_T}a).\\

We next turn to the `cool' regime, in which the wind dynamics, and hence also the enthalpy transport, differ significantly from the `hot' regime. This is best illustrated by the the outer atmosphere zonal and latitudinal enthalpy flux, which we plot in \autoref{fig:Enthalpy_Maps}e/g respectively. Here we find that the mean zonal and latitudinal enthalpy fluxes are approximately equal, and are very strongly shaped by the divergent component of the wind (\autoref{fig:Helmholtz}d) - with clear transport occurring from the hot day-side to the cooler night-side, both zonally and latitudinally across the poles, converging just west of the anti-stellar point, exactly where the deeper night-side hot-spot, and thermal inversion, is found (e.g. \autoref{fig:Wind_Temp}e/f/g). This divergence driven transport also explains why we do not detect/observe the day-side butterfly that is typically associated with hot Jupiter atmospheres: due to the relatively low influence that rotation plays on the dynamics, the day-night energy transport is primarily, isotropically, divergent, rather than the highly anisotropic (equatorial) transport found for the `hot' regime. As such the temperature structure in the outer atmosphere remains largely `locked' (and maybe a little broadened) until enough energy has been transported by the divergent flows to form a night-side hotspot (hence changing the tag applied by the CNN - see \autoref{fig:Categorisation_0192}b). \\
Moving deeper into the atmosphere, we start to see the impact that the night-side hot spot has on the mid-atmospheres horizontal enthalpy transport. Here we find that the zonal-enthalpy flux (\autoref{fig:Enthalpy_Maps}f), much like the rotational component of the zonal wind (\autoref{fig:Helmholtz}e), is dominated by an eastwards flow from the night-side hotspot towards the relatively cool day-side. This suggests that the divergent flows are dominant in the outer atmosphere, where the day-night forcing is strong, and `rotational' flows are dominate in the mid atmosphere, where the forcing has switched to being driven by the night-side hotspot, reinforced by the eddy winds (\autoref{fig:Helmholtz}f).  \\
Finally we come to the vertical enthalpy transport (see for example, \autoref{fig:Enthalpy_Maps}h), which is both weaker than the vertical energy transport found in the `hot' regime, as well as being more vertically confined. More specifically, we find that the main region of strong downward enthalpy transport is focused on the night-side around the hot-spot, and like the hot-spot itself, this downward transport does not extend into the deep atmosphere. Instead we find that vertical mixing is weak in the deep atmosphere, helping to explain the lack of observed deep heating (\autoref{fig:Longitudinal_T}b) and hence radius inflation. In fact, mixing is generally weak in all directions in the deep atmosphere, although it is slightly stronger in the zonal direction than vertically or latitudinally. This helps to explain the slight asymmetry seen in the deep atmosphere (and identified by the { thermal CNN}). Small amounts of thermal energy are transported to the deep atmosphere by vertical mixing (in an essentially random way since the flow is so weak), leading to slight temperature variations that become smoothed out longitudinally but not latitudinally - hence leaving us with a weak asymmetric thermal structure, as seen in \autoref{fig:Wind_Temp}h. Note that, unlike in the `hot' regime above, the particularly slow dynamical timescales of the deep atmosphere found here mean that we do not expect complete horizontal homogenisation to occur on any reasonable timescale. \\

\subsubsection{Difficulties, Warnings, and Advice for Using CNN(s) to Analyse Atmospheric Dynamics} \label{sec:AI_Characterisation_2}

Whilst the above results show some promising outcomes of pairing deep-learning neural networks with (atmospheric) simulations, this process is not without its own problems and limitations, which we discuss here. \\

One particular important factor that can impact the ability of CNN(s) to detect atmospheric features is the choice of colourmap used to visualise the data since this can significantly impact the ability of the neural-network to detect the edges and gradients that are key to encoding detectable features during the training process.
As part of this work, we tested training the thermal, and wind, CNN with every available colourmap included as part of MatPlotLib, using training data that was, other than the colourmap, identically tagged. Whilst the results were similar for many of the colourmaps chosen, a few stood out both positively and negatively. { Note that the effectiveness of each set of variable colourmap networks was evaluated using their confusion matrices, allowing for a direct comparison of the ability of each set of models to reproduce the set of known tags. } \\
For example, the colourmap `Jet', which used to be the default map used in MatPlotLib, and which remains a staple of astrophysics research to this day, resulted in particularly poor detection and characterisation of atmospheric features. This occurs for exactly the same reason why it was replaced as the default colourmap in MatPlotLib: due to rapid colour and brightness changes (i.e. the colourmap is not perceptionally uniform), `Jet' tends to both exaggerate and suppress small changes/gradients in the underlying data - whilst this can be an advantage when quickly inspecting data visually (but even this is in doubt), it also significantly hampers the ability of the multi-categorisation CNN to detect and isolate atmospheric features, either through mis-training, because the feature is hidden, or even because the colourmap has warped the morphology of the feature by exaggerating/suppressing gradients. \\
Interestingly, even the modern replacements, such as Viridis, Inferno, or Plasma, which were designed to solve this non-perceptional-uniformity problem are also poor choices for use with a CNN. This can be linked to their use of a large array of colours, which again can act to disguise, morph, and exaggerate/suppress features, as proved by the best colourmaps for use { with image-recognition algorithms}: monochromatic colourmaps. \\
Simply put, by using a monochromatic colourmap (such as Grays, Reds, Gist\_gray, etc), simulations with the same atmospheric feature (but with different magnitudes and horizontal gradients, due to, for example, differences in surface irradiation strength) are more likely to produce outputs that are visually near identical, making it easier for a CNN to both learn and detect said features. Furthermore, the use of monochromatic colourmaps can slightly reduce the memory footprint of the CNN by reducing the initial data-set-size by a factor of three, from RGB, to greyscale, thus, theoretically,  enabling either faster data processing whilst the simulations run, or larger and more complex CNNs (with either more complex layers, or more layers generally) to be developed in the same memory footprint. { However, as detailed in the Appendix, the CNN(s) we consider here are already fairly lightweight when compared to those used for, say, facial recognition, and so the possibilities for computational savings are minimal - for example, the `high-resolution', full RGB, network only take 4 minutes to train on four K80s (which themselves are relatively old). Furthermore, testing with training our thermal CNN using normalised raw data rather than images produced with matplotlib resulted in a only a small reduction, 0.3\% for our thermal CNN, in the size of the neural network. This occurred at the cost of introducing significant problems with the optimiser leading to significant oscillations during the validation phase of training the network. Whilst fixing such issues should be possible, it would require modifications to the structure of the network and individual layers, complicating issues for a non-AI expert all for a minimal reduction in computational cost and very limited potential improvement over a well chosen colormap.  }  \\
However this all comes at the cost of making the data harder to interpret visually/manually. The compromise, in terms of generating data that is suitable for use with both CNNs and humans, albeit at the cost of not reducing the memory footprint, is to use a diverging colourmap: i.e. a colour map with only two primary colours which diverge from a neutral colour, such as white, at either a fixed point (e.g. zero wind) or at the data mean, as used in, for example, \autoref{fig:Wind_Temp}. These colourmaps result in similar accuracy to monochromatic colourmaps for the { thermal CNN} when trying to recover training data, whilst also generating plots with which humans can easily visualise atmospheric features/dynamics. \\
Note that, whilst the above results are fairly robust, we advise that future studies, particularly those that intend to use { multi-categorisation} CNNs to analyse data unsupervised (i.e. during a simulations runtime), should fine tune both the colourmaps as well as the data boundaries in order to ensure suitability before committing to the run. \\

In addition to ensuring that the data is prepared in such a way as to be suitable for use with { a CNN}, we also have to ensure that the features being searched for are also detectable and are properly trained for. \\
This proved to be an issue here. As discussed in \autoref{sec:AI_categories_init}, our original plan included using a separate CNN to explore the presence of zonal jets/horizontal winds in our simulation data. This proved to be highly intractable for a variety of reasons. To start, we initially tried to train the CNN by using the, overlaid, arrow quivers shown in \autoref{fig:Wind_Temp}. However this ended up being nearly impossible since, at the resolution of the CNNs analysis, the arrows were essentially undetectable. Furthermore, even when we adjusted the plots to make the quivers more visible/bolder, { or increased the resolution of the initial layers of the network,} small changes in the horizontal wind structure lead to mis/failed detection of the jet - it appears that the localised nature of the quivers { not only} made the CNN very sensitive to changes in their structure { but also meant that the model struggled to detect large scale structure in the first place}. { It is possible that a significantly larger training data set, larger quivers, higher resolution images, or a different kernel size might solve this issue. } \\
{ Rather than testing the aforementioned ideas, which we leave to future work, we instead tried to avoid the quiver issue completely by replacing them} with horizontal wind maps, much like those used { with the thermal CNN}. This greatly improved the ability of our CNN to detect the zonal winds, but it also revealed two additional limitations. Firstly the zonal wind is not as horizontally homogenised as the zonal-mean zonal wind would suggest, thus making it trickier for the CNN to detect unless very carefully trained with a wide variety of zonal wind structures, and secondly, even when carefully trained, the CNN can have a difficult time detecting zonal jets due to their highly symmetric structure (similar difficulties are also faced when searching for the banded { structure with the thermal CNN}, however it is less of a problem there as alternative structures, which may lead to misidentification, are not so prevalent due to the relative quiescence of the deep atmosphere). More specifically, our CNNs can find it more difficult to isolate a band of temperature/wind than a more complex structure, such as a night-side hot-spot or a thermal butterfly. This is because said complex structures have more features (edges) for the neural-network to latch onto and learn, increasing the complexity of the neural-network and hence also its ability to internalise features. As a consequence of all of the above, we have not focused on the detection of zonal winds in this work, and we suggest that future studies which wish to implement data analysis via neural networks should focus on more detailed and/or derived atmospheric features, ranging from thermal structures to complex wind dynamics, such as vortices or (e.g. by using a Helmholtz decomposition) standing waves { (although here we again caution against using quivers due to their localised nature)}. \\

Another example of the possible difficulties faced when our { thermal CNN} to search for atmospheric features is the non-detection of the thermal butterfly structure at later times in our exemplary `hot' regime model (\autoref{fig:Categorisation_0021}). Comparisons between the training data, which was based upon early outputs of our full simulation set, and the steady-state outer atmosphere thermal profiles reveals the simple reason why this is the case: the strong equatorial jet and off-equator, low-latitude, counterflows has resulted in a butterfly structure that is much more longitudinally extended/stretched than any profiles included in the initial training data. As such, our trained { CNN was} unable to properly identify the evolved/advected feature. This serves as an important caveat of CNNs and neural networks in general: they are only as effective as their training data, and are not able to identify new or highly evolved/warped features. However, as previously discussed, this lack of detection itself can also be an interesting result, identifying new, uncommon, or unexpected features. Yet this is a cold comfort when trying to use a CNN, or AI more generally, to concurrently process an ongoing simulation - here the best solution is a broad training set which contains multiple examples of all the features expected. For example, now that we have steady-state training data for a wide variety of HD209458b-like atmospheric models at different orbital radii, and with equilibrium outer atmospheres, it should be possible to build CNNs to accurately, and quickly, analyse future simulations within, or near, this parameter space on the fly.

\section{Discussion and Conclusion} \label{sec:Conclusions}
In this paper, we have explored the role that AI driven image-classification, via the use of convolutional neural-networks, can have in the concurrent- and post-processing of simulations of planetary atmospheres, specifically HD209458b-like hot Jupiters at various orbital radii. \\
\subsection{Model and AI (CNN) Setup}
To that end, we started by running a series of HD209458b-like atmospheric models with different orbital radii, and hence different surface irradiation and synchronous planetary rotation rates. The orbital radii considered here varied between $0.012\textrm{au}$ and $0.334\textrm{au}$, but for the sake of brevity we chose to focus on models at two different orbital radii, with dynamics that are characteristic of their contemporaries. Specifically we focused on one model with an orbital radius of $0.021\textrm{au}$, which we refer to as our `hot' model, and one with an orbital radii of $0.192\textrm{au}$, which we refer to as our `cool' model. These simulations are based on those first presented in \citet{2019A&A...632A.114S}, but modified such that their outer atmosphere temperature forcing is derived from 1D models calculated using ATMO at every orbital radius of interest. From the first outputs of these simulations, we then selected and labelled a number of thermal and wind atmospheric features that we wished for our AI model to detect/characterise. These features included, the presence of a tidally locked day-side hot-spot, a horizontally advected day-side hot-spot (better known as a butterfly-like thermal structure), a latitudinally asymmetric deep atmosphere, a deep atmosphere which is fully longitudinally homogenised, and in which latitudinal temperature variations remains small (we we refer to as banded), and the presence of a super-rotating equatorial jet, although this latter tag/feature proved to be difficult to identify. \\
This hand labelled data-set was then fed into { a pair of multi-categorisation} convolutional neural-networks - i.e. a neural-network which is particularly suited to image recognition tasks (see \autoref{sec:AI} for a more detailed description) { and detecting multiple features non-exclusively}. Once trained, this neural network was then applied to the full time-series outputs of our exemplary simulations (which had been run to steady-state in the outer atmosphere - i.e a lower pressures). \\
\subsection{Identified Features, or Lack Thereof, with CNNs}
Applying the trained CNN(s) to our exemplary models revealed our first key result: at higher orbital radii, i.e. in the `cool' regime, { the thermal CNN multi-classification map} (which show the identified features verses both pressure and time) contained a region with no identified atmospheric features. The resulting analysis of this model, as well as other `cool' regime models, revealed a mid-pressure atmosphere (i.e. $0.05\textrm{bar}\rightarrow1\textrm{bar}$) that was behaving in a rather unusual way: The hottest region of the atmosphere had shifted from the irradiated day-side to the night-side, ending up just west of the anti-stellar point. Once this data was added { to the thermal CNN} (using interpolative oversampling to generate artificial training data, a necessity due to the small sample size available as this feature only occurs in the `cool' regime and over a limited pressure range), we were able to successfully identify this hot-spot, and its associated thermal inversion in all models with an orbital radii $>0.11$au - i.e. all `cool' regime models. \\

This is just one example of the differences observed between models in the `cool' and `hot' regimes, differences which extend throughout the model atmospheres and which appear to be highly linked to the mixing/transport/circulation regime that the models fall into. Only our analysis of the low-resolution and adiabatically initialised HD209458b model of \citet{2019A&A...632A.114S} revealed every, original (i.e. excluding the night-side hot-spot) feature, which was to be expected since it was HD209458b's dynamics that formed the basis for the original features selected for detection. \\
Starting in the `cool' regime, the identified features tended to correspond to weaker mixing and anisotropic horizontal energy transport - that is to say that the outer atmospheres dynamics remain highly radiatively forced (despite the relatively weak stellar irradiation), the mid-atmosphere is dominated by isotropic (i.e. divergent) energy transport from the day-side to the night-side, which eventually leads to the formation of a night-side hot-spot and associated thermal inversion, and the deep atmosphere is highly quiescent with weak mixing and deep heating that allow for slight latitudinal temperature gradients to develop and be maintained, which the { CNN identifies} as an asymmetric thermal structure. \\
On the other hand, for models that fall into the `hot' regime, we found that the identified dynamics correspond to strong zonal energy transport and significant horizontal, and vertical, mixing. For example, we find that in the outer atmosphere, once the radiative time-scale is long enough (i.e. not at very low pressures), the day-side hot-spot becomes significantly horizontally advected by both the equatorial jet as well as the associated mid-latitude counterflows, resulting in the well known butterfly-like thermal-structure on the day-side, with the exact shape depending upon the strength of the rotational influence, and hence the jet structure. This strong advection/mixing extends to the deep atmosphere, where not only do we find significant heating thanks to vertical potential temperature (enthalpy) transport, but also that the deep horizontal advection, which is strongest in the longitudinal direction, has resulted in strong zonal homogenisation paired with a weak latitudinal temperature gradient - this is the temperature structure we refer to as banded. \\
\subsection{Understanding the Different Dynamical Regimes}
In order to try and understand the differences between these two dynamical regimes, and also how the unusual night-side hot-spot and thermal inversion forms in the `cool' regime, we next explored the wind and energy transport (specifically enthalpy flux) in more detail. \\ 
Starting with the wind, we use a Helmholtz decomposition to split the horizontal wind into its divergent (i.e. `vorticity free'),  rotational (i.e. `divergence free'), and eddy (i.e. perturbations to the rotational wind) components. This reveals that, as expected, the wind dynamics differ significantly between the `cool' and `hot' regimes. \\
In the `hot' regime, strong stellar irradiation and rapid surface rotation mean that the wind dynamics are dominated by the rotational component of the wind. This is in agreement with \citet{2011ApJ...738...71S}, who suggest that the rotational component of the wind is correlated with standing Rossby and Kelvin waves that in turn drive a strong, highly advecting, equatorial jet: here the rotational component of the wind reveals both a strong equatorial jet as well as a significant, m=1, standing wave pattern.  \\
On the other hand, in the `cool' regime where the irradiation is weaker and the surface rotation is slower,the winds fall into a completely different dynamical regime: The divergent component of the wind dominates over the rotational component, although signs of the latter remain, exhibiting a weak, but stable, m=1, standing wave pattern that fails to drive a significant equatorial jet. As for the dominant divergent wind, this flows, isotropically, from the sub-stellar point to the unirradiated night-side, converging just west of the anti-stellar point. i.e. exactly the same location as the mid-atmosphere night-side hot-spot. \\

The horizontal enthalpy transport reveals that this is not a coincidence. For instance, our `cool' model reveals zonal and latitudinal enthalpy transport that is highly shaped by the divergent component of the wind, and hence, also converges just west of the anti-stellar point. This explains the formation of the night-side hot-spot at mid-pressures: In the very outer atmosphere, radiative forcing is too strong for significant advection to occur, however as we move deeper, the radiative time-scale lengthens and day-night advection starts to occur (isotropically thanks to the wind structure), leading to the formation of the night-side hotspot. This hot-spot then dominates the mid-atmospheres energy transport, leading to the unusual scenario that is night-day heat transport. However relative to the transport seen in HD209458b, or the `hot' regime, this transport is weak and does not extend into the deep atmosphere. this is reflected in the vertical enthalpy transport profile, which reveals that vertical advection is focused on maintaining the night-side hot-spot, leading to little to no deep heating and a quiescent deep atmosphere. This link between the night-side hot-spot and the divergent wind is further reinforced by more rapidly rotating `cool' regime models: As the influence that rotation has on the atmosphere rises, the location of both the hot-spot, as well as the divergent wind convergence point shifts westwards, likely as a result of the slight tilt introduced to the divergent wind by off-equator Coriolis forces. \\
A complementary result is found in the `hot' regime, although here, as discussed above, the enthalpy transport is controlled by the rotational wind, resulting in transport that is primarily driven by the zonal-jet, with weaker off-equator transport linked to the m=1 standing wave pattern. As such, we find significant horizontal advection near the equator, either eastwards where the jet dominates, or westwards off-equator where the standing-wave driven transport is strongest. Taken together, this transport results in the formation of the synonymous butterfly-like thermal structure. At higher latitudes, we also find evidence for the influence that Coriolis forces have on enthalpy transport, with a significant westward tilt developing as we move towards more rapid rotation.
As for the off-equator longitudinal and latitudinal enthalpy transport, this is highly correlated with the same Rossby and Kelvin standing wave pattern that drives the equatorial flows, including the significant westwards tilt as we move to higher latitudes, a tilt which is highly dependent upon the planetary rotation rate. Note however that, at very high rotation rates, off-equator heat transport is suppressed as the Coriolis force suppresses higher latitude winds. Finally, as suggested by \citet{2019A&A...632A.114S}, the strong zonal wind, and hence zonal enthalpy transport, that develops in this model results in significant vertical heat transport that extends from the outer atmosphere all the way to the bottom of the simulation domain, increasing the entropy of the internal adiabat, and hence driving significant radius inflation (as shown in \autoref{fig:Longitudinal_T}a).  \\

\subsection{Limitations and Advice for Future Pairings of CNNs with Atmospheric Models}
Whilst our above results show some promising outcomes of pairing neural networks with atmospheric simulations, it also reveals some of the limitations of this approach, as well as potential pitfalls and avenues for improvement. \\
To start, CNNs cannot detect any features for which they are not robustly trained. For example, our initial { training and validation data set} did not include the uncommon night-side hot-spot found in the `cool' regime, and thus, when we feed the data into our networks, no tags were assigned to the mid-atmosphere. More subtly, whilst we did train our { thermal CNN to detect butterfly-like} features, at rapid rotation rates the strong zonal jet and latitudinally compressed mid-latitude counterflows resulted in a butterfly-like thermal structure that was notably different from those included in our training set. Consequently, the { thermal CNN was} only able to assign the butterfly tag in our exemplary `hot' model near initialisation, before advection had significantly changed its structure. \\
Both of these examples emphasise how important a robust training set is when using neural-networks (and CNNs in particular) to analyse data. This is doubly true when trying to use a neural-network to concurrently analyse and process a running model, since there a result that is only revealed through further study of an unusual region of non-identification occurs too late to be of use in deciding which runs to continue, which to discard, or which have reached equilibrium. \\
However this does not mean that CNNs (and neural-networks more generally) cannot be used for concurrent-processing. One area for which they are particularly well suited is when paired with next-generation exascale super-computers. These next-generation machines will allow for incredibly high resolution, and long-timescale, simulations, albeit at the cost of vast amounts of computational resources that will be in high demand (and which carry a high cost both financially and environmentally). As such it is critical to ensure that the allocated computational resources are being used efficiently, as well as minimising researcher time required to analyse the vast outputs of these models. Thankfully, as part of the process of designing an exascale-calculation, it is typical to run a series of lower-cost (i.e. lower-resolution or shorter timescale), preliminary, simulations in order to both get a sense of the value/significance of a very-large-scale simulation, as well as the exact model parameters to use (for example, with DYNAMICO, it is import to recalibrate the diffusion time-scale since the models hyper-diffusion is resolution dependent). These preliminary models might provide a source of training data which can be used to generate a set of CNNs to analyse the final, production, simulation. In a similar vain, with our series of outer atmosphere equilibrium HD209458b-like models at different orbital radii, we are now have enough data to train { a robust thermal-data CNN} to analyse any models we want to run at intermediary orbital radii. For example, we showed that the networks trained here on our `high' resolution, HD209458-like models, could be applied to analyse the lower resolution HD209458b models of \citet{2019A&A...632A.114S}. { Furthermore, the validity of such a technique could be further verified, in future studies, by analysing the CNN(s) with a post-processing tool such as GradCAM \citep{2016arXiv161002391S}, which highlights important features in a sample image, allowing for confirmation that the CNN(s) predictions can be trusted. } \\

\subsection{Future Perspectives for Atmospheric Modelling}
As we have alluded to above, the primary differentiator between the two regimes discussed here appears to be the relative influence that rotation has on the dynamics, specifically on the horizontal wind structure. Furthermore understanding these differences their impacts may prove crucial to our understanding of hot Jupiters in longer orbits, orbits which are now becoming accessible to observations thanks to next generation telescopes, such as JWST (James Webb Space Telescope) or TESS (Transiting Exoplanet Survey Satellite). One example of this is the unusual feature we detected in the atmospheres of our more slowly rotating hot Jupiters models, a night-side hot-spot, which was first detected via our thermal CNN, thus reinforcing the value of pairing long-timescale, and computational complex models with trained networks for both concurrent- and post-processing. If this feature proves to be robust, it may have implications for our understanding of hot Jupiter atmospheric chemistry. For example, a thermal inversion on the night-side may significantly impact the distribution of chemical compounds by acting as a cold trap in which denser materials condense, essentially raining out of the outer atmosphere, thus becoming depleted.\\ 
As a result, we have a number of suggestions for future studies: Firstly we strongly encourage any future studies with high-resolution and long-time-scale simulations to consider pairing them with an AI model, if only to reduce the the initial analysis burden, with the AI helping to identify regions of interest or uncommon dynamics. Secondly, we propose a more indepth study isolating how rotation alone affects irradiated atmospheric dynamics, with a particular focus on day-night winds and energy transport in slowly irradiated Jupiters: i.e. on the unusual night-side hotspot. If this result still proves to be robust, we further suggest that this work be followed up with a next-generation GCM, which includes both non-equilibrium chemistry and robust radiative dynamics, to fully explore what possible impacts that a night-side hot-spot may have on observable dynamics. 

\appendix
\begin{figure*}[tbp] %
\begin{centering}
\includegraphics[width=0.6\textwidth]{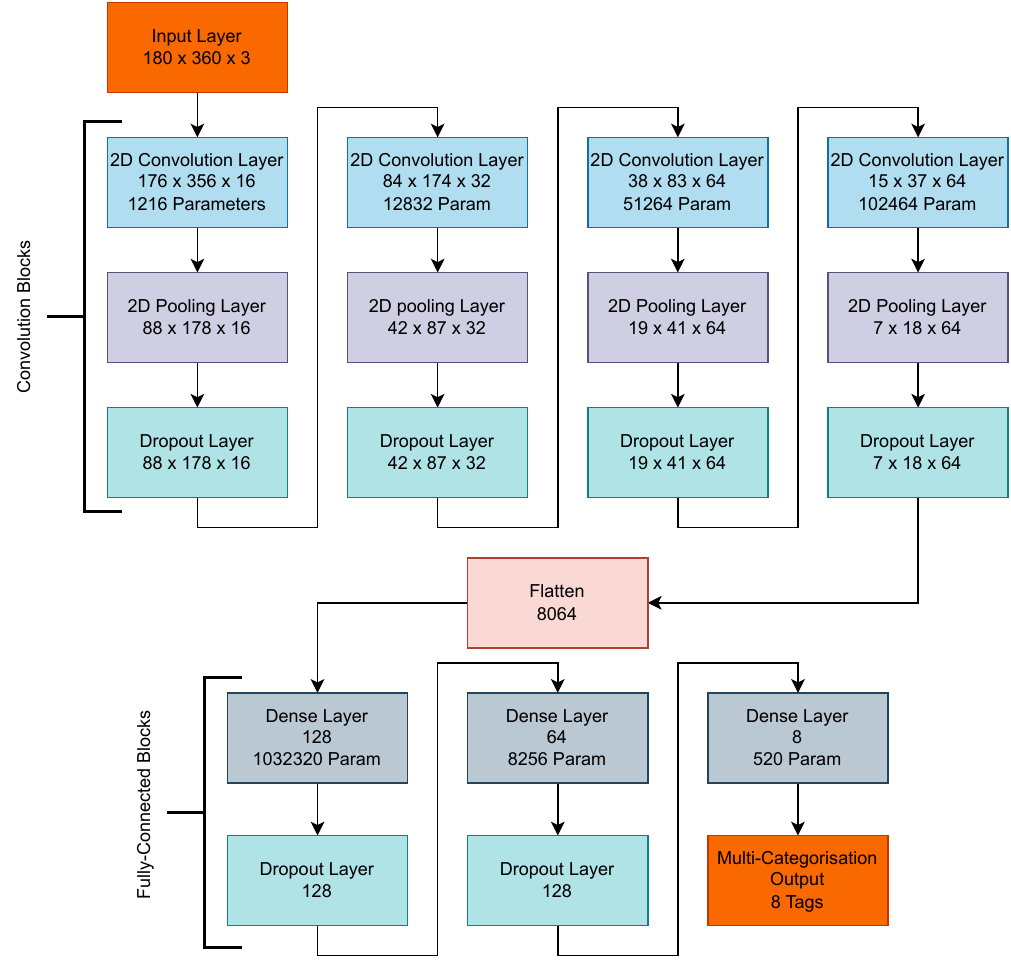}
\caption{ Flowchart showing the layers and blocks of a half resolution (initial image resolution is 180 by 360, half that produced by matplotlib for print quality figures) image recognition network designed to detect the 6 features of interest in the thermal structures of our HD209458b-like atmospheric models. Note that in each box we give the name of the layer, the output resolution of the layer, as well as the number of trainable parameters, when non-zero. \label{fig:Ai_flowchart} }
\end{centering}
\end{figure*}
{ Following and expanding upon the work of \citet{DeepLearningforSpatiallyExplicitPredictionofSynopticScaleFronts}, we implement a pair of multi-categorisation, lightweight, convolutional neural networks (CNN), one for thermal features and one for horizontal wind features. Here, as in the paper, we focus our discussion on model designed for thermal feature detection.\\

This model follows the structure shown in \autoref{fig:CNN} (and Figure 2 of \citealt{DeepLearningforSpatiallyExplicitPredictionofSynopticScaleFronts}). After input, the image data is analysed and reduced by a series of 4 convolution blocks of decreasing dimensionality and increasing complexity, as shown in \autoref{fig:Ai_flowchart}. Each convolution block consists of three layer: the 2D convolution layer itself, which has a kernel size of $5\times5$ and includes 16/32/64/64 filters in the first/second/third/forth block respectively, a 2D max pooling layer which downsamples the data, halving the resolution in both dimensions (i.e. latitudinally and longitudinally), and finally a dropout layer, which is only active during training, and which sets a fraction of the input arrays values to 0 (rescaling the remainder of the array), in order to reduce over-fitting. During the convolution process, this fraction is set to 0.25, whilst in the fully connected layers the fraction is increased to 0.5. Note that, in the convolution layers, the $5\times5$ kernel represents the matrix used to enhance features, i.e. detect edges. This kernel is swept across the entire input moving horizontally and vertically with a stride length of 1, hence reducing the dimensionality of the input by 4 in both the horizontal and vertical direction (since we do not include any padding). The exact form that this kernel matrix takes is a result of the training process undertaken, with the kernel being optimised to recover the feature of interest. Of course since we are looking for multiple features and since these features vary spatially, running a single kernel per convolution layer would be highly inefficient. Instead we consider and train multiple kernels per layer, with the number of kernels, referred to as the number of filters increasing as the dimensionality (and hence size) of the data set decreases: i.e. 16 filters in out first convolution block and 64 in the last.  \\
Once the convolution blocks have processed the data, and the total dimensionality has been reduced to $7\times18$ with 64 filters (i.e. $7\times18\times64$), the data-set is small enough that, after flattening (so that the dense layers connect all points), it can be fed into a series of `low-resolution' fully-connected blocks which consist of fully-connected dense layers, followed by over-fitting reducing dropout layers. The final dense layer returns 8 values corresponding to the probability of detection of each the trained features. This process is repeated for every pressure level and every time-averaged point in order to generate the multi-categorisation maps shown throughout this paper. It is important to note that, even after the dimensionality reduction associated with the convolution process, the first dense layer which fully connects the 8064 points in the flattened output of the convolution process with only 128 output points contains over a million weights, although this is small compared with the almost 25 million weights that would be required to fully-connect the initial image with a similar sized output (a process which is likely to lead to spurious outputs due to the massive single-step decrease in dimensionality from 194,000 to 128 points!).\\
In addition to the above it is import to note, for reproducibility, that: a) other than the final fully-connected dense layer (which uses a sigmoid function to generate the final probabilities), all the neural-network layers considered here include rectified linear unit (ReLU) activation, which is commonly used in CNNs, is believed to improve the efficiency of deep-learning \citep{10.5555/3104322.3104425}, and which essentially works by zeroing out any negative values in the output of the associated neural-network layer. b) the learning rate of the model was 0.001 and it made use of the 'adam' optimiser. c) the initial resolution of the network discussed here is half that of the input, print quality, image file - however tests with higher initial resolutions where performed, and whilst the accuracy slightly improved, the computational cost significantly ballooned. { And d) the training/validation data split was 80\%/20\% with pseudo-random assignment between the two categories (that is to say we made use of the random state parameter to ensure that the train/test split was consistent between models). } }

\begin{acknowledgements}
\nolinenumbers
F. Sainsbury-Martinez and P. Tremblin would like to acknowledge and thank the ERC for funding this work under the Horizon 2020 program project ATMO (ID: 757858). F. Sainsbury-Martinez would also like to thank UK Research and Innovation for additional support under grant number MR/T040726/1.\\
The authors also wish to thank Idris, CNRS, University Paris-Saclay, and MDLS for access to the supercomputer Ruche, without which the long time-scale calculations featured in this work would not have been possible. Additionally this work was granted access to the HPC resources of IDRIS (Jean-Zay) and CEA-TGCC (Irene/Joliot-Curie) under the 2020/2021 allocation - A0080410870 made as part of the GENCI Dari A8 call.
This work was supported by French government funding managed by the National Research Agency under the Investments for the Future program (PIA) grant ANR-21-ESRE-0030 (CONTINUUM).\\
{ Finally the authors with to thank the referee (and editor) for useful comments, questions, and suggestions which have significantly improved the readability of this manuscript. }
\end{acknowledgements}

\bibliography{papers}{}

\begin{thebibliography}{}
\expandafter\ifx\csname natexlab\endcsname\relax\def\natexlab#1{#1}\fi
\providecommand{\url}[1]{\href{#1}{#1}}
\providecommand{\dodoi}[1]{doi:~\href{http://doi.org/#1}{\nolinkurl{#1}}}
\providecommand{\doeprint}[1]{\href{http://ascl.net/#1}{\nolinkurl{http://ascl.net/#1}}}
\providecommand{\doarXiv}[1]{\href{https://arxiv.org/abs/#1}{\nolinkurl{https://arxiv.org/abs/#1}}}

\bibitem[{Abadi {et~al.}(2015)Abadi, Agarwal, Barham, Brevdo, Chen, Citro,
  Corrado, Davis, Dean, Devin, Ghemawat, Goodfellow, Harp, Irving, Isard, Jia,
  Jozefowicz, Kaiser, Kudlur, Levenberg, Man\'{e}, Monga, Moore, Murray, Olah,
  Schuster, Shlens, Steiner, Sutskever, Talwar, Tucker, Vanhoucke, Vasudevan,
  Vi\'{e}gas, Vinyals, Warden, Wattenberg, Wicke, Yu, \&
  Zheng}]{tensorflow2015-whitepaper}
Abadi, M., Agarwal, A., Barham, P., {et~al.} 2015, {TensorFlow}: Large-Scale
  Machine Learning on Heterogeneous Systems.
\newblock \url{https://www.tensorflow.org/}

\bibitem[{{Adams} {et~al.}(2018){Adams}, {Ford}, {Hambley}, {Hobson}, {Kavcic},
  {Maynard}, {Melvin}, {Mueller}, {Mullerworth}, {Porter}, {Rezny}, {Shipway},
  \& {Wong}}]{2018arXiv180907267A}
{Adams}, S.~V., {Ford}, R.~W., {Hambley}, M., {et~al.} 2018, arXiv e-prints,
  arXiv:1809.07267.
\newblock \doarXiv{1809.07267}

\bibitem[{Albawi {et~al.}(2017)Albawi, Mohammed, \& Al-Zawi}]{8308186}
Albawi, S., Mohammed, T.~A., \& Al-Zawi, S. 2017, in 2017 International
  Conference on Engineering and Technology (ICET), 1--6

\bibitem[{{Amundsen} {et~al.}(2016){Amundsen}, {Mayne}, {Baraffe}, {Manners},
  {Tremblin}, {Drummond}, {Smith}, {Acreman}, \&
  {Homeier}}]{2016A&A...595A..36A}
{Amundsen}, D.~S., {Mayne}, N.~J., {Baraffe}, I., {et~al.} 2016, \aap, 595,
  A36, \dodoi{10.1051/0004-6361/201629183}

\bibitem[{{Beatty} {et~al.}(2017){Beatty}, {Madhusudhan}, {Tsiaras}, {Zhao},
  {Gilliland}, {Knutson}, {Shporer}, \& {Wright}}]{2017AJ....154..158B}
{Beatty}, T.~G., {Madhusudhan}, N., {Tsiaras}, A., {et~al.} 2017, \aj, 154,
  158, \dodoi{10.3847/1538-3881/aa899b}

\bibitem[{{B{\"o}ker} {et~al.}(2022){B{\"o}ker}, {Arribas}, {L{\"u}tzgendorf},
  {Alves de Oliveira}, {Beck}, {Birkmann}, {Bunker}, {Charlot}, {de Marchi},
  {Ferruit}, {Giardino}, {Jakobsen}, {Kumari}, {L{\'o}pez-Caniego}, {Maiolino},
  {Manjavacas}, {Marston}, {Moseley}, {Muzerolle}, {Ogle}, {Pirzkal},
  {Rauscher}, {Rawle}, {Rix}, {Sabbi}, {Sargent}, {Sirianni}, {te Plate},
  {Valenti}, {Willott}, \& {Zeidler}}]{2022A&A...661A..82B}
{B{\"o}ker}, T., {Arribas}, S., {L{\"u}tzgendorf}, N., {et~al.} 2022, \aap,
  661, A82, \dodoi{10.1051/0004-6361/202142589}

\bibitem[{{Castelli} \& {Kurucz}(2003)}]{2003IAUS..210P.A20C}
{Castelli}, F., \& {Kurucz}, R.~L. 2003, in Modelling of Stellar Atmospheres,
  ed. N.~{Piskunov}, W.~W. {Weiss}, \& D.~F. {Gray}, Vol. 210, A20.
\newblock \doarXiv{astro-ph/0405087}

\bibitem[{{Charbonneau} {et~al.}(2000){Charbonneau}, {Brown}, {Latham}, \&
  {Mayor}}]{2000ApJ...529L..45C}
{Charbonneau}, D., {Brown}, T.~M., {Latham}, D.~W., \& {Mayor}, M. 2000, \apjl,
  529, L45, \dodoi{10.1086/312457}

\bibitem[{Cullen \& Brown(2009)}]{doi:10.1098/rsta.2008.0268}
Cullen, M., \& Brown, A. 2009, Philosophical Transactions of the Royal Society
  A: Mathematical, Physical and Engineering Sciences, 367, 2947,
  \dodoi{10.1098/rsta.2008.0268}

\bibitem[{{Deitrick} {et~al.}(2022){Deitrick}, {Heng}, {Schroffenegger},
  {Kitzmann}, {Grimm}, {Malik}, {Mendon{\c{c}}a}, \&
  {Morris}}]{2022MNRAS.512.3759D}
{Deitrick}, R., {Heng}, K., {Schroffenegger}, U., {et~al.} 2022, \mnras, 512,
  3759, \dodoi{10.1093/mnras/stac680}

\bibitem[{Developers(2023)}]{tensorflow_developers_2023_8118033}
Developers, T. 2023, TensorFlow, v2.12.1,  Zenodo,
  \dodoi{10.5281/zenodo.8118033}

\bibitem[{{Drummond} {et~al.}(2016){Drummond}, {Tremblin}, {Baraffe},
  {Amundsen}, {Mayne}, {Venot}, \& {Goyal}}]{2016A&A...594A..69D}
{Drummond}, B., {Tremblin}, P., {Baraffe}, I., {et~al.} 2016, \aap, 594, A69,
  \dodoi{10.1051/0004-6361/201628799}

\bibitem[{Dubos {et~al.}(2015)Dubos, Dubey, Tort, Mittal, Meurdesoif, \&
  Hourdin}]{gmd-8-3131-2015}
Dubos, T., Dubey, S., Tort, M., {et~al.} 2015, Geoscientific Model Development,
  8, 3131, \dodoi{10.5194/gmd-8-3131-2015}

\bibitem[{{Dubos} \& {Voitus}(2014)}]{2014JAtS...71.4621D}
{Dubos}, T., \& {Voitus}, F. 2014, Journal of Atmospheric Sciences, 71, 4621,
  \dodoi{10.1175/JAS-D-14-0080.1}

\bibitem[{Dutton(1986)}]{dutton1986ceaseless}
Dutton, J. 1986, The Ceaseless Wind: An Introduction to the Theory of
  Atmospheric Motion, Dover books on earth sciences (Dover Publications).
\newblock \url{https://books.google.fr/books?id=g7URAQAAIAAJ}

\bibitem[{{Ferruit} {et~al.}(2022){Ferruit}, {Jakobsen}, {Giardino}, {Rawle},
  {Alves de Oliveira}, {Arribas}, {Beck}, {Birkmann}, {B{\"o}ker}, {Bunker},
  {Chariot}, {de Marchi}, {Franx}, {Henry}, {Karakla}, {Kassin}, {Kumari},
  {L{\'o}pez-Caniego}, {L{\"u}tzgendorf}, {Maiolino}, {Manjavacas}, {Marston},
  {Moseley}, {Muzerolle}, {Pirzkal}, {Rauscher}, {Rix}, {Sabbi}, {Sirianni},
  {te Plate}, {Valenti}, {Willott}, \& {Zeidler}}]{2022A&A...661A..81F}
{Ferruit}, P., {Jakobsen}, P., {Giardino}, G., {et~al.} 2022, \aap, 661, A81,
  \dodoi{10.1051/0004-6361/202142673}

\bibitem[{{Fortney} {et~al.}(2008){Fortney}, {Lodders}, {Marley}, \&
  {Freedman}}]{2008ApJ...678.1419F}
{Fortney}, J.~J., {Lodders}, K., {Marley}, M.~S., \& {Freedman}, R.~S. 2008,
  \apj, 678, 1419, \dodoi{10.1086/528370}

\bibitem[{Fukushima \& Miyake(1982)}]{fukushima1982neocognitron}
Fukushima, K., \& Miyake, S. 1982, in Competition and cooperation in neural
  nets (Springer), 267--285

\bibitem[{{Gandhi} \& {Madhusudhan}(2019)}]{2019MNRAS.485.5817G}
{Gandhi}, S., \& {Madhusudhan}, N. 2019, \mnras, 485, 5817,
  \dodoi{10.1093/mnras/stz751}

\bibitem[{Guerlet {et~al.}(2014)Guerlet, Spiga, Sylvestre, Indurain, Fouchet,
  Leconte, Millour, Wordsworth, Capderou, Bézard, \& Forget}]{GUERLET2014110}
Guerlet, S., Spiga, A., Sylvestre, M., {et~al.} 2014, Icarus, 238, 110 ,
  \dodoi{https://doi.org/10.1016/j.icarus.2014.05.010}

\bibitem[{{Guillot} \& {Showman}(2002)}]{2002A&A...385..156G}
{Guillot}, T., \& {Showman}, A.~P. 2002, \aap, 385, 156,
  \dodoi{10.1051/0004-6361:20011624}

\bibitem[{{Hammond} \& {Lewis}(2021)}]{2021PNAS..11822705H}
{Hammond}, M., \& {Lewis}, N.~T. 2021, Proceedings of the National Academy of
  Science, 118, e2022705118, \dodoi{10.1073/pnas.2022705118}

\bibitem[{Hubeny {et~al.}(2003)Hubeny, Burrows, \& Sudarsky}]{Hubeny_2003}
Hubeny, I., Burrows, A., \& Sudarsky, D. 2003, The Astrophysical Journal, 594,
  1011, \dodoi{10.1086/377080}

\bibitem[{{Jakobsen} {et~al.}(2022){Jakobsen}, {Ferruit}, {Alves de Oliveira},
  {Arribas}, {Bagnasco}, {Barho}, {Beck}, {Birkmann}, {B{\"o}ker}, {Bunker},
  {Charlot}, {de Jong}, {de Marchi}, {Ehrenwinkler}, {Falcolini}, {Fels},
  {Franx}, {Franz}, {Funke}, {Giardino}, {Gnata}, {Holota}, {Honnen}, {Jensen},
  {Jentsch}, {Johnson}, {Jollet}, {Karl}, {Kling}, {K{\"o}hler}, {Kolm},
  {Kumari}, {Lander}, {Lemke}, {L{\'o}pez-Caniego}, {L{\"u}tzgendorf},
  {Maiolino}, {Manjavacas}, {Marston}, {Maschmann}, {Maurer}, {Messerschmidt},
  {Moseley}, {Mosner}, {Mott}, {Muzerolle}, {Pirzkal}, {Pittet}, {Plitzke},
  {Posselt}, {Rapp}, {Rauscher}, {Rawle}, {Rix}, {R{\"o}del}, {Rumler},
  {Sabbi}, {Salvignol}, {Schmid}, {Sirianni}, {Smith}, {Strada}, {te Plate},
  {Valenti}, {Wettemann}, {Wiehe}, {Wiesmayer}, {Willott}, {Wright}, {Zeidler},
  \& {Zincke}}]{2022A&A...661A..80J}
{Jakobsen}, P., {Ferruit}, P., {Alves de Oliveira}, C., {et~al.} 2022, \aap,
  661, A80, \dodoi{10.1051/0004-6361/202142663}

\bibitem[{Krizhevsky {et~al.}(2012)Krizhevsky, Sutskever, \&
  Hinton}]{NIPS2012_c399862d}
Krizhevsky, A., Sutskever, I., \& Hinton, G.~E. 2012, in Advances in Neural
  Information Processing Systems, ed. F.~Pereira, C.~Burges, L.~Bottou, \&
  K.~Weinberger, Vol.~25 (Curran Associates, Inc.).
\newblock
  \url{https://proceedings.neurips.cc/paper/2012/file/c399862d3b9d6b76c8436e924a68c45b-Paper.pdf}

\bibitem[{Lagerquist {et~al.}(2019)Lagerquist, McGovern, \&
  II}]{DeepLearningforSpatiallyExplicitPredictionofSynopticScaleFronts}
Lagerquist, R., McGovern, A., \& II, D. J.~G. 2019, Weather and Forecasting,
  34, 1137 , \dodoi{10.1175/WAF-D-18-0183.1}

\bibitem[{LeCun {et~al.}(1989)LeCun, Boser, Denker, Henderson, Howard, Hubbard,
  \& Jackel}]{lecun1989backpropagation}
LeCun, Y., Boser, B., Denker, J.~S., {et~al.} 1989, Neural computation, 1, 541

\bibitem[{Lecun {et~al.}(1998{\natexlab{a}})Lecun, Bottou, Bengio, \&
  Haffner}]{726791}
Lecun, Y., Bottou, L., Bengio, Y., \& Haffner, P. 1998{\natexlab{a}},
  Proceedings of the IEEE, 86, 2278, \dodoi{10.1109/5.726791}

\bibitem[{Lecun {et~al.}(1998{\natexlab{b}})Lecun, Bottou, Bengio, \&
  Haffner}]{yann1998}
---. 1998{\natexlab{b}}, Proceedings of the IEEE, 86, 2278 ,
  \dodoi{10.1109/5.726791}

\bibitem[{{Lee} {et~al.}(2021){Lee}, {Parmentier}, {Hammond}, {Grimm},
  {Kitzmann}, {Tan}, {Tsai}, \& {Pierrehumbert}}]{2021MNRAS.506.2695L}
{Lee}, E. K.~H., {Parmentier}, V., {Hammond}, M., {et~al.} 2021, \mnras, 506,
  2695, \dodoi{10.1093/mnras/stab1851}

\bibitem[{{Lothringer} {et~al.}(2018){Lothringer}, {Barman}, \&
  {Koskinen}}]{2018ApJ...866...27L}
{Lothringer}, J.~D., {Barman}, T., \& {Koskinen}, T. 2018, \apj, 866, 27,
  \dodoi{10.3847/1538-4357/aadd9e}

\bibitem[{Madhusudhan \& Seager(2010)}]{Madhusudhan_2010}
Madhusudhan, N., \& Seager, S. 2010, The Astrophysical Journal, 725, 261,
  \dodoi{10.1088/0004-637x/725/1/261}

\bibitem[{{Maynard} {et~al.}(2020){Maynard}, {Melvin}, \&
  {M{\"u}ller}}]{2020QJRMS.146.3917M}
{Maynard}, C., {Melvin}, T., \& {M{\"u}ller}, E.~H. 2020, Quarterly Journal of
  the Royal Meteorological Society, 146, 3917, \dodoi{10.1002/qj.3880}

\bibitem[{{Mayne} {et~al.}(2014{\natexlab{a}}){Mayne}, {Baraffe}, {Acreman},
  {Smith}, {Wood}, {Amundsen}, {Thuburn}, \& {Jackson}}]{2014GMD.....7.3059M}
{Mayne}, N.~J., {Baraffe}, I., {Acreman}, D.~M., {et~al.} 2014{\natexlab{a}},
  Geoscientific Model Development, 7, 3059, \dodoi{10.5194/gmd-7-3059-2014}

\bibitem[{{Mayne} {et~al.}(2019){Mayne}, {Drummond}, {Debras}, {Jaupart},
  {Manners}, {Boutle}, {Baraffe}, \& {Kohary}}]{2019ApJ...871...56M}
{Mayne}, N.~J., {Drummond}, B., {Debras}, F., {et~al.} 2019, \apj, 871, 56,
  \dodoi{10.3847/1538-4357/aaf6e9}

\bibitem[{{Mayne} {et~al.}(2014{\natexlab{b}}){Mayne}, {Baraffe, I.}, {Acreman,
  D. M.}, {Smith, C.}, {Browning, M. K.}, {Amundsen, D. Sk\aa{}lid}, {Wood,
  N.}, {Thuburn, J.}, \& {Jackson, D. R.}}]{Mayne_2014}
{Mayne}, N.~J., {Baraffe, I.}, {Acreman, D. M.}, {et~al.} 2014{\natexlab{b}},
  A\&A, 561, A1, \dodoi{10.1051/0004-6361/201322174}

\bibitem[{{Miesch}(2005)}]{2005LRSP....2....1M}
{Miesch}, M.~S. 2005, Living Reviews in Solar Physics, 2, 1,
  \dodoi{10.12942/lrsp-2005-1}

\bibitem[{{Molli{\`e}re} {et~al.}(2015){Molli{\`e}re}, {van Boekel},
  {Dullemond}, {Henning}, \& {Mordasini}}]{2015ApJ...813...47M}
{Molli{\`e}re}, P., {van Boekel}, R., {Dullemond}, C., {Henning}, T., \&
  {Mordasini}, C. 2015, \apj, 813, 47, \dodoi{10.1088/0004-637X/813/1/47}

\bibitem[{Nair \& Hinton(2010)}]{10.5555/3104322.3104425}
Nair, V., \& Hinton, G.~E. 2010, in Proceedings of the 27th International
  Conference on International Conference on Machine Learning, ICML'10 (Madison,
  WI, USA: Omnipress), 807–814

\bibitem[{{O'Shea} \& {Nash}(2015)}]{2015arXiv151108458O}
{O'Shea}, K., \& {Nash}, R. 2015, arXiv e-prints, arXiv:1511.08458.
\newblock \doarXiv{1511.08458}

\bibitem[{{Pontoppidan} {et~al.}(2022){Pontoppidan}, {Blome}, {Braun}, {Brown},
  {Carruthers}, {Coe}, {DePasquale}, {Espinoza}, {Garcia Marin}, {Gordon},
  {Henry}, {Hustak}, {James}, {Koekemoer}, {LaMassa}, {Law}, {Lockwood},
  {Moro-Martin}, {Mullally}, {Pagan}, {Player}, {Proffitt}, {Pulliam},
  {Ramsay}, {Ravindranath}, {Reid}, {Robberto}, {Sabbi}, \&
  {Ubeda}}]{2022arXiv220713067P}
{Pontoppidan}, K., {Blome}, C., {Braun}, H., {et~al.} 2022, arXiv e-prints,
  arXiv:2207.13067.
\newblock \doarXiv{2207.13067}

\bibitem[{{Ranjan} {et~al.}(2011){Ranjan}, {Charbonneau}, {Deming}, {Agol},
  {Burrows}, {Clampin}, {Desert}, {Gilliland}, {Knutson}, {Madhusudhan},
  {Mandell}, {Seager}, \& {Showman}}]{2011ESS.....2.1206R}
{Ranjan}, S., {Charbonneau}, D., {Deming}, D., {et~al.} 2011, in AAS/Division
  for Extreme Solar Systems Abstracts, Vol.~2, AAS/Division for Extreme Solar
  Systems Abstracts, 12.06

\bibitem[{Rauscher \& Menou(2010)}]{Rauscher_2010}
Rauscher, E., \& Menou, K. 2010, The Astrophysical Journal, 714, 1334,
  \dodoi{10.1088/0004-637x/714/2/1334}

\bibitem[{{Sainsbury-Martinez} {et~al.}(2021){Sainsbury-Martinez}, {Casewell},
  {Lothringer}, {Phillips}, \& {Tremblin}}]{2021A&A...656A.128S}
{Sainsbury-Martinez}, F., {Casewell}, S.~L., {Lothringer}, J.~D., {Phillips},
  M.~W., \& {Tremblin}, P. 2021, \aap, 656, A128,
  \dodoi{10.1051/0004-6361/202141637}

\bibitem[{{Sainsbury-Martinez} {et~al.}(2019){Sainsbury-Martinez}, {Wang},
  {Fromang}, {Tremblin}, {Dubos}, {Meurdesoif}, {Spiga}, {Leconte}, {Baraffe},
  {Chabrier}, {Mayne}, {Drummond}, \& {Debras}}]{2019A&A...632A.114S}
{Sainsbury-Martinez}, F., {Wang}, P., {Fromang}, S., {et~al.} 2019, \aap, 632,
  A114, \dodoi{10.1051/0004-6361/201936445}

\bibitem[{Sainsbury-Martinez {et~al.}(2023)Sainsbury-Martinez, Tremblin,
  Schneider, Carone, Baraffe, Chabrier, Helling, Decin, \&
  Jørgensen}]{10.1093/mnras/stad1905}
Sainsbury-Martinez, F., Tremblin, P., Schneider, A.~D., {et~al.} 2023, Monthly
  Notices of the Royal Astronomical Society, stad1905,
  \dodoi{10.1093/mnras/stad1905}

\bibitem[{{Schneider} {et~al.}(2022){Schneider}, {Carone}, {Decin},
  {J{\o}rgensen}, \& {Helling}}]{2022A&A...666L..11S}
{Schneider}, A.~D., {Carone}, L., {Decin}, L., {J{\o}rgensen}, U.~G., \&
  {Helling}, C. 2022, \aap, 666, L11, \dodoi{10.1051/0004-6361/202244797}

\bibitem[{{Selvaraju} {et~al.}(2016){Selvaraju}, {Cogswell}, {Das}, {Vedantam},
  {Parikh}, \& {Batra}}]{2016arXiv161002391S}
{Selvaraju}, R.~R., {Cogswell}, M., {Das}, A., {et~al.} 2016, arXiv e-prints,
  arXiv:1610.02391, \dodoi{10.48550/arXiv.1610.02391}

\bibitem[{Showman {et~al.}(2008)Showman, Cooper, Fortney, \&
  Marley}]{Showman_2008}
Showman, A.~P., Cooper, C.~S., Fortney, J.~J., \& Marley, M.~S. 2008, The
  Astrophysical Journal, 682, 559, \dodoi{10.1086/589325}

\bibitem[{{Showman} \& {Polvani}(2011)}]{2011ApJ...738...71S}
{Showman}, A.~P., \& {Polvani}, L.~M. 2011, \apj, 738, 71,
  \dodoi{10.1088/0004-637X/738/1/71}

\bibitem[{{Showman} {et~al.}(2020){Showman}, {Tan}, \&
  {Parmentier}}]{2020SSRv..216..139S}
{Showman}, A.~P., {Tan}, X., \& {Parmentier}, V. 2020, \ssr, 216, 139,
  \dodoi{10.1007/s11214-020-00758-8}

\bibitem[{Spiegel {et~al.}(2009)Spiegel, Silverio, \& Burrows}]{Spiegel_2009}
Spiegel, D.~S., Silverio, K., \& Burrows, A. 2009, The Astrophysical Journal,
  699, 1487, \dodoi{10.1088/0004-637x/699/2/1487}

\bibitem[{{Tremblin} {et~al.}(2015){Tremblin}, {Amundsen}, {Mourier},
  {Baraffe}, {Chabrier}, {Drummond}, {Homeier}, \&
  {Venot}}]{2015ApJ...804L..17T}
{Tremblin}, P., {Amundsen}, D.~S., {Mourier}, P., {et~al.} 2015, \apjl, 804,
  L17, \dodoi{10.1088/2041-8205/804/1/L17}

\bibitem[{{Tremblin} {et~al.}(2017){Tremblin}, {Chabrier}, {Mayne}, {Amundsen},
  {Baraffe}, {Debras}, {Drummond}, {Manners}, \&
  {Fromang}}]{2017ApJ...841...30T}
{Tremblin}, P., {Chabrier}, G., {Mayne}, N.~J., {et~al.} 2017, Astrophysical
  Journal, 841, 30

\bibitem[{{Tsai} {et~al.}(2014){Tsai}, {Dobbs-Dixon}, \&
  {Gu}}]{2014ApJ...793..141T}
{Tsai}, S.-M., {Dobbs-Dixon}, I., \& {Gu}, P.-G. 2014, \apj, 793, 141,
  \dodoi{10.1088/0004-637X/793/2/141}

\bibitem[{Vallis(2006)}]{Vallis17}
Vallis, G.~K. 2006, Atmospheric and Oceanic Fluid Dynamics: Fundamentals and
  Large-Scale Circulation, 2nd edn. (Cambridge, U.K.: Cambridge University
  Press), 946

\bibitem[{Williamson(2007)}]{WILLIAMSON2007}
Williamson, D.~L. 2007, Journal of the Meteorological Society of Japan. Ser.
  II, 85B, 241, \dodoi{10.2151/jmsj.85B.241}

\bibitem[{{Zahnle} {et~al.}(2009){Zahnle}, {Marley}, {Freedman}, {Lodders}, \&
  {Fortney}}]{2009ApJ...701L..20Z}
{Zahnle}, K., {Marley}, M.~S., {Freedman}, R.~S., {Lodders}, K., \& {Fortney},
  J.~J. 2009, \apjl, 701, L20, \dodoi{10.1088/0004-637X/701/1/L20}

\end{thebibliography}
\bibliographystyle{aasjournal}

\end{document}